\documentclass[aps,prl,twocolumn,superscriptaddress]{revtex4-1}

\makeatletter
\def\@fnsymbol#1{\ensuremath{\ifcase#1\or \dagger\or *\or \ddagger\or
   \mathsection\or \mathparagraph\or \|\or **\or \dagger\dagger
   \or \ddagger\ddagger \else\@ctrerr\fi}}
\makeatother

\usepackage{dcolumn}
\usepackage{amsfonts}
\usepackage{amsmath}
\usepackage{amssymb}
\usepackage{amsthm}
\usepackage{mathrsfs}
\usepackage{amsbsy}
\usepackage{amstext}
\usepackage{siunitx}
\usepackage{fixltx2e}
\usepackage{upgreek,xspace}
\usepackage{multirow}
\usepackage{wasysym}
\usepackage{mathtools}
\usepackage{bbold,bm}
\usepackage{graphicx}
\usepackage[colorlinks=true,citecolor=blue,linkcolor=blue,urlcolor=blue]{hyperref}
\usepackage{bm}
\usepackage{color}
\usepackage[latin1]{inputenc}

\newcommand{\beginsupplement}{%
        \setcounter{table}{0}
        \renewcommand{\thetable}{S\arabic{table}}%
        \setcounter{figure}{0}
        \renewcommand{\thefigure}{S\arabic{figure}}%
        \setcounter{equation}{0}
        \renewcommand{\theequation}{S\arabic{equation}}%
     }

\begin{document}

\title{\Large Exciton-driven antiferromagnetic metal in a \\ correlated van der Waals insulator}

\author{Carina A. Belvin}
\altaffiliation{These authors contributed equally to this work.}
\affiliation{Department of Physics, Massachusetts Institute of Technology, Cambridge, Massachusetts 02139, USA}

\author{Edoardo Baldini}
\altaffiliation{These authors contributed equally to this work.}
\affiliation{Department of Physics, Massachusetts Institute of Technology, Cambridge, Massachusetts 02139, USA}

\author{Ilkem Ozge Ozel}
\affiliation{Department of Physics, Massachusetts Institute of Technology, Cambridge, Massachusetts 02139, USA}

\author{Dan Mao}
\affiliation{Department of Physics, Massachusetts Institute of Technology, Cambridge, Massachusetts 02139, USA}

\author{Hoi Chun Po}
\affiliation{Department of Physics, Massachusetts Institute of Technology, Cambridge, Massachusetts 02139, USA}

\author{Clifford J. Allington}
\affiliation{Department of Physics, Massachusetts Institute of Technology, Cambridge, Massachusetts 02139, USA}

\author{Suhan Son}
\affiliation{Center for Correlated Electron Systems, Institute for Basic Science, Seoul 08826, Korea}
\affiliation{Center for Quantum Materials, Department of Physics and Astronomy, Seoul National University, Seoul 08826, Korea}

\author{Beom Hyun Kim}
\affiliation{Korea Institute for Advanced Study, Seoul 02455, Korea}

\author{Jonghyeon Kim}
\affiliation{Department of Physics, Yonsei University, Seoul 03722, Korea}

\author{Inho Hwang}
\affiliation{Center for Correlated Electron Systems, Institute for Basic Science, Seoul 08826, Korea}
\affiliation{Center for Quantum Materials, Department of Physics and Astronomy, Seoul National University, Seoul 08826, Korea}

\author{Jae Hoon Kim}
\affiliation{Department of Physics, Yonsei University, Seoul 03722, Korea}

\author{Je-Geun Park}
\altaffiliation{Email: jgpark10@snu.ac.kr}
\affiliation{Center for Correlated Electron Systems, Institute for Basic Science, Seoul 08826, Korea}
\affiliation{Center for Quantum Materials, Department of Physics and Astronomy, Seoul National University, Seoul 08826, Korea}

\author{T. Senthil}
\affiliation{Department of Physics, Massachusetts Institute of Technology, Cambridge, Massachusetts 02139, USA}

\author{Nuh Gedik}
\altaffiliation{Email: gedik@mit.edu}
\affiliation{Department of Physics, Massachusetts Institute of Technology, Cambridge, Massachusetts 02139, USA}
	
\date{June 15, 2021}

\begin{abstract}
\noindent \normalsize Collective excitations of bound electron-hole pairs---known as excitons---are ubiquitous in condensed matter, emerging in systems as diverse as band semiconductors, molecular crystals, and proteins. Recently, their existence in strongly correlated electron materials has attracted increasing interest due to the excitons' unique coupling to spin and orbital degrees of freedom. The non-equilibrium driving of such dressed quasiparticles offers a promising platform for realizing unconventional many-body phenomena and phases beyond thermodynamic equilibrium. Here, we achieve this in the van der Waals correlated insulator NiPS$_3$ by photoexciting its newly discovered spin--orbit-entangled excitons that arise from Zhang-Rice states. By monitoring the time evolution of the terahertz conductivity, we observe the coexistence of itinerant carriers produced by exciton dissociation and the long-wavelength antiferromagnetic magnon that coherently precesses in time. These results demonstrate the emergence of a transient metallic state that preserves long-range antiferromagnetism, a phase that cannot be reached by simply tuning the temperature. More broadly, our findings open an avenue toward the exciton-mediated optical manipulation of magnetism.
\end{abstract}

\maketitle

The prototypical correlated system, known as a Mott insulator, possesses a ground state in which the electrons are localized, one per site, due to their strong Coulomb interaction \cite{khomskii2014transition}. Consequently, the electron spins are also localized and order antiferromagnetically at low temperature due to entropy and energy considerations. Starting from this insulating antiferromagnetic ground state, the only phases that can be reached by tuning the temperature are a paramagnetic insulator or a paramagnetic metal. Owing to their inherently intertwined degrees of freedom, correlated materials can display excitons that, unlike their analog in band insulators, couple to spin and orbital degrees of freedom \cite{kim2014excitonic,warzanowski2020multiple} or form more exotic composite quasiparticles \cite{jeckelmann2003optical}. Therefore, driving these excitons with light offers a playground for exploring many-body phenomena beyond the physics of band insulators, especially the promise of reaching new phases of matter.

\begin{figure*}[t!]
\includegraphics[width=0.75\textwidth]{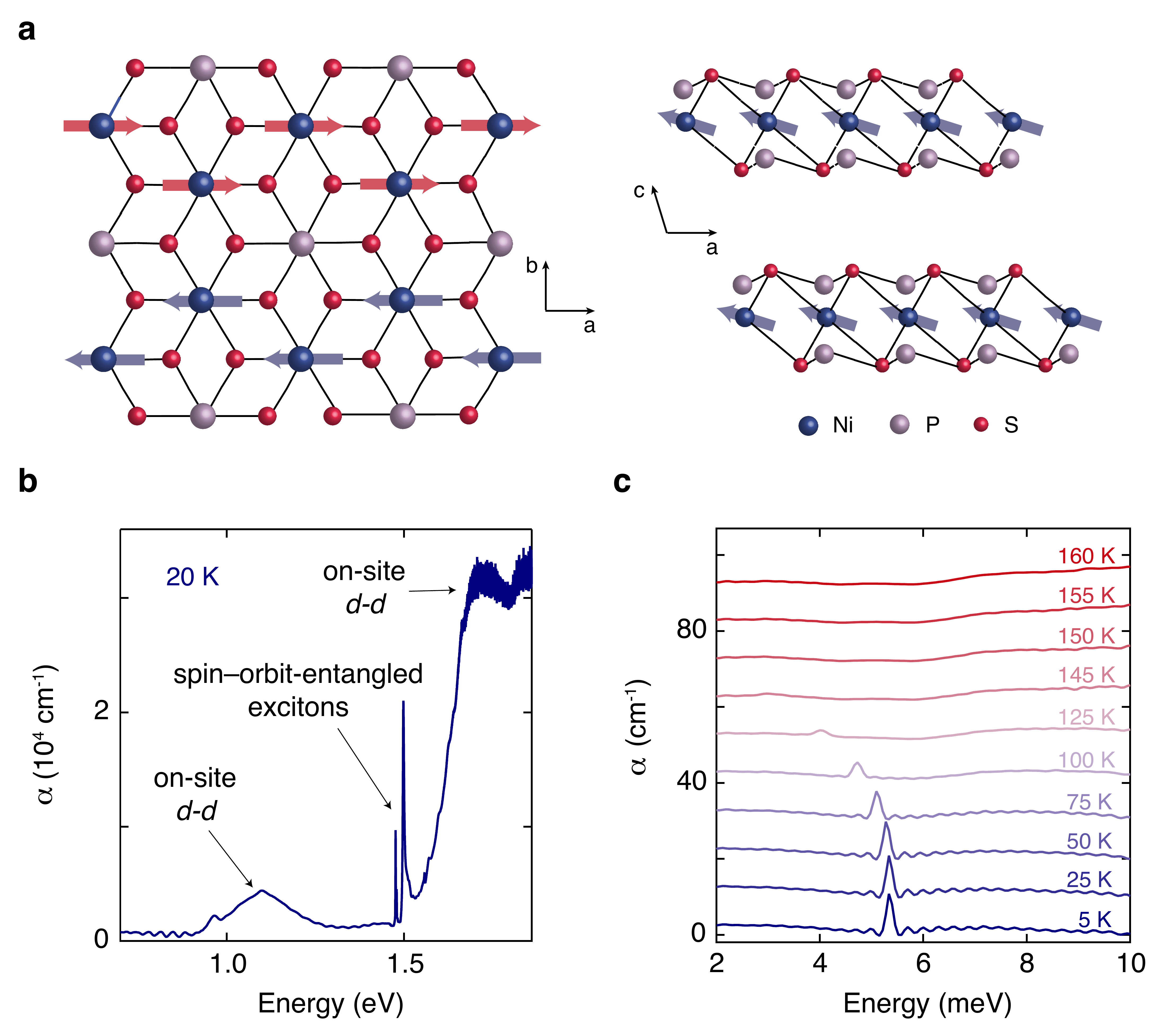}
\caption{\textbf{Crystal structure and optical properties of NiPS$_3$.} \textbf{a}, Crystal and magnetic structure of NiPS$_3$. The Ni atoms (blue spheres) are arranged in a two-dimensional honeycomb lattice in the $ab$-plane. The magnetic moments (red and blue arrows), which point mostly along the $a$-direction, form a zigzag antiferromagnetic pattern with weak ferromagnetic coupling in the $c$-direction. \textbf{b}, Optical absorption ($\alpha$) of NiPS$_3$ below the charge-transfer gap at 20~K. The features around 1.5~eV are the spin--orbit-entangled exciton transitions reported in Ref. \cite{kang2020coherent}. The broad structures around 1.1 and 1.7~eV are on-site $d$-$d$ transitions. The data is adapted from Ref. \cite{kang2020coherent}. \textbf{c}, Absorption spectrum in the THz range at various temperatures. The traces are offset vertically by 10~cm$^{-1}$ for clarity. The low-temperature curves show an absorption peak corresponding to the lowest-energy magnon resonance. The magnon energy softens with increasing temperature and the mode disappears above $T_N$ = 157~K.}
\label{fig:Fig1}
\end{figure*}

One example that holds intriguing possibilities for exploring the effects of a non-equilibrium population of excitons in a correlated insulator is the recently discovered spin--orbit-entangled excitons in the van der Waals antiferromagnet NiPS$_3$ \cite{kang2020coherent}. In this compound, the Ni atoms are arranged in a two-dimensional honeycomb lattice and their magnetic moments exhibit a zigzag antiferromagnetic order below the N\'eel temperature $T_N=157$ K \cite{wildes2015magnetic} (Fig.~\ref{fig:Fig1}a). Optically, NiPS$_3$ has a charge-transfer gap of $\sim\,$1.8~eV \cite{kim2018charge} at low temperature. Below this charge excitation lies a rich spectrum of sub-gap absorption resonances (Fig.~\ref{fig:Fig1}b), including on-site $d$-$d$ transitions around 1.1 and 1.7~eV and the complex of spin--orbit-entangled excitons around 1.5~eV \cite{kang2020coherent}. These excitons are based on Zhang-Rice states \cite{zhang1988effective}, which consist of a hole spin in a localized Ni 3$d$ orbital and a hole spin shared by the 3$p$ orbitals of its surrounding S ligands. Through an extensive characterization of their equilibrium properties using multiple spectroscopic probes, these peculiar excitons were assigned to the transition from a Zhang-Rice triplet to singlet and were found to exhibit intrinsic coupling to the long-range antiferromagnetic order \cite{kang2020coherent}. This inherent connection between the excitons and antiferromagnetism presents fascinating prospects for manipulating both the spins and charge carriers of the system upon non-equilibrium driving, an avenue that has not been explored to date.

Here, we photoexcite NiPS$_3$ close to its spin--orbit-entangled exciton transitions well below the charge-transfer gap. By monitoring the low-energy electrodynamics of the system in the terahertz (THz) range, we reveal the presence of an itinerant conductivity due to the dissociation of excitons into mobile carriers, accompanied by the excitation of a long-wavelength antiferromagnetic spin precession. These observations indicate a transient conducting antiferromagnetic state that cannot be achieved by tuning the temperature of the system.

\section*{Results}
\noindent\textbf{Terahertz probing of electronic and magnetic dynamics.} To probe the antiferromagnetic order in NiPS$_3$, we focus on its lowest-energy magnon at the Brillouin zone center \cite{lanccon2018magnetic} whose real-space spin precession is depicted in Supplementary Fig.~\ref{fig:FigS16}. First, we reveal the equilibrium lineshape of this magnon mode in the THz absorption ($\alpha$) (Fig.~\ref{fig:Fig1}c). At 5~K, the magnon has an extremely sharp linewidth of $\sim\,$0.1~meV, which indicates a long damping time. Above $T_N$, the mode is no longer present. Its low-temperature energy of 5.3~meV softens with increasing temperature following an order-parameter-like dependence (violet circles in Fig.~\ref{fig:Fig3}d). 

\begin{figure*}[t!]
\includegraphics[width=0.97\textwidth]{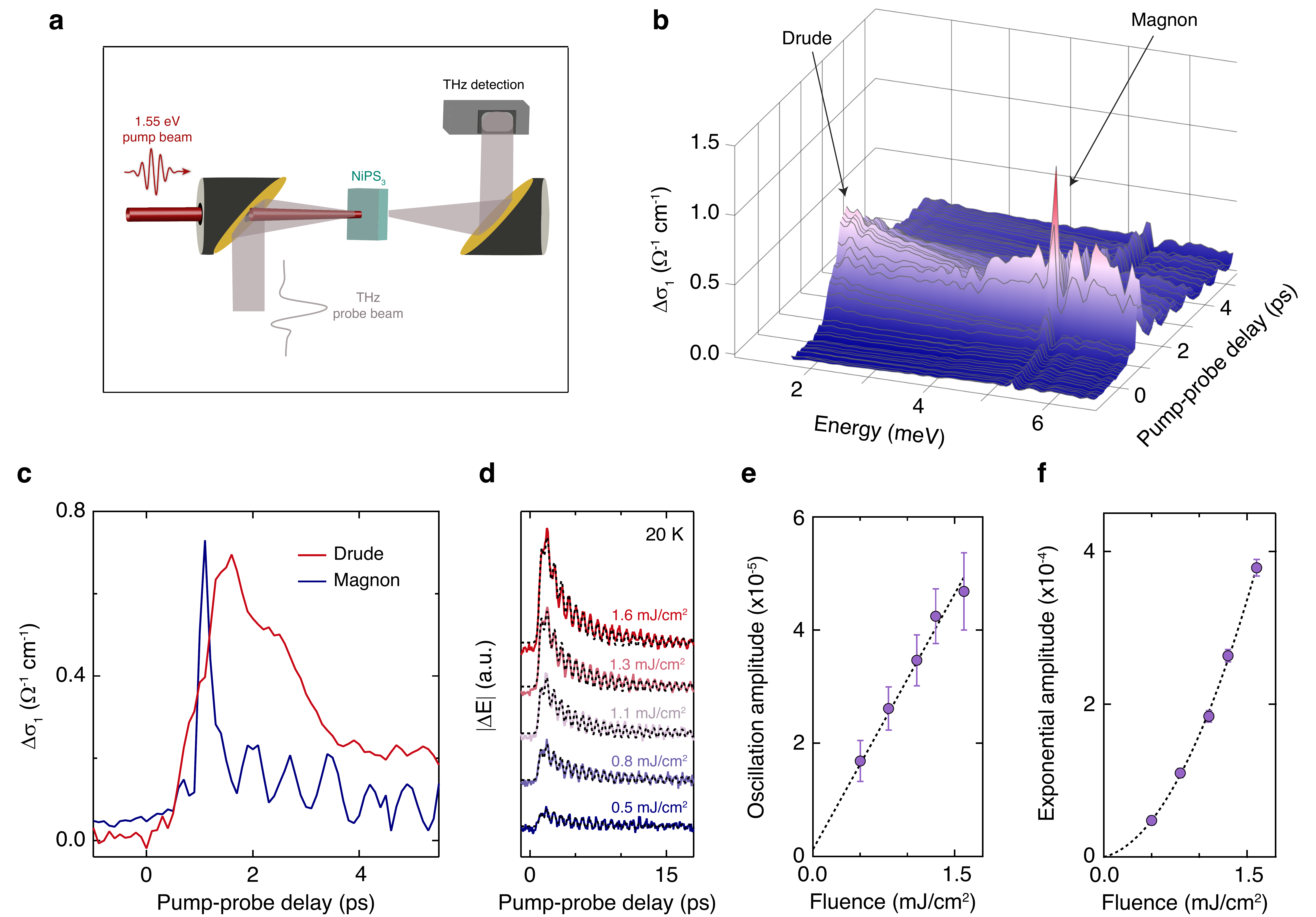}
\caption{\textbf{Observation of an itinerant conductivity and a coherent magnon upon photoexcitation of the spin--orbit-entangled excitons.} \textbf{a}, Schematic of the experimental setup for the THz transmission measurements. \textbf{b}, Spectro-temporal evolution of the pump-induced change in the real part of the optical conductivity ($\Delta\sigma_1$). The temperature is 20~K and the absorbed pump fluence is 1.3~mJ/cm$^2$. The two features present in the data are a Drude-like response at low energies and the first derivative of a Lorentzian lineshape around the magnon energy. \textbf{c}, Temporal evolution of $\Delta\sigma_1$ showing coherent magnon oscillations that begin during the rise of the Drude. \textbf{d}, Temporal evolution of the pump-induced change in the THz electric field ($\Delta E$) of the spectrally-integrated measurement as a function of absorbed pump fluence at 20~K. The black dashed lines are fits to the sum of a damped oscillation and an exponential background. The traces are offset vertically for clarity. \textbf{e}, The amplitude of the oscillation as a function of fluence extracted from the fits. The oscillation amplitude varies linearly with fluence, indicating that it is proportional to the density of photogenerated excitons. \textbf{f}, The amplitude of the exponential as a function of fluence. The quadratic behavior is due to exciton dissociation. The error bars in \textbf{e} and \textbf{f} represent the 95\% confidence interval for the corresponding fit parameters.}
\label{fig:Fig2}
\end{figure*}

\begin{figure*}[t!]
\includegraphics[width=0.92\textwidth]{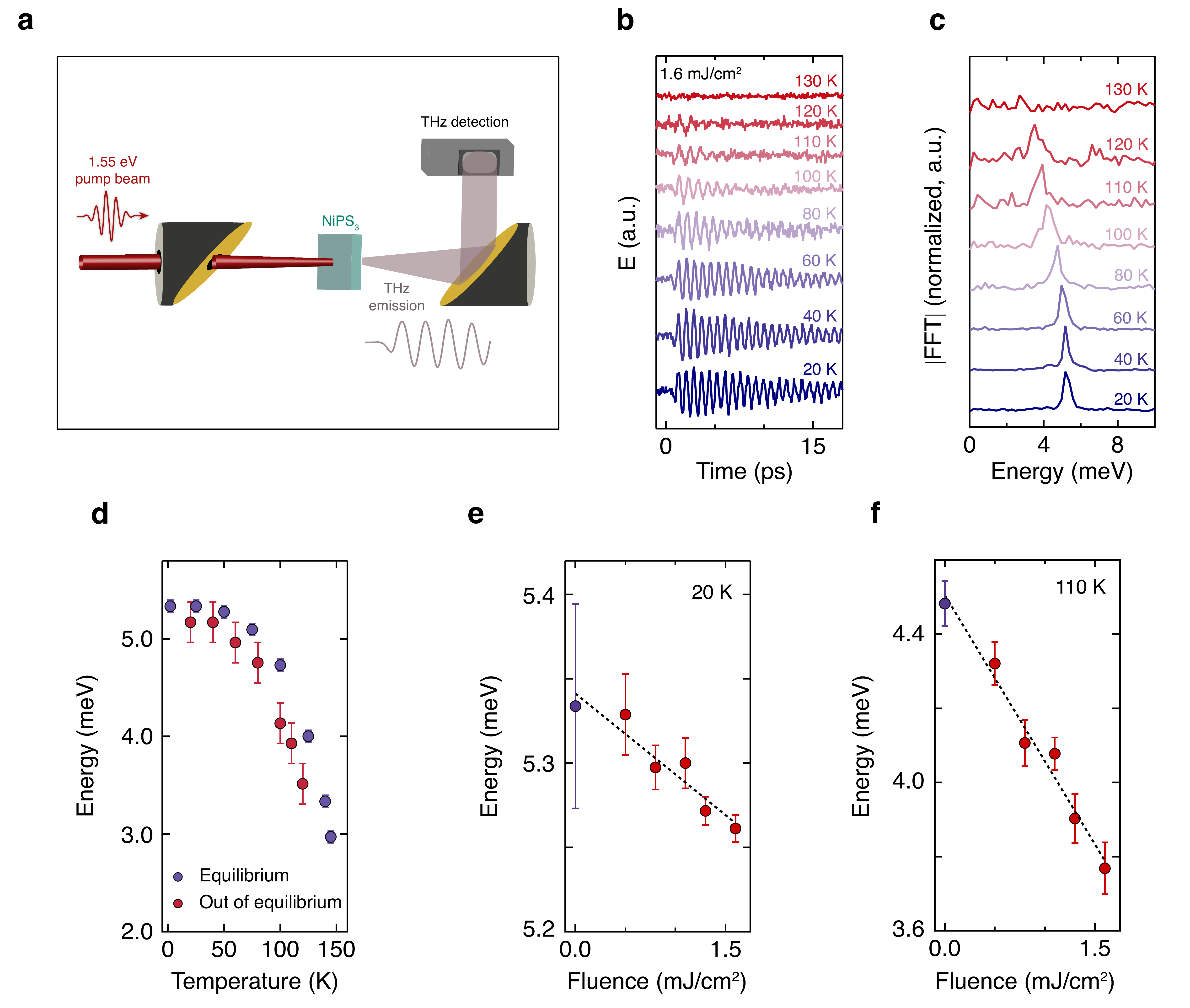}
\caption{\textbf{Evidence for the involvement of the spin--orbit-entangled excitons in the coherent magnon generation.} \textbf{a}, Schematic of the experimental setup for the THz emission measurements. \textbf{b}, Temporal evolution of the emitted THz electric field ($E$) at various temperatures showing coherent magnon oscillations. The absorbed pump fluence is 1.6~mJ/cm$^2$. \textbf{c}, Fourier transform of the traces in \textbf{b}. A softening of the energy is observed as the temperature approaches $T_N$. The traces in \textbf{b} and \textbf{c} are offset vertically for clarity. \textbf{d}, Comparison of the magnon energies from the measurements performed in equilibrium (Fig.~\ref{fig:Fig1}c) and out of equilibrium (\textbf{c}), revealing that the energy of the driven system is redshifted. The error bars for the violet points represent the energy resolution of the time-domain THz experiment, and those for the red points indicate the resolution of the Fourier transform of the THz emission traces. \textbf{e},\textbf{f}, Absorbed pump fluence dependence of the energy of the magnon oscillations at 20~K (\textbf{e}) and 110~K (\textbf{f}) obtained from fits to the THz emission data (red points). The violet points correspond to the magnon energies in equilibrium at these two temperatures. The energy decreases linearly in both cases. The error bars in \textbf{e} and \textbf{f} denote the 95\% confidence interval of the fits.}
\label{fig:Fig3}
\end{figure*}

Next, we photoexcite the system using a near-infrared laser pulse whose photon energy lies in the spectral region of the spin--orbit-entangled excitons of NiPS$_3$. We simultaneously track the electronic and magnetic degrees of freedom by mapping the change in the low-energy optical conductivity with a weak, delayed THz probe pulse (Fig.~\ref{fig:Fig2}a). Figure~\ref{fig:Fig2}b shows the spectro-temporal evolution of the pump-induced change in the real part of the optical conductivity ($\Delta\sigma_1$) at low temperature. The spectral response contains two features: a Drude-like behavior, indicating the presence of itinerant electronic carriers, and the first derivative of a Lorentzian lineshape around the magnon energy. The positive lobe of the narrow derivative-like shape is at lower energies, which signifies that the magnon energy is slightly redshifted compared to its equilibrium value (by only $\sim\,$2\%). No sizeable broadening of the lineshape occurs, meaning that the collective mode does not evolve into a paramagnon and hence the long-range antiferromagnetic order is preserved \cite{gretarsson2017raman}. In Fig.~\ref{fig:Fig2}c, we compare the temporal evolution of the Drude conductivity and the magnon by integrating the response over their characteristic energy regions (1.6-1.8~meV and 5.2-5.4~meV, respectively). We find that the Drude contribution persists for several picoseconds and the magnon exhibits coherent oscillations at the redshifted energy (due to THz emission, which we discuss later). Notably, these oscillations begin during the rise of the Drude response, indicating that a non-thermal mechanism is responsible for launching the magnon coherently. Additionally, we infer the nature of the carriers participating in the itinerant response by extracting the Drude scattering rate ($\gamma$) and plasma frequency ($\omega_p$) from the conductivity spectrum (see Supplementary Note 4E for additional details). At the peak of the Drude signal (Supplementary Fig.~\ref{fig:FigS10}), we obtain $\gamma = 4$~meV and $\omega_p = 4.7$~meV, which corresponds to a carrier mobility ($\mu$) in the range $\mu \sim 1100-2300$~cm$^2$/(Vs). By tracking the time evolution of the conductivity, we find that the scattering rate remains constant (Supplementary Fig.~\ref{fig:FigS11}). This implies that the carriers are cold and lie around the band edges \cite{zielbauer1996ultrafast}.\\

\begin{figure*}[t!]
\includegraphics[width=0.87\textwidth]{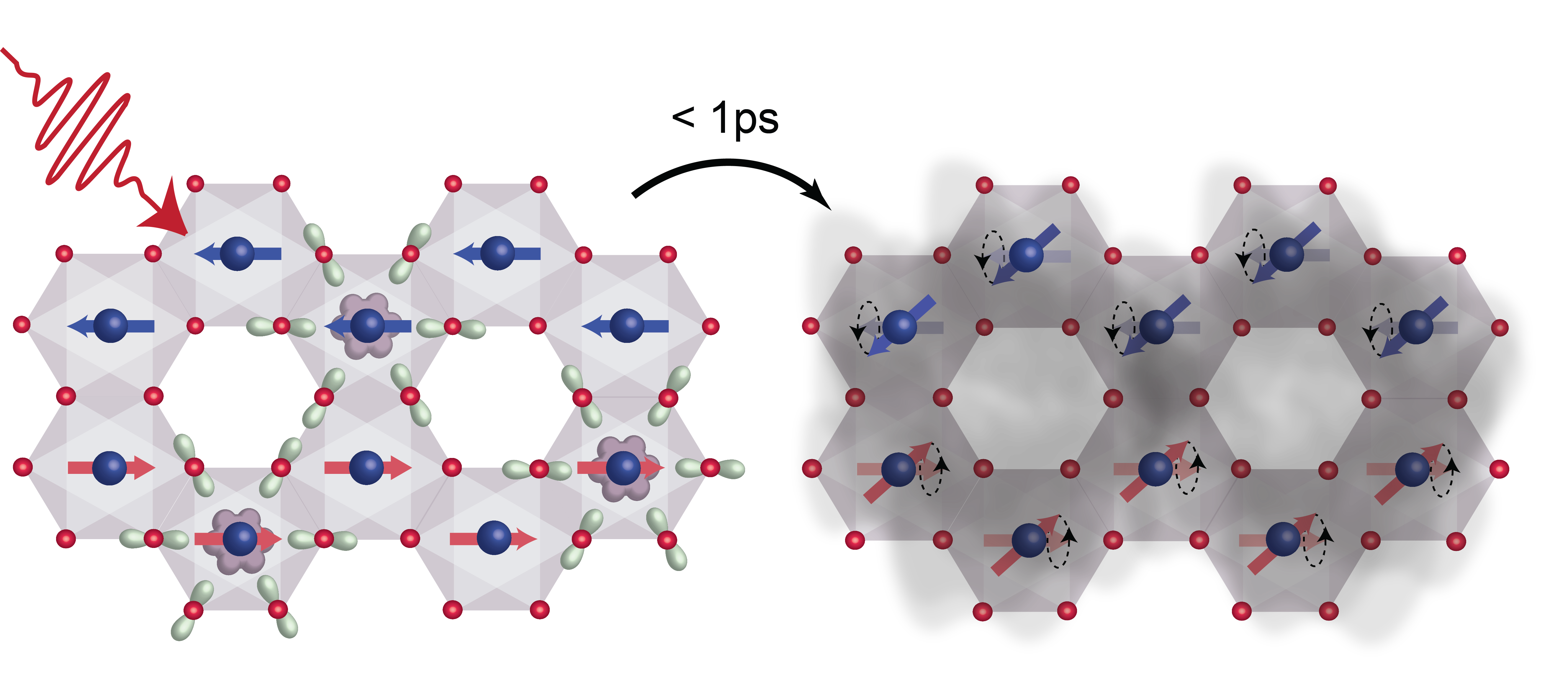}
\caption{\textbf{Cartoon of the photoinduced dynamics.} A near-infrared pump pulse excites the spin--orbit-entangled excitons in NiPS$_3$ (left panel). The purple shape around the Ni atoms (blue spheres) represents the region where the electron density is confined, whereas the light green structures around the S atoms (red spheres) define where the hole density is distributed. These wavefunctions are computed considering the difference in electron density between the ground state with $^3T_{2g}$ symmetry and the excited state with $^1A_{1g}$ symmetry using configuration interaction theory. Through their coupling to the underlying antiferromagnetic order, the excitons launch a coherent spin precession (denoted by the circular motion of the red and blue arrows, right panel). Within 1~ps, the excitons also dissociate into itinerant carriers yielding a conducting state (gray shading) on top of the persisting long-range antiferromagnetic order.}
\label{fig:Fig4}
\end{figure*}

\noindent\textbf{Role of spin--orbit-entangled excitons in the photoinduced state.} In the following, we show that the excitons are solely responsibly for creating this anomalous antiferromagnetic conducting state. First, we examine how the differential signal varies with the absorbed pump laser fluence. In order to maintain a high accuracy, we focus on the spectrally-integrated response by fixing the THz time at the peak of the THz waveform and scanning only the pump delay stage. The pump-induced change in the THz electric field ($\Delta E$) shows the magnon oscillation on top of an exponential relaxation due to the presence of the itinerant carriers (Fig.~\ref{fig:Fig2}d). Fitting each trace to the sum of a damped oscillation and an exponential background (black dashed lines) reveals that the amplitude of the oscillation varies linearly with absorbed fluence (Fig.~\ref{fig:Fig2}e) while the amplitude of the Drude scales quadratically (Fig.~\ref{fig:Fig2}f). The linear scaling of the magnon oscillation amplitude indicates that it is proportional to the density of photoexcited excitons. On the other hand, the quadratic dependence of the Drude response is rather anomalous. Two different phenomena can generate itinerant carriers with such a fluence dependence \cite{castro1971photoconduction}: direct two-photon absorption \cite{boyd2003nonlinear} or exciton dissociation \cite{braun1968singlet,bergman1974photoconductivity,lee1993transient,sun2014observation}. Given that our pump photon energy lies in the vicinity of exciton transitions, two-photon absorption through a virtual state is very unlikely. This is confirmed by the linear scaling of the magnon amplitude on the absorbed laser fluence. Further evidence against two-photon absorption is found by extending our experiments to FePS$_3$ and MnPS$_3$, two other compounds closely related to NiPS$_3$. Pumping below the optical gap but far from any exciton transition yields no photoinduced THz signal (see Supplementary Note 1). Thus, exciton dissociation is the source of the quadratic Drude response in NiPS$_3$. Possible microscopic mechanisms for the dissociation process are annihilation via exciton-exciton interactions \cite{braun1968singlet,lee1993transient,sun2014observation} and pump-induced exciton photoionization \cite{bergman1974photoconductivity}.

Second, we elucidate the ultrafast dynamics of the coherent magnon and establish its connection to the photogenerated spin--orbit-entangled excitons. We detect the THz radiation emitted by the sample in the absence of an incident THz probe (Fig.~\ref{fig:Fig3}a). This signal yields direct signatures of coherent magnons when dipoles in the sample oscillate perpendicular to the direction of light propagation \cite{mikhaylovskiy2015ultrafast}, as is the case here, and is devoid of any electronic response. Figure~\ref{fig:Fig3}b shows the emitted THz electric field ($E$) as a function of time after the pump pulse arrival at various temperatures. The signal consists of a single-frequency oscillation that persists for a long time at low temperature and becomes increasingly more damped as the magnetic transition is approached. A Fourier transform analysis reveals that the oscillation energy softens towards $T_N$ (Fig.~\ref{fig:Fig3}c), confirming the magnetic origin of this coherent collective mode. Comparing this scaling of the magnon energy to that of the equilibrium case at all temperatures reveals that the energy in the driven system is redshifted (Fig.~\ref{fig:Fig3}d), which corroborates our finding in Fig.~\ref{fig:Fig2}b. Furthermore, the phase of the oscillation is independent of the linear or circular polarization state of the pump (see Supplementary Fig.~\ref{fig:FigS6} and Supplementary Note 4A), ruling out a conventional mechanism for coherent magnon generation based on impulsive stimulated Raman scattering \cite{kirilyuk2010ultrafast}. Fixing the temperature, we examine the dependence of the coherent magnon energy on the absorbed pump fluence. The results are shown in Figs.~\ref{fig:Fig3}e and~\ref{fig:Fig3}f for 20~K and 110~K, respectively. We observe that the magnon energy decreases linearly as a function of fluence, with the data points at 110~K exhibiting a steeper slope due to the more fragile nature of the antiferromagnetic order closer to $T_N$. This linear dependence on laser fluence suggests that the magnon energy is also proportional to the density of photogenerated excitons. Since the redshift is non-thermal in nature (see Supplementary Fig.~\ref{fig:FigS7} and Supplementary Note 4B), the excitons are responsible for the coherent magnon dynamics. We remark that the lack of broadening of the magnon lineshape (Fig.~\ref{fig:Fig2}b) as well as the scaling of the magnon amplitude with absorbed fluence (Figs.~\ref{fig:Fig2}e) rules out the possibility that the signal originates from the nucleation of conducting patches within the antiferromagnetic insulating background \cite{battisti2017universality,gretarsson2017raman} (see Supplementary Note 6 for details regarding the scenario of phase separation).

\section*{Discussion}
Combining all the observables monitored by our experiment, we can now describe the ultrafast dynamics following photoexcitation of the spin--orbit-entangled excitons in NiPS$_3$ (Fig.~\ref{fig:Fig4}). Upon light absorption, excitons are created (the calculated exciton wavefunction is depicted in purple and green) and their strong coupling to the antiferromagnetic background is key to launching long-wavelength coherent magnon oscillations. Simultaneously, the excitons dissociate into mobile carriers with small excess energies. This results in a homogeneous itinerant conductivity while the underlying long-range antiferromagnetic order is preserved. Such a direct observation of both mobile carriers and long-range spin correlations is uniquely possible in our time-resolved THz experiment due to the concomitant mapping of the Drude response and the zone-center magnon lineshape with high energy resolution. 

This emergence of a photoinduced conducting antiferromagnetic state is highly unusual and has not previously been reported in an undoped Mott insulator, where such a phase does not exist at any temperature in equilibrium. Our photoexcitation scheme of pumping below-gap spin--orbit-entangled excitons also profoundly differs from one that relies on photodoping hot particle-hole pairs across the Mott gap. In the latter case, the injected quasiparticles release their excess energy through the emission of hot optical phonons and high-frequency magnons, which leads to the melting of long-range antiferromagnetic correlations while transient metallicity is established \cite{dean2016ultrafast,afanasiev2019ultrafast,yang2020ultrafast}. This quench of the long-range magnetic order proceeds through the destruction of the local moments, the reduction of the effective exchange interaction, or the thermal transfer of energy from the electronic carriers and phonons to the spins \cite{balzer2015nonthermal}. In contrast, the spin--orbit-entangled excitons in NiPS$_3$ display an extremely narrow linewidth \cite{kang2020coherent}, implying that they retain a high degree of coherence for at least several picoseconds. Also, as mentioned above, our data show that the free carriers generated by exciton dissociation are cold before becoming trapped at impurity centers. All of these phenomena prevent the transfer of energy from hot optical phonons to the magnetic order through phonon-spin relaxation channels and thus protect long-range antiferromagnetism. Our results demonstrate the need for the future development of advanced theoretical methods to investigate the detailed interactions of a non-equilibrium population of excitons in a correlated environment. Currently, the study of excitonic interactions in the dynamics of Mott insulators is still in its infancy.

This mechanism of pumping spin--orbit-entangled excitons in NiPS$_3$ and investigating the resulting magnetic and electronic responses offers a powerful protocol for studying many-body exciton physics. Since excitons living in a correlated environment can also experience strong coupling to degrees of freedom other than spin, this mechanism can be extended to realize other exotic types of collective coherent control or novel phases of matter. While in NiPS$_3$ the photoexcitation of excitons coupled to the magnetic order triggers a coherent magnon, in correlated materials with other types of exciton couplings it will lead to the tailored modulation of, for example, orbital order where promising candidates are the Mott-Hubbard excitons of titanates \cite{gossling2008mott} and vanadates \cite{novelli2012ultrafast}.

\section*{Methods}

\begin{figure}[htb!]
\includegraphics[width=0.86\columnwidth]{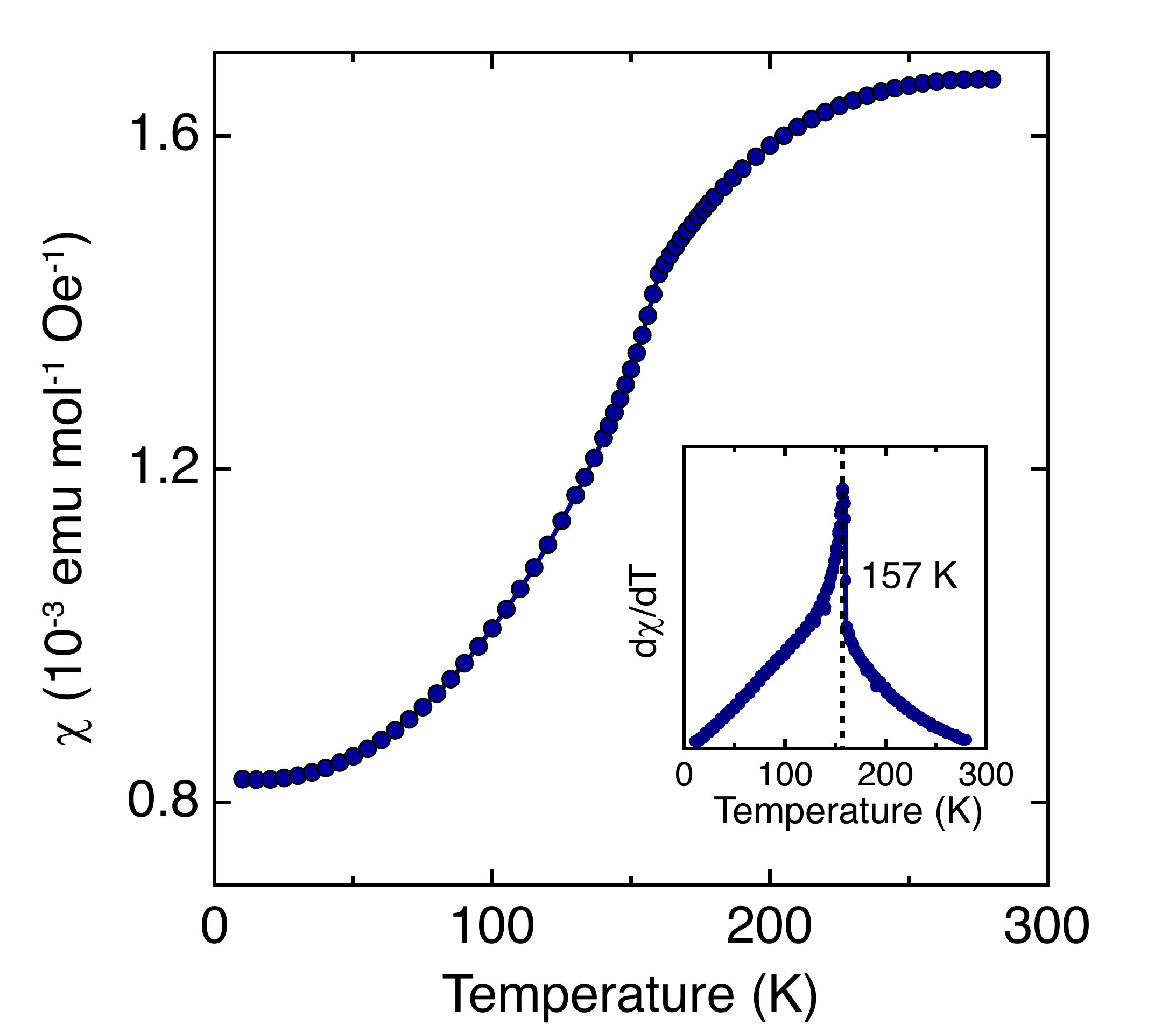}
\caption{\textbf{Magnetic susceptibility of our NiPS$_3$ crystal.} Magnetic susceptibility $\chi$ as a function of temperature. The inset shows the derivative of $\chi$ with respect to temperature, with the anomaly indicating that $T_N \sim$ 157~K for this crystal.}
\label{fig:FigS1}
\end{figure}

\noindent\textbf{Single crystal growth and characterization.} High-quality single crystals of NiPS$_3$, FePS$_3$, and MnPS$_3$ were grown by chemical vapor transport, as described previously \cite{kuo2016exfoliation}. The NiPS$_3$ single crystal used in the THz measurements had a thickness of 1.2~mm and lateral dimensions of $\sim\,$5 by 5~mm. To characterize our NiPS$_3$ crystal, we measured the in-plane magnetic susceptibility as a function of temperature (Supplementary Fig.~\ref{fig:FigS1}). The derivative of the susceptibility reveals that the magnetic transition temperature for our sample is $T_N \sim$ 157~K, which is very similar to that of previous reports \cite{kim2018charge,kim2019suppression}. We also performed heat capacity measurements as a function of temperature (Supplementary Fig.~\ref{fig:FigS2}). Likewise, the heat capacity exhibits an anomaly at $T_N$.\\

\noindent\textbf{Time-domain and ultrafast THz spectroscopy.} We used a Ti:Sapphire amplified laser system that emits 100~fs pulses at a photon energy of 1.55~eV and a repetition rate of 5 kHz to generate THz pulses via optical rectification in a ZnTe crystal. For the time-domain THz spectroscopy experiment, we used electro-optic sampling in a second ZnTe crystal to detect the THz signal transmitted through the sample by overlapping the THz field with a gate pulse at 1.55~eV. We determined the frequency-dependent complex transmission coefficient by comparing the THz electric field through our NiPS$_3$ crystal to that through a reference aperture of the same size (the apertures used for the sample and reference were each 2~mm in diameter). From the measured complex transmission coefficient, the complex optical parameters of NiPS$_3$ were then extracted numerically using the method of Ref. \cite{duvillaret1996reliable}.

For the out-of-equilibrium experiments (THz transmission and THz emission), a fraction of the laser output was used as a pump beam at 1.55~eV. In the THz emission measurements, the THz field spontaneously radiated by the sample upon photoexcitation was subsequently detected using the scheme described above. In contrast, in the THz transmission experiment, a delayed, weak THz probe was transmitted through the photoexcited sample and then detected. The spot size of the pump beam was 4~mm in diameter at the sample position to ensure a uniform excitation of the area probed by the THz pulse ($\sim\,$1~mm in diameter). The time delay between pump and probe (pump delay stage) and the time delay between the THz probe and the gate pulse (gate delay stage) could be varied independently. To measure the spectrally-integrated response, the THz time was fixed at the peak of the THz waveform and the pump delay stage was scanned. The spectrally-resolved measurements were obtained by synchronously scanning the pump and gate delay stages such that the pump-probe delay time remained fixed using the method reported in Ref. \cite{beard2000transient}, in order to avoid any frequency-dependent artifacts in the non-equilibrium spectra. An optical chopper was placed in the pump beam path to allow for the measurement of the THz electric field in the presence/absence of the pump pulse. The data analysis procedure for extracting the pump-induced change in the optical properties of NiPS$_3$ is discussed in Supplementary Note 3.\\

\noindent\textbf{Theoretical calculations.} To characterize the real-space spin precession involved in the lowest-energy magnon mode at the Brillouin zone center, we considered an effective spin model of spin-1 Ni sites with no orbital degeneracy. Assuming the interlayer coupling is small, for each layer the effective spin Hamiltonian consists of $XXZ$ terms up to third-nearest neighbors and single-ion anisotropy along the $a$- and the $c$-axis. We obtained the magnon dispersion relation by applying the Holstein-Primakoff transformation and block-diagonalizing the resulting Hamiltonian. The details of the computation are presented in Supplementary Note 7.

To identify many-body states involved in the spin--orbit-entangled excitons, we performed a configuration interaction calculation of an NiS$_6$ cluster. The calculation included all possible states with $d^8$, $d^9\underline{L}^1$, and $d^{10}\underline{L}^2$ configurations, where \underline{$L$} refers to the ligand S 3$p$-hole orbitals. We took into account the cubic crystal field and the on-site Coulomb interaction for Ni 3$d$ orbitals, which are parameterized with Slater-Condon parameters ($F^0$, $F^2$, $F^4$). We assumed that all the ligand 3$p$ orbitals have the same energy levels characterized by the charge-transfer energy, and the hopping integrals between the Ni 3$d$ and S 3$p$ orbitals were determined with two parameters $V_{pd\sigma}$ and $V_{pd\pi}$ using the Slater-Koster theory. The spatial distribution of the spin--orbit-entangled excitons was computed from the difference in electron density between the ground state with $^3T_{2g}$ symmetry and the excited state with $^1A_{1g}$ symmetry.
\vspace{-5pt}

\section*{Data Availability} 
\vspace{-4pt}
\noindent The data that support the findings of this study are available from the corresponding authors upon reasonable request.

\section*{Acknowledgments}
\noindent We acknowledge useful discussions with Jonathan Pelliciari and Ki Hoon Lee. Work at MIT was supported by the US Department of Energy, BES DMSE, and by the Gordon and Betty Moore Foundation's EPiQS Initiative grant GBMF4540. C.A.B. and E.B. acknowledge additional support from the National Science Foundation Graduate Research Fellowship under Grant No. 1122374 and the Swiss National Science Foundation under fellowships P2ELP2-172290 and P400P2-183842, respectively. D.M. and T.S. are supported by a US Department of Energy grant DE-SC0008739, and T.S., in part, by a Simons Investigator award from the Simons Foundation. H.C.P. is supported by a Pappalardo Fellowship at MIT and a Croucher Foundation Fellowship. Work at IBS-CCES was supported by the Institute for Basic Science (IBS) in Korea (Grant No. IBS-R009-G1) and work at CQM was supported by the Leading Researcher Program of the National Research Foundation of Korea (Grant No. 2020R1A3B2079375). J.H.K. acknowledges support from the National Research Foundation of Korea (NRF) grants funded by the Ministry of Science, ICT, and Future Planning (MSIP) of Korea (NRL Program No. NRF-2019R1I1A2A01062306, SRC Program No. NRF-2017R1A5A1014862). 

\section*{Author contributions}
\noindent C.A.B., E.B., I.O.O., and C.J.A. performed the THz experiments. C.A.B. and E.B. analyzed the experimental data. S.S., I.H., and J.-G.P. grew the single crystals. J.K. and J.H.K. carried out the optical transmission measurements. D.M., H.C.P., and T.S. performed the magnon calculations, and B.H.K. performed the exciton calculations. C.A.B., E.B., J.-G.P., and N.G. wrote the manuscript with crucial input from all other authors. This project was supervised by N.G.

\section*{Competing interests}
\noindent The authors declare no competing interests.

\section*{Materials and correspondence}
\noindent Correspondence and requests for materials should be addressed to N.G. and J.-G.P.

\clearpage
\beginsupplement
\onecolumngrid
\begin{center}
\textbf{\Large Supplementary Information for ``Exciton-driven antiferromagnetic metal in a correlated van der Waals insulator''}
\end{center}\hfill\break
\twocolumngrid

\subsection{\normalsize Supplementary Note 1: MnPS$_3$ and FePS$_3$}

To better understand the behavior observed in NiPS$_3$, it is useful to compare the results on NiPS$_3$ presented in the main text to those of closely related van der Waals antiferromagnets. In this respect, we focus on the compounds MnPS$_3$ and FePS$_3$, whose N\'{e}el temperatures are $\sim\,$78~K and $\sim\,$120~K, respectively. Inelastic neutron scattering measurements show that there are no magnon resonances in our observable energy range in either material (the lowest-energy magnon resonances were found to be $\sim\,$0.5~meV \cite{wildes1998spin} and $\sim\,$15~meV \cite{lanccon2016magnetic} for MnPS$_3$ and FePS$_3$, respectively). We confirm the absence of any magnon modes by measuring the equilibrium THz absorption as a function of temperature.

We further note that neither MnPS$_3$ nor FePS$_3$ shows any pump-probe signal (Drude absorption) when we excite each system with an ultrashort near-infrared pump pulse, in contrast to NiPS$_3$. The lack of a non-equilibrium response can be understood by examining optical absorption data for these compounds. Unlike NiPS$_3$, which exhibits sharp spin--orbit-entangled excitons in the vicinity of our pump photon energy \cite{kang2020coherent}, MnPS$_3$ and FePS$_3$ are purely transparent at 1.55~eV, with on-site $d$-$d$ transitions lying elsewhere and the optical charge-transfer gap at higher energies ($E_{\text{gap}}$ $\sim\,$2.9~eV for MnPS$_3$ and $\sim\,$1.8~eV for FePS$_3$ at low temperature) \cite{brec1979physical,grasso1986optical,banda1986opticalFePS3,grasso1991optical,joy1992optical,gnatchenko2011exciton}.

\begin{figure}[htb!]
\includegraphics[width=0.85\columnwidth]{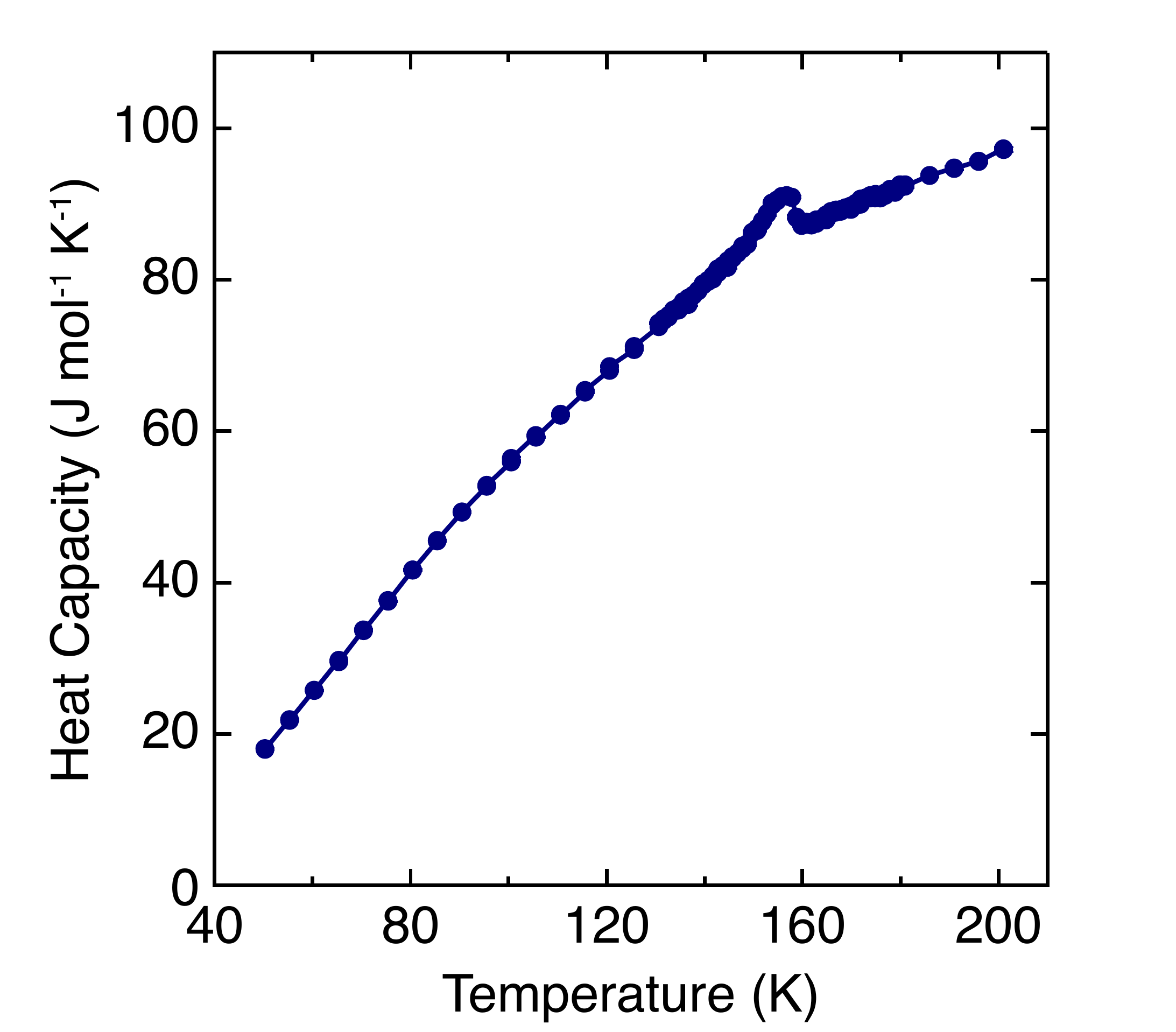}
\caption{\textbf{Heat capacity of our NiPS$_3$ crystal.} Heat capacity as a function of temperature. The curve shows an anomaly at $T_N$ similar to the magnetic susceptibility data.}
\label{fig:FigS2}
\end{figure}

\subsection{\normalsize Supplementary Note 2: Details of the photoexcitation scheme in our ultrafast NiPS$_3$ measurements}

\begin{figure}[htb!]
\includegraphics[width=\columnwidth]{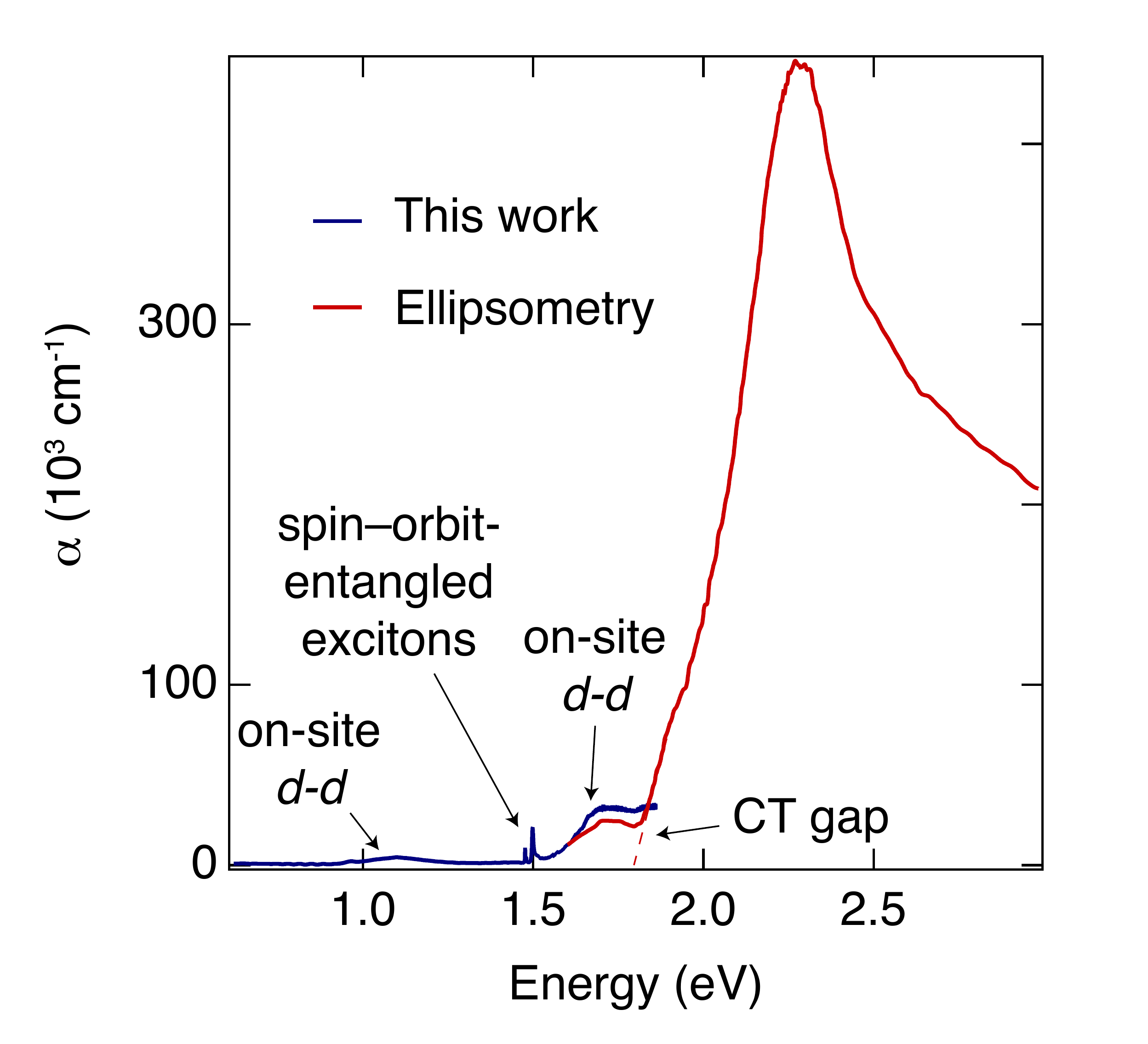}
\caption{\textbf{Optical absorption of NiPS$_3$ over an extended energy range.} Absorption coefficient ($\alpha$) in equilibrium. The blue curve is the absorption data shown in Fig.~\ref{fig:Fig1}b in the main text and the red curve is the absorption coefficient determined from ellipsometry measurements reported in Ref. \cite{kim2018charge}. The former includes the two broad on-site $d$-$d$ transitions around 1.1 and 1.7~eV and the narrow spin--orbit-entangled excitons around 1.5~eV, and the latter shows the energy of the charge-transfer (CT) gap. The absorption plotted over this extended energy range highlights the fact that our pump pulse energy of 1.55~eV lies well below the tail of the CT gap.}
\label{fig:FigS3}
\end{figure}

In our ultrafast THz experiments, we photoexcite NiPS$_3$ in the spectral region of the spin--orbit-entangled excitons. We use a pump photon energy of 1.55~eV, which lies well below the tail of the charge-transfer (CT) gap. To verify this, in Figure~\ref{fig:FigS3} we show the optical absorption of NiPS$_3$ in equilibrium over an extended energy range compared to that of Fig.~\ref{fig:Fig1}b in the main text. The blue curve is the data from Fig.~\ref{fig:Fig1}b, and the red curve is determined from ellipsometry data presented in Ref. \cite{kim2018charge}. This plot of the absorption coefficient demonstrates that our pump photon energy of 1.55~eV is indeed well below the CT gap.

\subsection{\normalsize Supplementary Note 3: Data analysis procedure for the NiPS$_3$ ultrafast spectrally-resolved THz transmission measurement}

In order to fully characterize the low-energy response of NiPS$_3$ to photoexcitation at 1.55~eV, we measure the complete time- and frequency-resolved change in the THz transmission (commonly referred to as a ``two-dimensional (2D) scan''). The experimental details of this technique are presented in the Methods section. Here, we describe our data analysis procedure for extracting the pump-induced changes in the optical properties of NiPS$_3$.

The most general approach for obtaining the complex optical parameters from a THz transmission experiment is the transfer matrix method \cite{furman1992basics,karlsen2019approximations}. This method calculates the transmission coefficient through a multilayer structure as a product of matrices. The transfer matrix of light at normal incidence passing through the $j^{\text{th}}$ layer with complex refractive index $\tilde{n}_j$ and thickness $d_j$ is given by 
\begin{equation}
    M_j = \left(
    \begin{array}{cc}
        \text{cos}\left(\dfrac{\omega \tilde{n}_j d_j}{c}\right) & -\dfrac{i}{\tilde{n}_j}\text{sin}\left(\dfrac{\omega \tilde{n}_j d_j}{c}\right) \\
       -i\tilde{n}_j\text{sin}\left(\dfrac{\omega \tilde{n}_j d_j}{c}\right) & \text{cos}\left(\dfrac{\omega \tilde{n}_j d_j}{c}\right)
    \end{array}
    \right),
\end{equation}
where $c$ is the speed of light. To obtain the total transfer matrix of the system, we take the product of the transfer matrices of each individual layer:
\begin{equation}
M_\text{tot} = \prod_{j=1}^N M_j = \left(
\begin{array}{cc}
m_{11} & m_{12} \\
m_{21} & m_{22}
\end{array}
\right),
\end{equation}
where $N$ is the total number of layers in the structure. We can then derive the transmission coefficient of the system in terms of the matrix elements of $M_\text{tot}$:
\begin{equation}
t = \dfrac{2\tilde{n}_i}{\tilde{n}_im_{11}+\tilde{n}_i\tilde{n}_fm_{12}+m_{21}+\tilde{n}_fm_{22}},
\end{equation}
\noindent where the indices $i$ and $f$ refer to the initial and final layers of the stack.

In our ultrafast THz measurement, we detect the THz electric field transmitted through the photoexcited sample ($E_\text{pump}(t)$) and compare it to that transmitted through the sample in equilibrium ($E_0(t)$) when the pump pulse is blocked by the optical chopper. By taking the difference $\Delta E(t) = E_\text{pump}(t) - E_0(t)$ and performing a Fourier transform, we can relate these measured quantities to the change in the complex transmission coefficient by
\begin{equation}
\dfrac{\Delta E(\omega)}{E_0(\omega)} = \dfrac{t_\text{pump}(\omega)-t_0(\omega)}{t_0(\omega)},
\end{equation}
where $t_\text{pump}(\omega)$ and $t_0(\omega)$ are the transmission coefficients through the photoexcited and equilibrium samples, respectively. From this, we obtain the measured change in the transmission coefficient.

The next step is to calculate the transmission through the photoexcited and equilibrium samples using the transfer matrix method as described above. The photoinduced change in the complex refractive index of the material is expected to decay exponentially in the direction of light propagation $z$ as $\tilde{n}(\omega,z) = \tilde{n}_0(\omega) + \Delta\tilde{n}(\omega)e^{-z/d_p}$, where $\tilde{n}_0(\omega)$ is the index of the sample in equilibrium and $d_p$ is the penetration depth of the pump beam. Therefore, the sample can be partitioned into $N$ homogeneous layers of thickness $d$ with refractive index $\tilde{n}(\omega,z)$ determined by the distance $z$ of each layer. Applying the transfer matrix method, one can calculate the change in the transmission coefficient, and by comparing this to the measured value, the change in the complex refractive index $\Delta \tilde{n}(\omega)=\Delta n(\omega)-i\Delta\kappa(\omega)$ can be extracted for each frequency in the THz probe spectrum. From this quantity, we can obtain the photoinduced changes in other relevant optical parameters such as the absorption coefficient $\Delta\alpha(\omega)$ and the complex optical conductivity $\Delta\sigma(\omega)=\Delta\sigma_1(\omega)+i\Delta\sigma_2(\omega)$.

\begin{figure}[htb!]
\includegraphics[width=0.8\columnwidth]{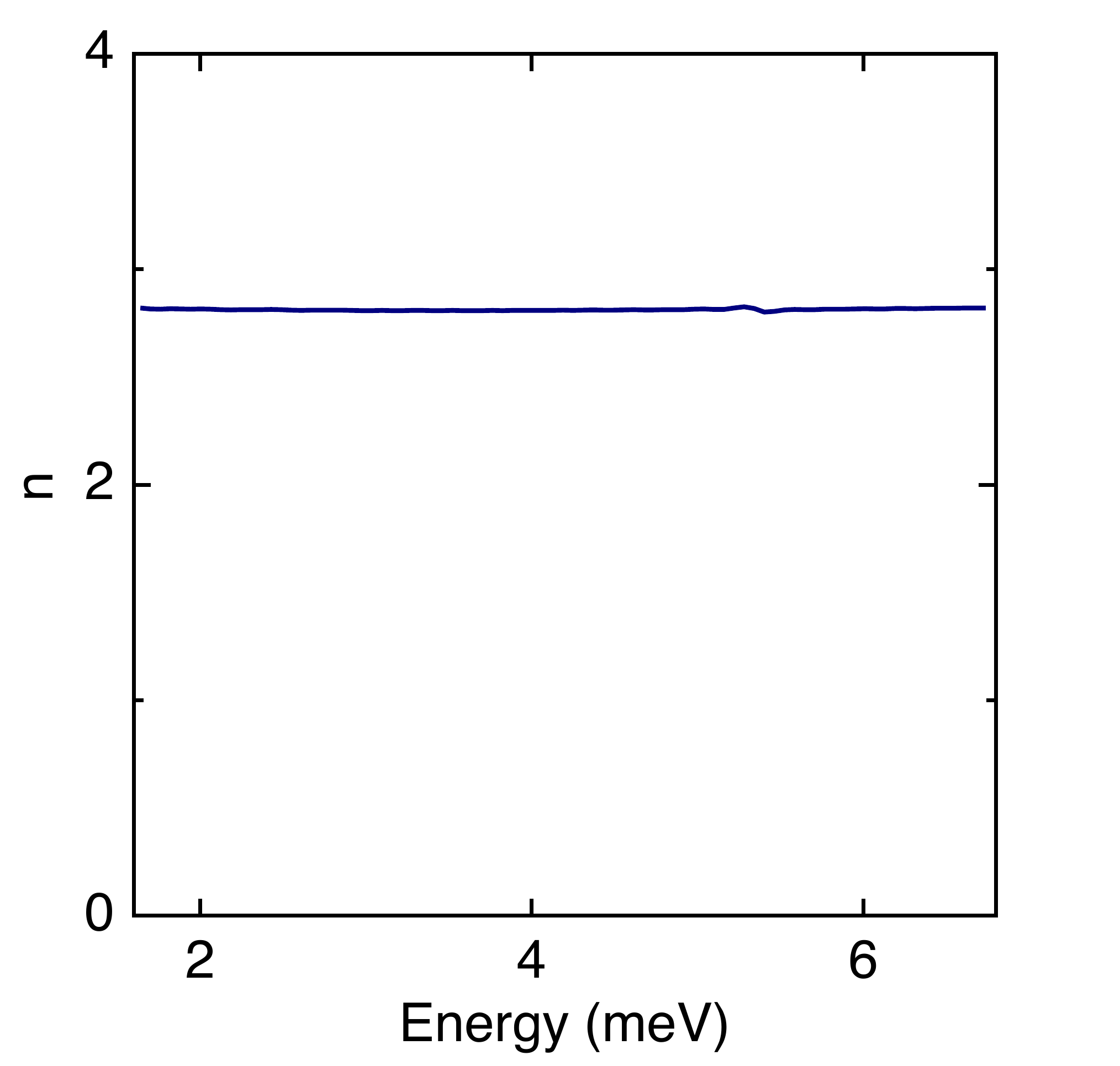}
\caption{\textbf{Equilibrium refractive index of NiPS$_3$.} Real part of the refractive index in equilibrium as a function of frequency measured by time-domain THz spectroscopy. The index shows almost no frequency dependence throughout the measured spectrum.}
\label{fig:FigS4}
\end{figure}

\begin{figure}[htb!]
\includegraphics[width=1.1\columnwidth]{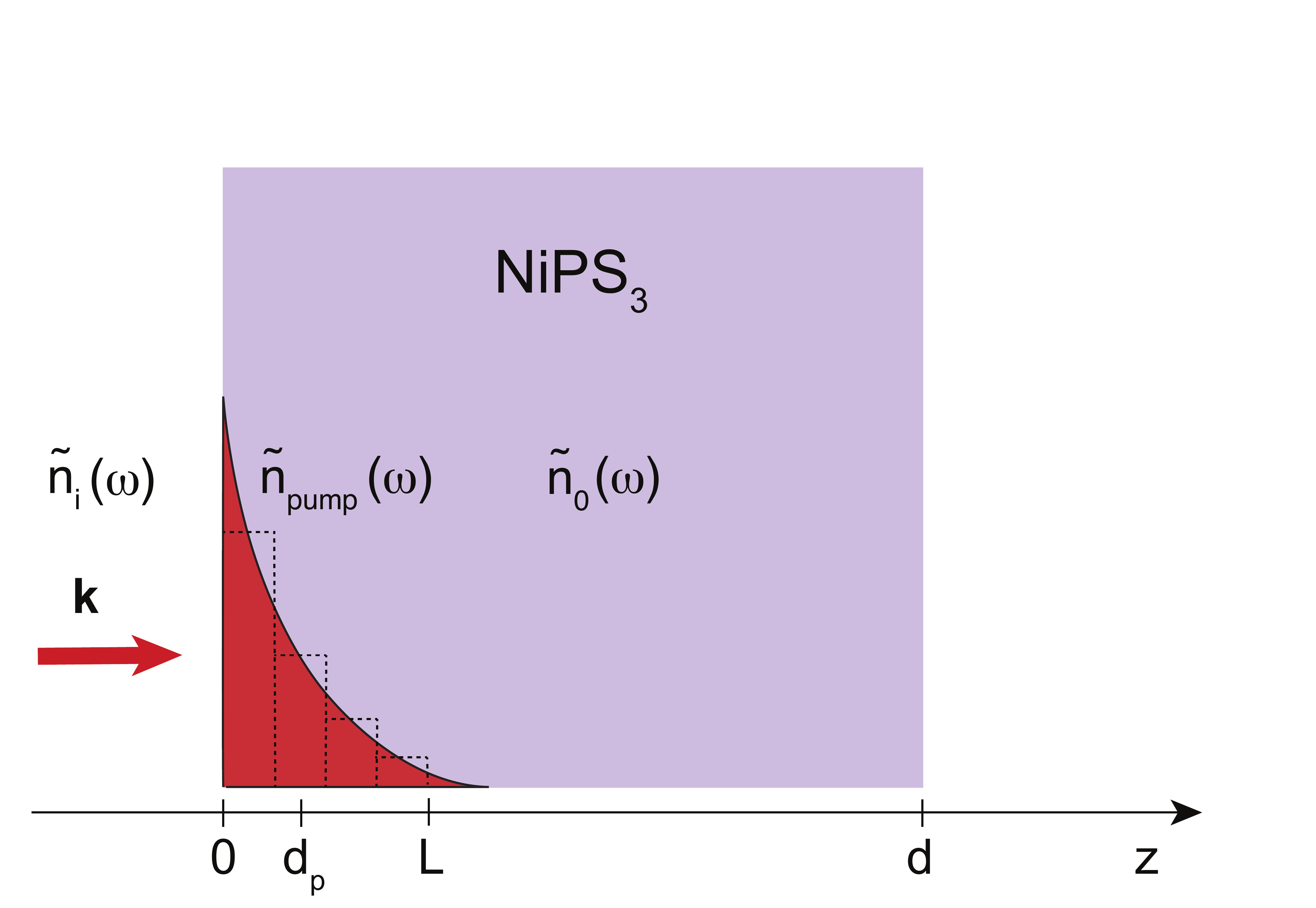}
\caption{\textbf{Illustration of the length scales involved in our ultrafast THz transmission experiment.} The sample thickness is $d=$ 1.2~mm and the penetration depth of the 1.55~eV pump pulse is $d_p=$ 4.28~$\mu$m. We use a total thickness $L=3d_p$ to represent the photoexcited region, which is partitioned into 4 layers (black dashed rectangles) and has a complex refractive index $\tilde{n}_\text{pump}(\omega)$. The complex index of the sample in equilibrium is denoted by $\tilde{n}_0(\omega)$. The light propagates through the multilayer stack in the $z$ direction at normal incidence starting from air ($\tilde{n}_i(\omega)=1$).}
\label{fig:FigS5}
\end{figure}

In the case of our ultrafast THz transmission experiment on NiPS$_3$, the penetration depth of the 1.55~eV pump pulse is $d_p=$ 4.28~$\mu$m, whereas the sample thickness is $d=$1.2~mm and the penetration depth of the THz probe beam is $>\,$2~mm for all frequencies in our THz spectrum. A common approximation that is made in ultrafast THz transmission measurements is to treat the photoexcited region as a single, homogeneous layer of thickness $d_p$ \cite{beard2001subpicosecond,ulbricht2011carrier}. However, a recent study that analyzed various approximations used in ultrafast THz measurements found that one should be cautious when applying such approximations as the calculated optical quantities can deviate significantly from the actual ones \cite{karlsen2019approximations}. Therefore, we analyzed our data with the full transfer matrix method by partitioning the photoexcited region into multiple layers. We used a total thickness of $L=3d_p=$ 12.84~$\mu$m (corresponding to the distance at which the photoinduced change in the index drops to $e^{-3}$, i.e. 0.05 times its initial value) to represent the photoexcited region rather than the entire sample thickness, as the sample thickness is much larger than $d_p$ and it becomes computationally costly to perform the transfer matrix calculation for a large number of layers. Figure~\ref{fig:FigS5} depicts a diagram of our sample dimensions including the photoexcited region (with index $\tilde{n}_{pump}(\omega)$) and the unexcited part of the sample (with index $\tilde{n}_0(\omega)$, which was determined from our equilibrium time-domain THz spectroscopy measurement). The photoinduced change in the index is given by $\Delta\tilde{n}(\omega)=\tilde{n}_{pump}(\omega)-\tilde{n}_0(\omega)$. The transmission coefficient through the sample following photoexcitation is
\begin{equation}
t_\text{pump}(\omega) = \dfrac{2}{m_{11}+\tilde{n}_0m_{12}+m_{21}+\tilde{n}_0m_{22}}.
\end{equation}
The matrix elements are calculated from
\begin{align}
\begin{split}
M_\text{tot} &= \left(
\begin{array}{cc}
m_{11} & m_{12} \\
m_{21} & m_{22}
\end{array}
\right) \\
&= \prod_{j=1}^N \left(
\begin{array}{cc}
        \text{cos}\left(\dfrac{\omega \tilde{n}_j d_j}{c}\right) & -\dfrac{i}{\tilde{n}_j}\text{sin}\left(\dfrac{\omega \tilde{n}_j d_j}{c}\right) \\
       -i\tilde{n}_j\text{sin}\left(\dfrac{\omega \tilde{n}_j d_j}{c}\right) & \text{cos}\left(\dfrac{\omega \tilde{n}_j d_j}{c}\right)
    \end{array}
\right),
\end{split}
\end{align}
where $\tilde{n}_j=\tilde{n}_0+\Delta\tilde{n}e^{-(j-0.5)d_j/d_p}$ (we use the $z$ value at the center of each layer) and $d_j=L/N=3d_p/N$. The transmission coefficient in the absence of the pump pulse is given by
\begin{equation}
t_0(\omega)=\dfrac{2}{1+\tilde{n}_0}e^{-i\omega\tilde{n}_0L/c}.
\end{equation}
To determine the number of layers $N$ needed to obtain a solution with minimal error, we computed $\Delta\sigma_1(\omega)$ at a pump-probe delay of $t=1.6$~ps (the time delay at which the Drude response is a maximum - see Fig.~\ref{fig:Fig2}c in the main text) for several values of $N$. We found that the extracted optical parameters converge rapidly with increasing values of $N$. Given this and the fact that it is computationally time consuming to analyze the entire 2D map for large values of $N$, we have used $N=4$ to produce the plots in Fig.~\ref{fig:Fig2} in the main text. We note that the real part of the refractive index in equilibrium ($n_0$) is nearly constant across the measured THz spectrum (Fig.~\ref{fig:FigS4}) and it exhibits almost no change upon photoexcitation.

When determining the photoinduced changes in the optical parameters of the system, it is very important to carefully take into account the mismatch between the penetration depths of the pump and probe and their relation to the THz wavelengths over the entire probe spectrum. If we neglect the penetration depth mismatch and instead assume that the pump excitation is uniform across the entire sample thickness, we find that this analysis underestimates $\Delta\sigma_1$ by two orders of magnitude.

\begin{figure*}[htb!]
\includegraphics[width=0.8\textwidth]{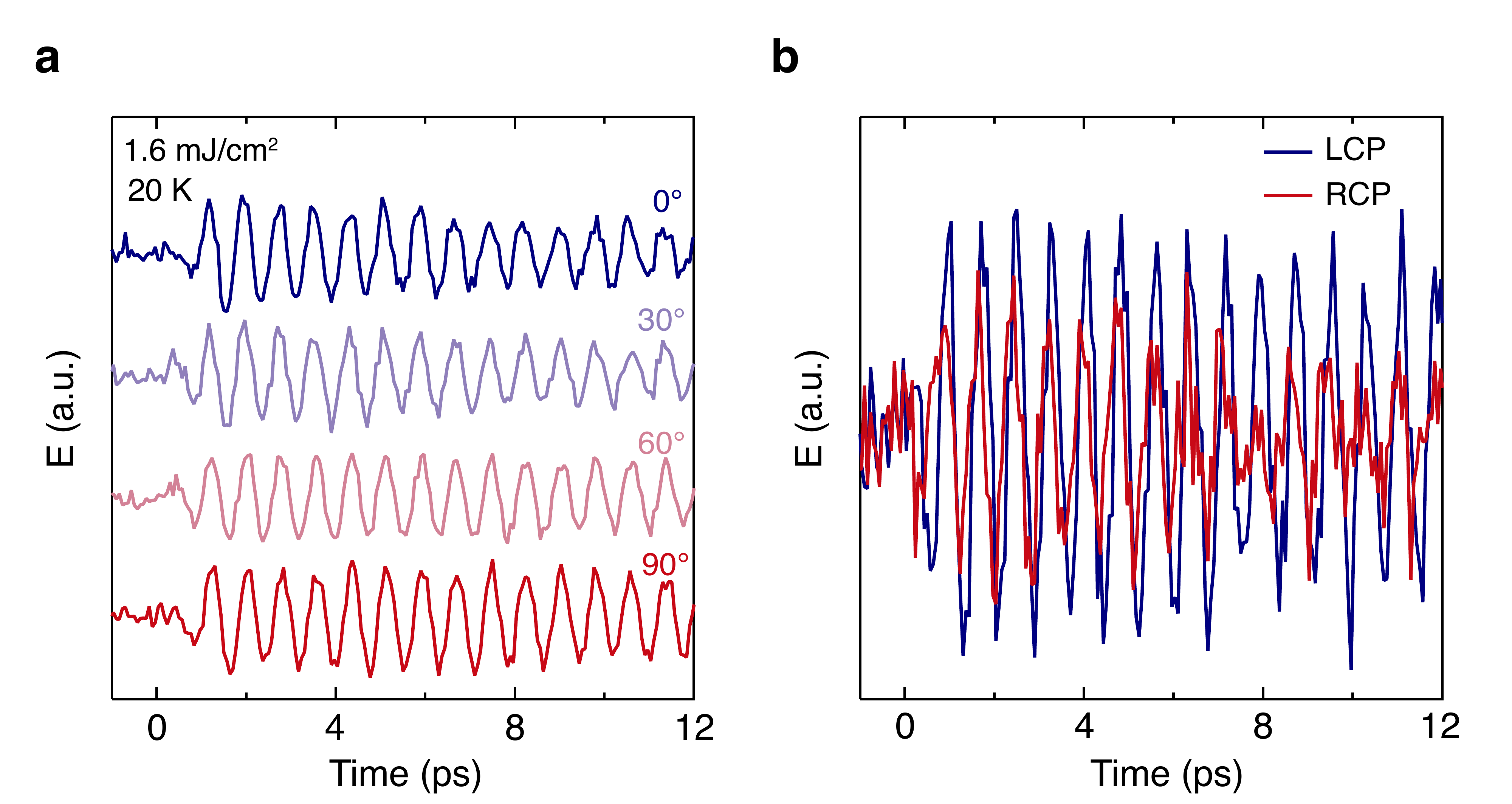}
\caption{\textbf{Pump polarization dependence for NiPS$_3$.} \textbf{a}, Dependence of the emitted THz electric field ($E$) on various linear polarization directions of the pump beam. The temperature is 20~K and the absorbed fluence is 1.6~mJ/cm$^2$. There is no change in phase of the magnon oscillations, ruling out the inverse Cotton-Mouton effect as the mechanism responsible for launching the coherent magnon. \textbf{b}, Dependence of the THz emission signal on left (blue) and right (red) circularly polarized light. Again, the oscillation phase remains unchanged, this time indicating that the magnon excitation is not due to the inverse Faraday effect.}
\label{fig:FigS6}
\end{figure*}

We note that all of the above analysis can only be applied to measure the quasi-steady state response of the system, i.e. pump-induced changes that vary on a timescale that is slow compared to the duration of the THz pulse ($\sim\,$1~ps). This is indeed the situation in NiPS$_3$, where the non-equilibrium Drude response and redshifted magnon persist for more than 15~ps (see Fig.~\ref{fig:Fig2}d in the main text). Moreover, the presence of coherent magnon oscillations has been independently verified in our THz emission measurements (see Supplementary Note 4C). When analyzing the dynamics of a system at early pump probe delays ($<1$~ps), one needs to utilize a more involved approach, such as the finite-difference time-domain method \cite{larsen2011finite}, which takes into account the response function of the experimental setup. However, this method requires a model of the photoinduced carrier dynamics to be known a priori.

\subsection{\normalsize Supplementary Note 4: Additional pump-probe data for NiPS$_3$}

\subsubsection{\normalsize A. Pump polarization dependence}

To establish that the coherent magnon is launched via the photogenerated spin--orbit-entangled excitons, we rule out other conventional mechanisms that are involved in coherent magnon generation in solids. This is achieved through a detailed pump polarization dependence. Figure~\ref{fig:FigS6}a shows the emitted THz electric field ($E$) for different linear polarization directions of the pump pulse. There is no change in the phase of the magnon oscillations, which rules out the inverse Cotton-Mouton effect as the generation mechanism \cite{kalashnikova2007impulsive}. Figure~\ref{fig:FigS6}b compares the magnon oscillations when the pump pulse is right and left circularly polarized. The same oscillation phase is again observed, this time ruling out the inverse Faraday effect \cite{kimel2005ultrafast}.

\subsubsection{\normalsize B. Pump fluence dependence of the magnon energy}

\begin{figure}[htb!]
\includegraphics[width=0.9\columnwidth]{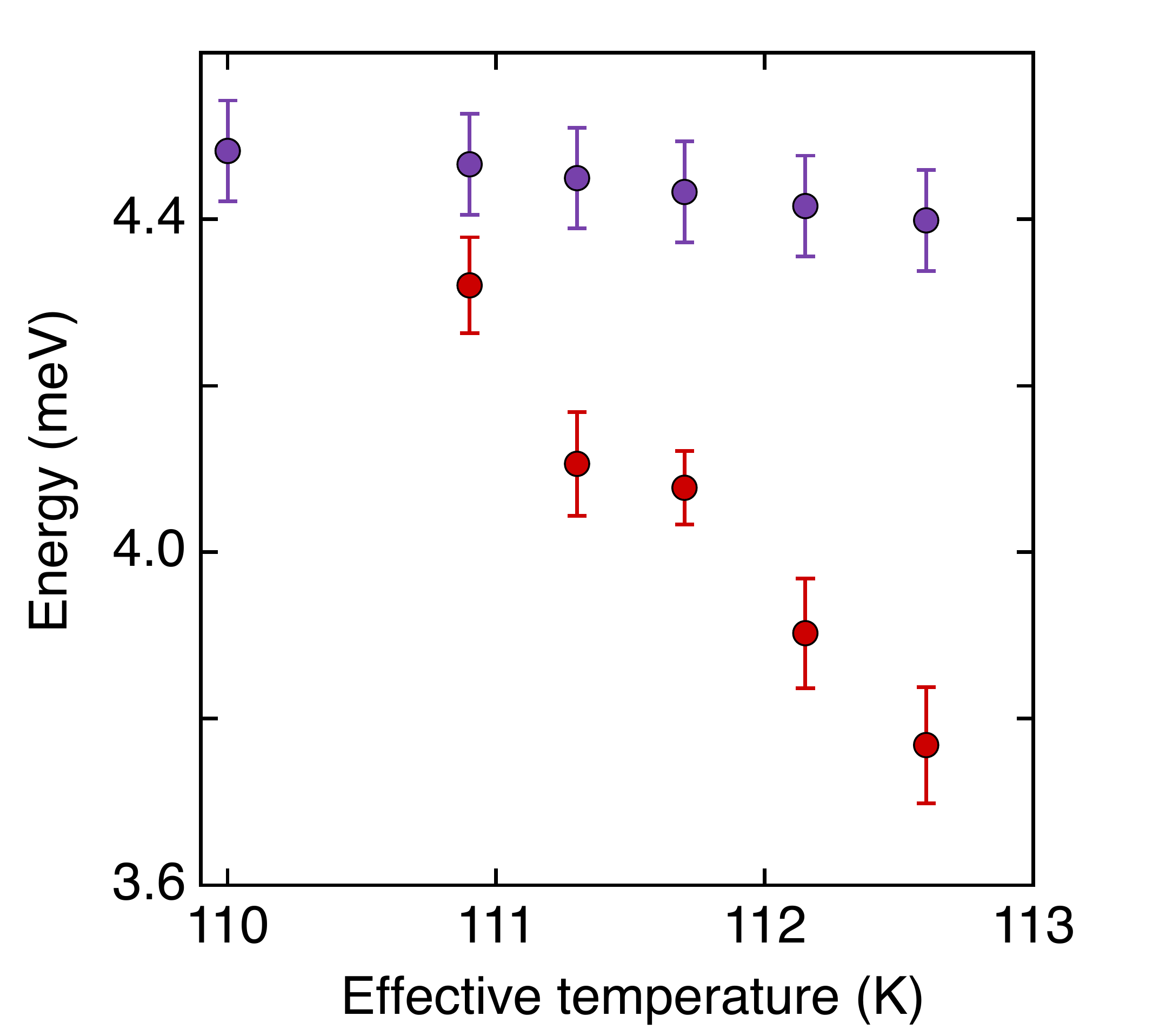}
\caption{\textbf{Non-thermal nature of the fluence dependence of the magnon energy.} Dependence of the magnon energy on the effective temperature of the sample. The red points are extracted from the THz emission measurements upon photoexcitation at 110~K (same as those shown in Fig.~\ref{fig:Fig3}f in the main text). The violet points represent the thermal change in the magnon energy and are calculated from the heat deposited in the sample for each absorbed pump fluence and the heat capacity data. The violet data point at 110~K corresponds to the magnon energy in equilibrium at this temperature. This confirms that our photoexcitation mechanism is non-thermal in nature. The error bars are similarly defined as in Figs.~\ref{fig:Fig3}d,f in the main text.}
\label{fig:FigS7}
\end{figure}

In this section, we demonstrate that the dependence of the magnon energy on the absorbed pump laser fluence (shown in Figs.~\ref{fig:Fig3}e,f in the main text) is non-thermal. Figure~\ref{fig:FigS7} presents the data from Fig.~\ref{fig:Fig3}f at 110~K (red points) along with the change in energy that would occur thermally from the increase in the lattice temperature due to the pump laser at each fluence (violet points). The change in temperature is calculated from the heat deposited into the sample by the pump at a given fluence and the heat capacity of NiPS$_3$ (Fig.~\ref{fig:FigS2}). Then this calculated temperature change is added to the starting temperature to obtain the effective temperature ($x$ axis of Fig.~\ref{fig:FigS7}) of the sample during the measurement at a particular fluence. The data point at 110~K corresponds to the magnon energy in equilibrium (no pump beam) and is obtained by interpolating the order-parameter-like dependence of the energy as a function of temperature in equilibrium (Fig.~\ref{fig:Fig3}d in the main text, violet points). Thus, from Fig.~\ref{fig:FigS7} we can see that the decrease in energy observed in the THz emission experiment (red points) is substantially larger than what would be detected if the action of the pump was purely thermal in nature (violet points). Further evidence that the mechanism responsible for launching the coherent magnon is non-thermal is that the magnon oscillations begin during the rise of the Drude, as discussed in the main text.

\subsubsection{\normalsize C. THz transmission and THz emission comparison}

\begin{figure}[htb!]
\includegraphics[width=0.9\columnwidth]{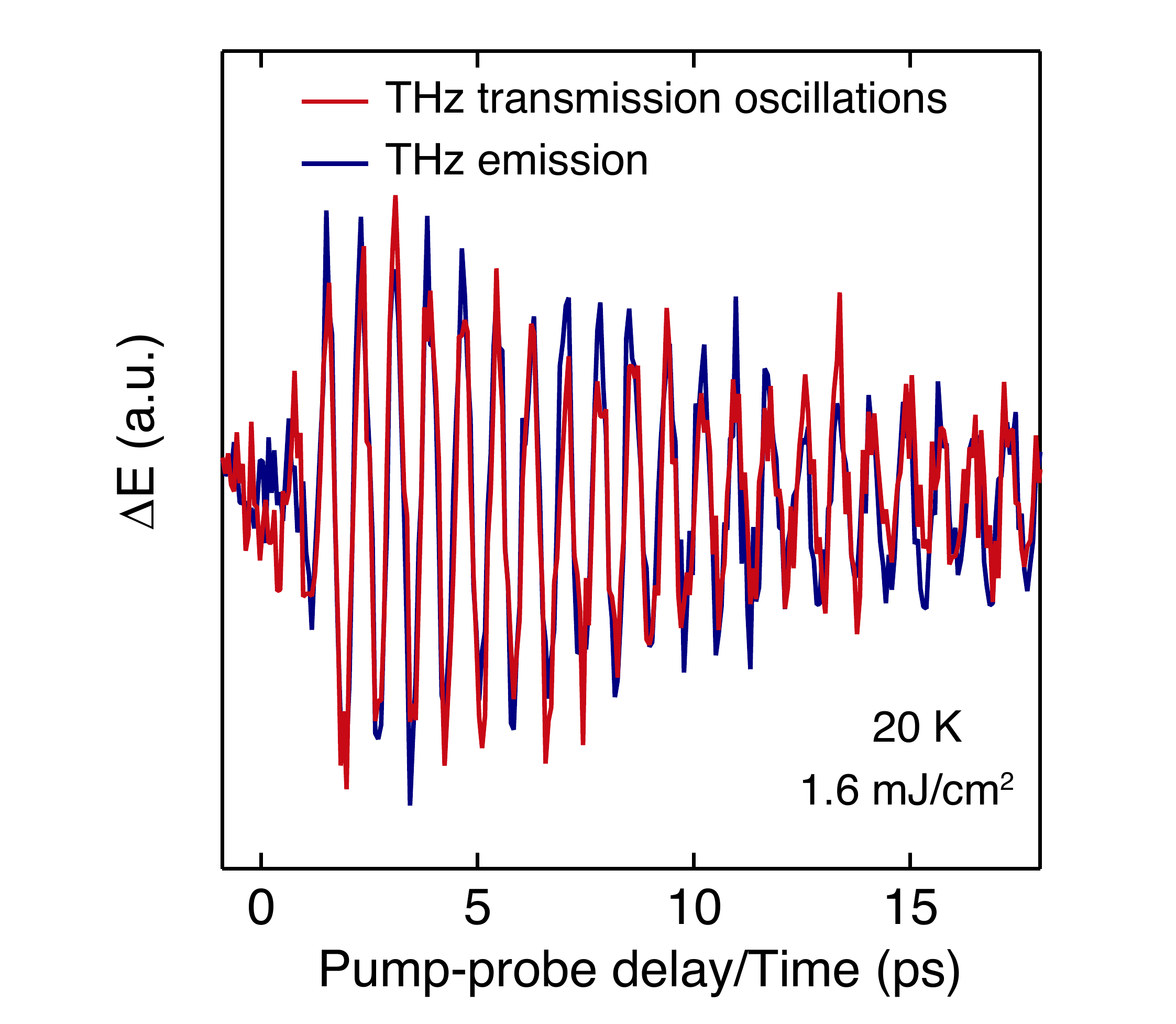}
\caption{\textbf{Comparison of magnon oscillations in the THz transmission and emission experiments.} Oscillations from the THz transmission experiment (red) plotted with the THz emission signal (blue), both at 20~K and with an absorbed fluence of 1.6~mJ/cm$^2$. The traces are nearly identical, confirming the magnon origin of the THz transmission oscillations.}
\label{fig:FigS8}
\end{figure}

In this section, we verify that the oscillations observed in the THz transmission experiment (Fig.~\ref{fig:Fig2}a in the main text) are the same magnon oscillations detected with THz emission (Fig.~\ref{fig:Fig3}a in the main text). Figure~\ref{fig:FigS8} displays the THz transmission oscillations (the 1.6~mJ/cm$^2$ curve in Fig.~\ref{fig:Fig2}d in the main text after subtracting the exponential background) along with the results of a THz emission measurement with the same experimental parameters. The two traces are nearly identical, confirming the magnon origin of the coherent oscillations present in the THz transmission data.

\subsubsection{\normalsize D. Rise time of the pump-probe traces}

\begin{figure}[htb!]
\includegraphics[width=0.9\columnwidth]{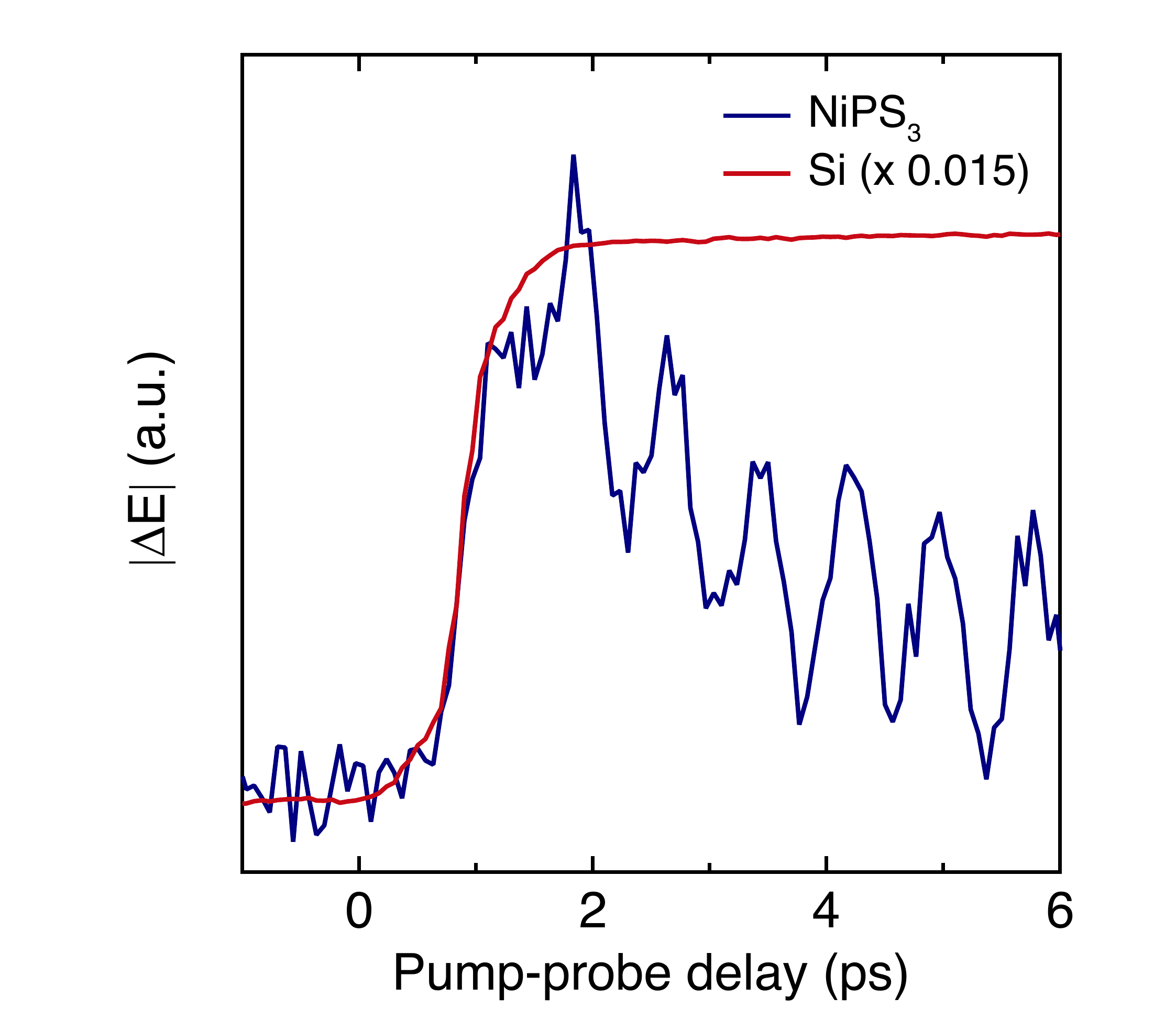}
\caption{\textbf{Resolution-limited rise time of the THz transmission experiment.} Comparison of the pump-induced change in the THz electric field ($\Delta E$) in NiPS$_3$ and a test sample of high-resistivity silicon. The rise of the pump-induced THz electric field in both cases is identical, indicating that it is limited by the intrinsic time resolution of our setup.}
\label{fig:FigS9}
\end{figure}

Figure~\ref{fig:FigS9} shows a representative trace of the pump-induced change in the THz electric field ($\Delta E$) in NiPS$_3$ along with the pump-probe signal of a test sample of high-resistivity silicon (Si). The signal of Si stems from the free-carrier absorption in the THz range that follows above-gap excitation at 1.55~eV. The rise in both curves is identical, indicating that the rise time is limited by the intrinsic time resolution of our THz setup. In this plot, the absorbed fluence is 0.3~mJ/cm$^2$ for Si and 0.5~mJ/cm$^2$ for NiPS$_3$, but all fluences showed the same resolution-limited rise time.

\begin{figure*}[htb!]
\includegraphics[width=0.8\textwidth]{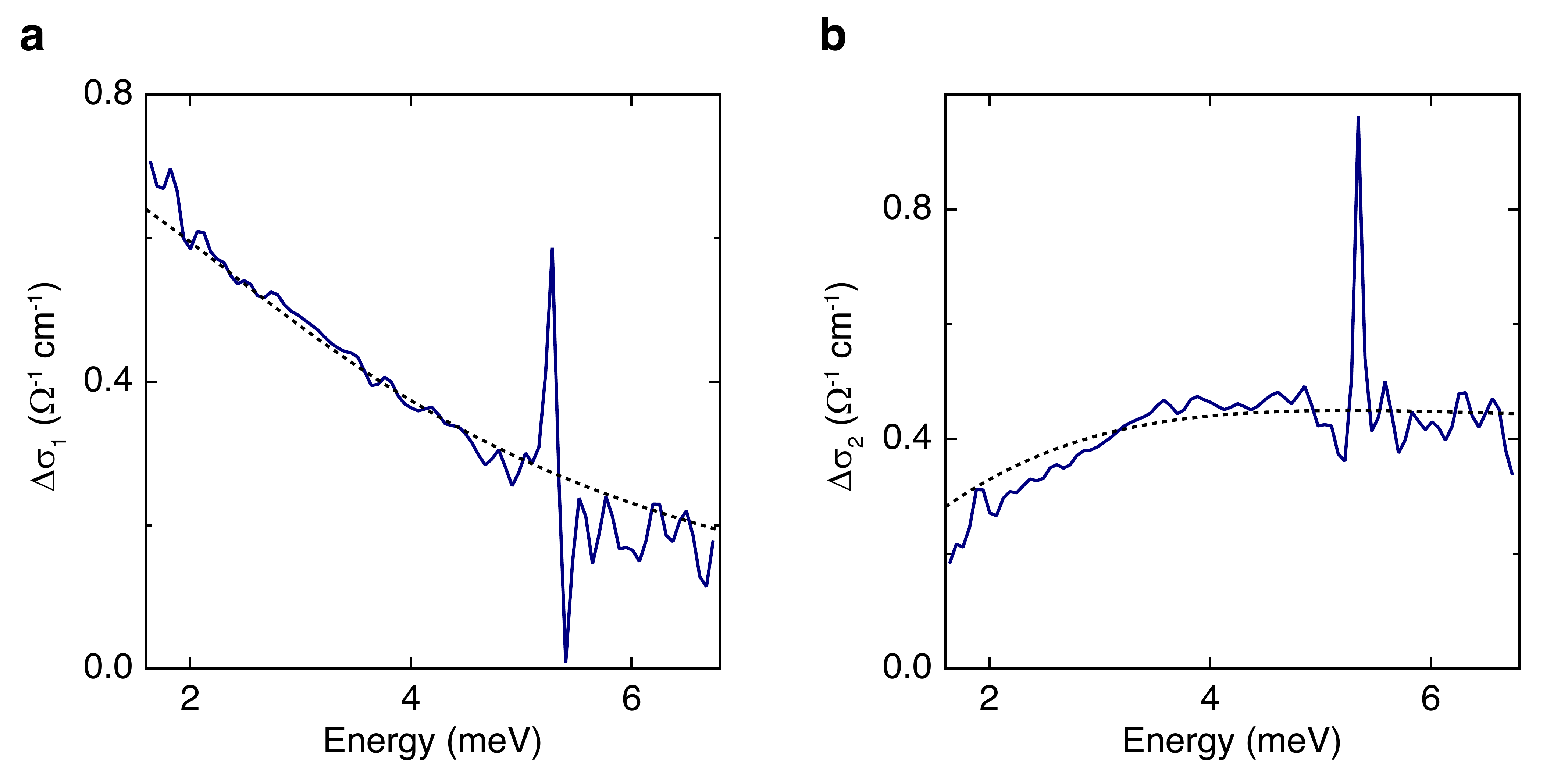}
\caption{\textbf{Fit of the photoinduced Drude response.} Pump-induced change in the real ($\Delta\sigma_1$, \textbf{a}) and imaginary ($\Delta\sigma_2$, \textbf{b}) parts of the optical conductivity as a function of frequency at a pump-probe delay of 1.6~ps. The blue curves represent the data and the black dashed lines are fits to a Lorentzian centered at zero frequency.}
\label{fig:FigS10}
\end{figure*}

\subsubsection{\normalsize E. Nature of the photoinduced conducting state}

To obtain more quantitative information about the exciton-driven conducting state in NiPS$_3$, we examine the pump-induced change in the optical conductivity as a function of frequency at the pump-probe delay of the maximum Drude response. Figures~\ref{fig:FigS10}a and \ref{fig:FigS10}b show the real ($\Delta\sigma_1$) and imaginary ($\Delta\sigma_2$) parts, respectively, of the change in conductivity at a time delay of 1.6~ps. The conductivity values that we observe are comparable to those reported in previous experiments on other materials in which excitons dissociate \cite{kaindl2003ultrafast,huber2006stimulated}. Ignoring the magnon part of the spectrum, the two curves can be accurately fit simultaneously to a Drude response using the program RefFIT \cite{kuzmenko2005kramers}. As mentioned in the main text, we extracted a plasma frequency of $\omega_p=4.7$~meV and a total scattering rate of $\gamma=4$~meV. From these values, we can estimate the carrier mobility to be $\mu\sim1800$~cm$^2$/(Vs), relying on the bare electron mass. This value changes to $\mu\sim 1100-2300$~cm$^2$/(Vs) when taking into account the carrier effective mass estimated from electronic structure calculations ($m^*\sim0.78-1.6\,m_e$) \cite{lane2020thickness}.

\begin{figure*}[htb!]
\includegraphics[width=0.9\textwidth]{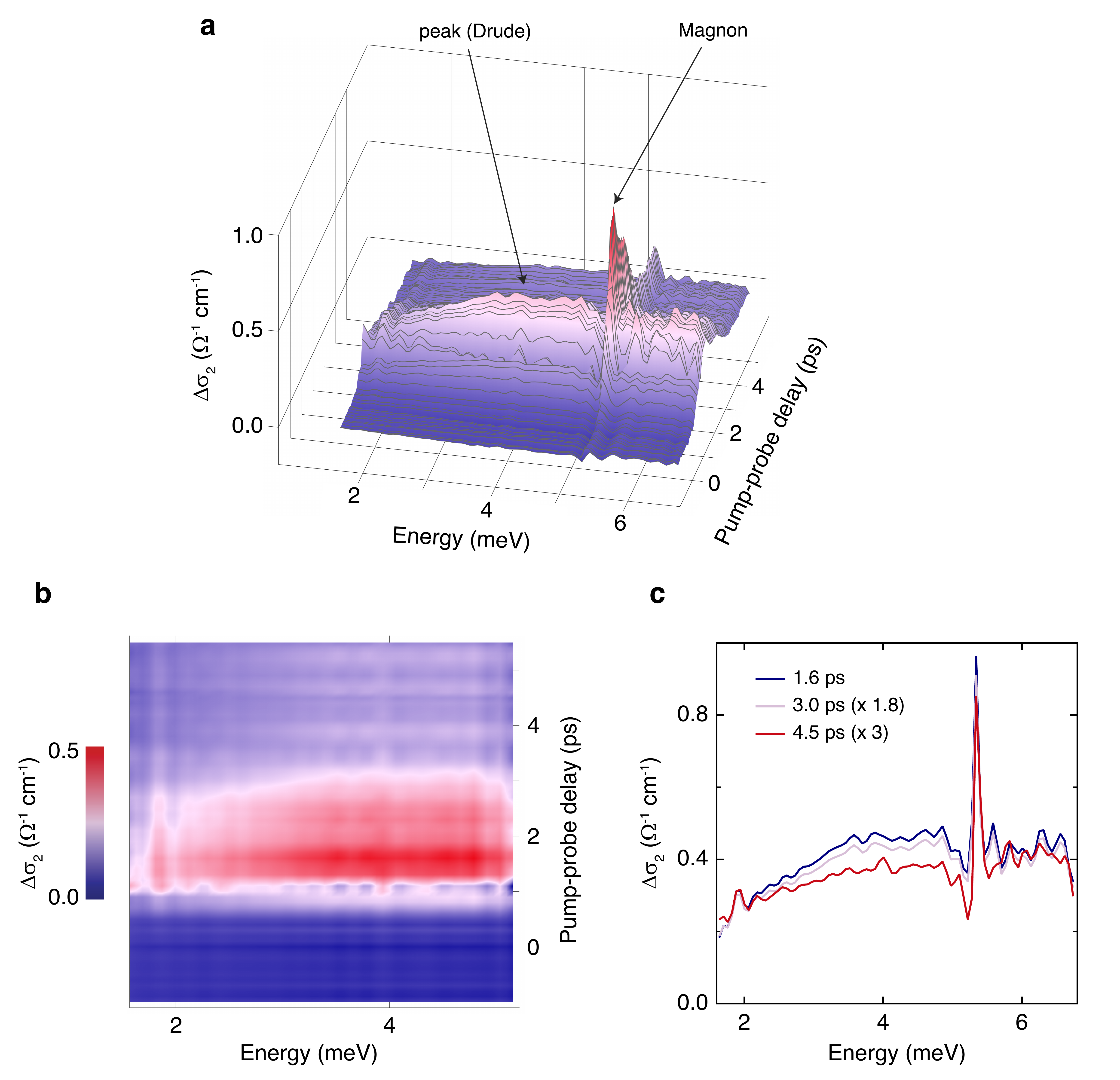}
\caption{\textbf{Time independence of the Drude scattering rate.} \textbf{a}, Spectro-temporal evolution of the pump-induced change in the imaginary part of the optical conductivity ($\Delta\sigma_2$). The temperature is 20~K and the absorbed pump fluence is 1.3~mJ/cm$^2$. The two features present are a broad peak around 4 meV associated with the Drude response and a Lorentzian lineshape around the magnon energy. \textbf{b}, Top view of \textbf{a} showing only the Drude contribution. \textbf{c}, Spectral dependence of $\Delta\sigma_2$ at a few representative pump-probe delay times. We see that the peak of the $\Delta\sigma_2$ signal does not vary over time, indicating a constant scattering rate.}
\label{fig:FigS11}
\end{figure*}

Next, we establish that the time evolution of the Drude signal is solely governed by the change in $\omega_p$ and not by the variation of $\gamma$. This information is encoded in $\Delta\sigma_2$ (Fig.~\ref{fig:FigS11}) since the peak associated with the reactive part of the Drude response provides an estimate of the total scattering rate $\gamma$ \cite{beard2000transient,beard2001subpicosecond}. Figure~\ref{fig:FigS11}a shows the full spectro-temporal evolution of $\Delta\sigma_2$, and Fig.~\ref{fig:FigS11}b is a top view of Fig.~\ref{fig:FigS11}a zoomed into lower energies corresponding to the Drude contribution. We observe that the peak in the $\Delta\sigma_2$ signal remains constant around 4~meV over time. Another visualization of this feature is offered in Fig.~\ref{fig:FigS11}c, where the $\Delta\sigma_2$ spectrum is compared at a few representative pump-probe delays. The observation that $\gamma$ does not vary in time indicates that the mobile carriers produced by exciton dissociation are cold and lie around the band edges \cite{zielbauer1996ultrafast}. This is in stark contrast to the cooling processes that are still active after 1~ps in materials where high-energy bosons are emitted upon injection of hot carriers \cite{kampfrath2005strongly}. Consequently, the exciton dissociation in NiPS$_3$ results in mobile carriers that do not possess enough excess energy to emit hot optical phonons and high-frequency magnons, which is a crucial aspect in the preservation of the underlying antiferromagnetic order at all time delays (signaled by the persistence of the sharp $q$ = 0 magnon mode). Further evidence that the decay of the Drude response is due to a decrease in $\omega_p$ is given by the depletion in the spectral weight of $\Delta\sigma_1$ over time (Fig.~\ref{fig:FigS14}), as this quantity is directly proportional to $\omega_p^2$ by the optical sum rule.

\begin{figure*}[htb!]
\includegraphics[width=0.8\textwidth]{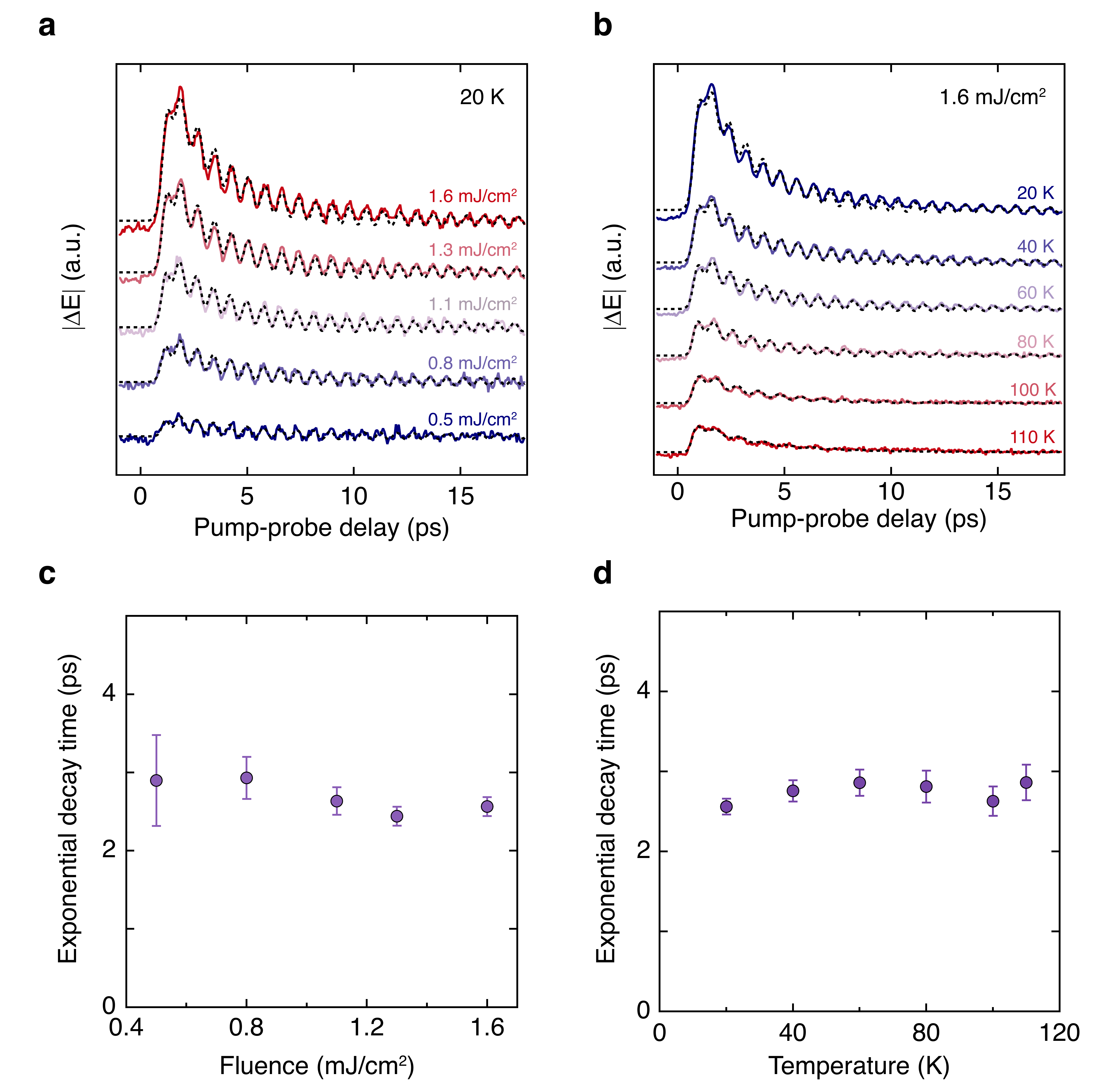}
\caption{\textbf{Fluence and temperature independence of the Drude decay time.} \textbf{a},\textbf{b}, Temporal evolution of the pump-induced change in the THz electric field ($\Delta E$) of the spectrally-integrated measurement as a function of absorbed pump fluence at 20~K (\textbf{a}, same as Fig.~\ref{fig:Fig2}d in the main text) and as a function of temperature at an absorbed pump fluence of 1.6~mJ/cm$^2$ (\textbf{b}). \textbf{c},\textbf{d}, Dependence of the exponential decay time on the absorbed pump fluence (\textbf{c}) and on temperature (\textbf{d}) extracted from the fits to the data in \textbf{a} and \textbf{b}, respectively. There is almost no variation in the decay time in either case, demonstrating that the itinerant carrier recombination dynamics is dominated by their localization in deep traps and that carriers are not subsequently thermally activated. The error bars in \textbf{c} and \textbf{d} represent the 95\% confidence interval of the fits.}
\label{fig:FigS12}
\end{figure*}

For these reasons, we can relate the decay of the Drude conductivity to the recombination dynamics of the mobile carriers. To unravel which recombination mechanisms contribute to the decay, we measure the ultrafast THz transmission as a function of absorbed fluence (Fig.~\ref{fig:FigS12}a, same as Fig.~\ref{fig:Fig2}d in the main text) and temperature (Fig.~\ref{fig:FigS12}b). We observe that all the temporal traces can be fit with only one exponential function, indicating the presence of a single recombination pathway for the mobile carriers. The fluence dependence of the exponential decay time is shown in Fig.~\ref{fig:FigS12}c. The decay time is nearly independent of fluence, which suggests that the carrier recombination dynamics is governed by their trapping or self-trapping at deep impurity centers or defects as described by Shockley-Read-Hall (SRH) theory \cite{hall1952electron,shockley1952statistics}. The general equation governing the lifetime $\tau$ of free carriers in semiconductors is $\dfrac{1}{\tau}=A+Bn+Cn^2$, where $n$ is the carrier density and $A$, $B$, and $C$ are the coefficients for nonradiative recombination (SRH), radiative recombination, and Auger recombination, respectively \cite{linnros1998carrier}. Here, the first term is responsible for the observed nearly fluence-independent decay time. In Fig.~\ref{fig:FigS12}d, we show the temperature dependence of the exponential decay time. The decay time is also constant with temperature, confirming that the trapped levels are deep and no thermal activation of carriers occurs after their localization \cite{lui2003fluence}. Therefore, the depletion of the THz spectral weight is accompanied by its transfer to another spectral region outside of our measured range, most likely in the mid- or near-infrared (i.e. where impurity centers and small polarons absorb light) \cite{okamoto2011photoinduced}.

We also verified that the observed Drude signal stems from the transport of quasi-free carriers and it is not due to the band-like motion of large polaronic carriers \cite{hendry2004electron,dean2011polaronic}. In order for a large polaron to form in a deformable medium, the long-range Coulomb potential $V^{LR}$(\textbf{r}) between an excess carrier and the ionic lattice must be large. This potential can be expressed as
\begin{equation}
V^{LR}(\textbf{r}) = - \Bigg[ \frac{1}{\epsilon_r(\infty)} - \frac{1}{\epsilon_r(0)}\Bigg] \frac{e^2}{|\textbf{r}| \epsilon_0},
\end{equation}
where \textbf{r} is the vector between an electron and an ionic site, $e$ is the electron charge, $\epsilon_0$ is the vacuum permittivity, and $\epsilon_r(0)$ and $\epsilon_r(\infty)$ are the static and high-frequency dielectric constants, respectively. The form of the expression ensures that the fast electronic contribution to the polarizability is canceled and only the nuclear contribution is taken into account. As established in polaron theory \cite{emin2013polarons}, large polaron formation is possible only when $\epsilon_r(0)$ is at least twice the value of $\epsilon_r(\infty)$. In NiPS$_3$, our static THz data show that $\epsilon_r(0)$ = 7.9. In contrast, $\epsilon_r(\infty)$ can be extracted from optical spectroscopy data in a frequency range that is above the relevant longitudinal optical phonon energies but below the energy of the lowest interband transitions. A reasonable choice of the photon energy in NiPS$_3$ is in the range $0.3-1$~eV. At these energies $\epsilon_r(\infty)$ = $9.3-10$ \cite{piacentini1982optical,kim2018charge}, i.e. a value that is even larger than $\epsilon_r(0)$. For this reason, we can rule out a large polaron origin for the coherent Drude response detected in our ultrafast THz transmission experiment and conclude that the carriers involved in the itinerant transport have a quasi-free character. This indicates that the effective mass that is relevant to estimate the carrier mobility is the one extracted from electronic structure calculations that do not account for electron-phonon coupling \cite{lane2020thickness}. Indeed, no mass enhancement due to a Fr\"ohlich-type electron-phonon coupling is expected for NiPS$_3$. These arguments confirm that the carrier mobility lies in the 10$^3$~cm$^2$/(Vs) range, i.e. well above the typical carrier mobilities $>$1~cm$^2$/(Vs) expected from the large polaron scenario (which would require an unreasonable $m^*>1000m_e$).

Finally, we remark that the existence of the itinerant conductivity, caused by the dissociation of the photogenerated excitons, is not associated with those excitons having a coupling to magnetic degrees of freedom. Indeed, bare excitons (regular electron-hole bound states) that dissociate through exciton-exciton interactions or pump-induced exciton photoionization can give rise to photoconductivity as has been found in a variety of band semiconductors \cite{braun1968singlet,bergman1974photoconductivity,catalano1975luminescence,lee1993transient,sun2014observation}. We do not foresee any limitation in the extension of this phenomenon to excitons in correlated systems.

\subsection{\normalsize Supplementary Note 5: Significance of our non-equilibrium results and comparison to band semiconductors}

In this section, we describe how the current results on NiPS$_3$ represent an anomalous phenomenon that has no counterpart in the physics of photoexcited band semiconductors.

In pristine semiconductors devoid of strong electronic correlations, the charge gap arises because of band theory arguments. Photoexciting electron-hole pairs above this band gap (up to carrier densities of 10$^{19}$-10$^{21}$~cm$^{-3}$) typically results in the appearance of a Drude response in the THz or far-infrared range due to free-carrier (intraband) absorption \cite{ulbricht2011carrier}. Therefore, during the time that precedes complete electron-hole recombination (varying from several picoseconds to nanoseconds depending on the system), the initially undoped band semiconductor transiently develops metallic conductivity, signaled by the appearance of a Drude response \cite{hase2003birth}. This effect has been widely studied in semiconductor physics to estimate the carrier transport and mobility in materials of optoelectronic interest \cite{ulbricht2011carrier}, to realize switchable optical components \cite{almeida2004all,rivas2006optically}, or to clarify the optical nonlinearities affecting a semiconductor's exciton resonances \cite{haug1985basic}.

It is well known that the above-gap photoexcitation process described above also often results in the generation of phonons that coherently evolve as a function of time \cite{ishioka2010coherent}. These phonons can be optical or acoustic, and they can be generated at $q=0$ or at finite momentum \cite{garrett1997ultrafast,liao2016photo}. In band semiconductors, coherent phonons are usually triggered via a non-thermal deformation potential coupling (i.e. the so-called displacive excitation in the case of optical phonons) \cite{zeiger1992theory,stevens2002coherent}, but also other mechanisms (e.g., thermoelasticity, piezoelectric coupling, etc.) can play a role \cite{ruello2015physical}. These coherent phonons coexist and interact with the electron-hole plasma created by the pump pulse. The real (frequency) and imaginary (decay rate) parts of the self-energy are therefore profoundly affected by this electron-phonon interaction \cite{hase2003birth,ishioka2008ultrafast,liao2016photo}. Nevertheless, the phonons still represent well-defined collective modes of the lattice at all momenta because the crystal has neither undergone a structural phase transition nor has been melted by the photoexcitation.

The scenario emerging in our experiments on NiPS$_3$ is radically different and cannot be viewed as a simple extension of the previous case to coherent magnons. In a Mott insulator, the charge gap stems from strong electronic correlations and it is not a result of band theory \cite{khomskii2014transition}. Antiferromagnetic order typically develops at low temperature to reduce the ground-state energy of the system, with magnons emerging as well-defined collective modes. When above-gap light excitation photodopes a Mott insulator, the presence of the itinerant carriers is always accompanied by the melting of the long-range antiferromagnetic order \cite{dean2016ultrafast,afanasiev2019ultrafast,yang2020ultrafast}. At short timescales, this melting can proceed through different mechanisms, such as the collapse of the magnetic moments, the quench of the exchange interaction, or the direct injection of energy from the hot photocarriers into the spin system \cite{balzer2015nonthermal}. Irrespective of the detailed pathway, the long-wavelength magnon---which signals the presence of long-range antiferromagnetic order---always collapses by broadening and losing intensity. This is also observed in the insulator-to-metal transition realized by chemical doping a Mott insulator \cite{gretarsson2017raman}. Consequently, only short-range magnetic correlations survive and are reflected in the persistence of paramagnons at short-wavelength (finite $q$) and high energy \cite{dean2016ultrafast,yang2020ultrafast}. For these reasons, the coexistence of a Drude response and a very sharp $q=0$ magnon in NiPS$_3$ is an unexpected and hitherto-unobserved finding, markedly different from the melting dynamics reported thus far in photoexcited Mott insulators.

\subsection{\normalsize Supplementary Note 6: Discussion regarding the possibility of phase separation in the photoinduced state}

In the main text, we indicate that our results preclude an interpretation of the antiferromagnetic conducting state as a phase-separated state composed of metallic patches embedded in an antiferromagnetic insulator. Below we explain in detail the features of our data that rule out this alternative scenario.

\begin{figure}[htb!]
\includegraphics[width=1\columnwidth]{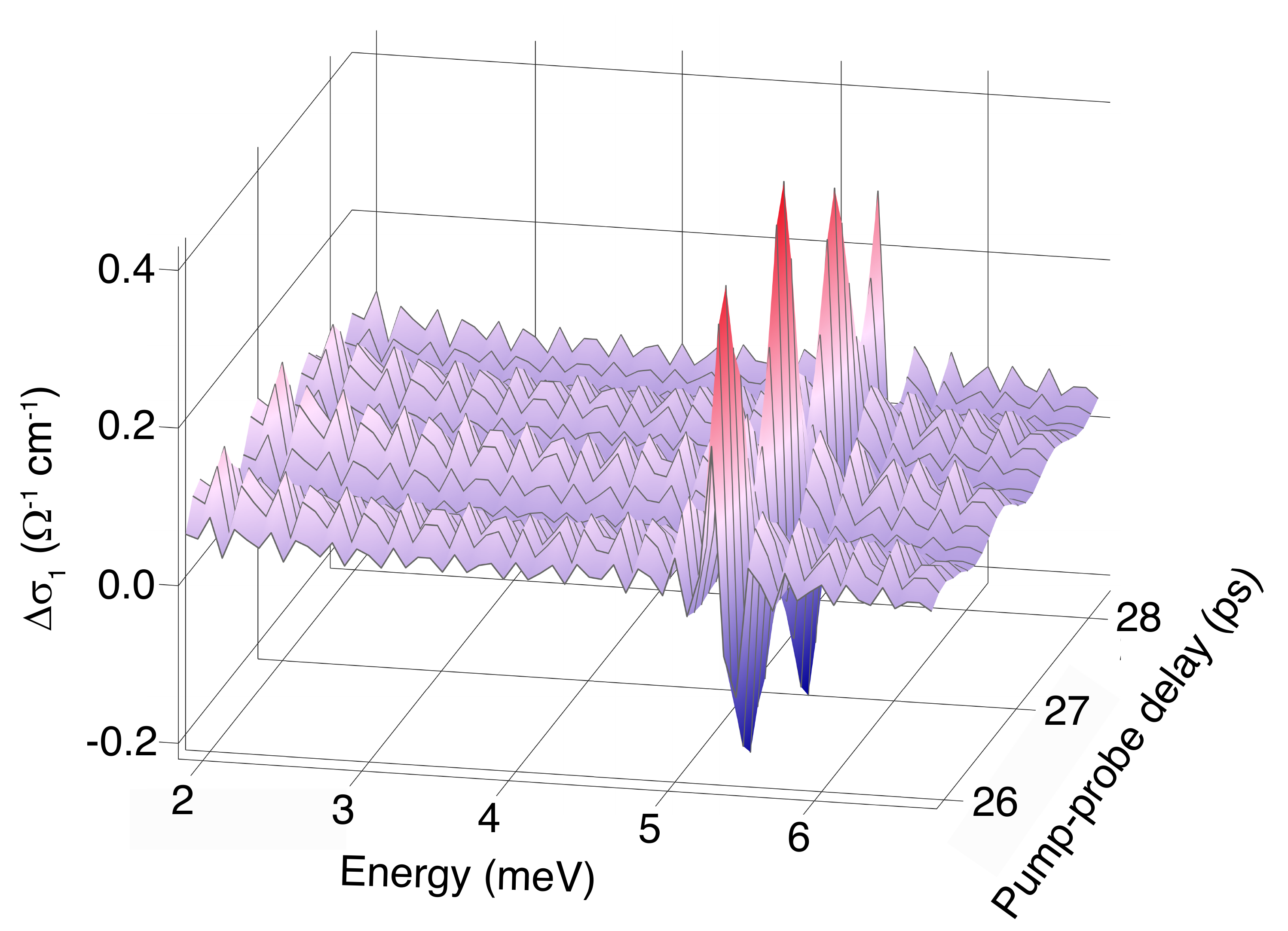}
\caption{\textbf{Spectro-temporal evolution of the conductivity at later delay times.} Spectro-temporal evolution of the pump-induced change in the real part of the optical conductivity ($\Delta\sigma_1$) for pump-probe delay times ranging from 26 to 28~ps. The temperature is 20~K and the absorbed pump fluence is 1.3~mJ/cm$^2$. In contrast to Fig.~\ref{fig:Fig2}b in the main text, at these later delay times the Drude response has disappeared and only the first-derivative-like shape around the magnon energy remains. The coherent magnon oscillations as a function of time are still present as well.}
\label{fig:FigS13}
\end{figure}

\begin{figure}[htb!]
\includegraphics[width=0.99\columnwidth]{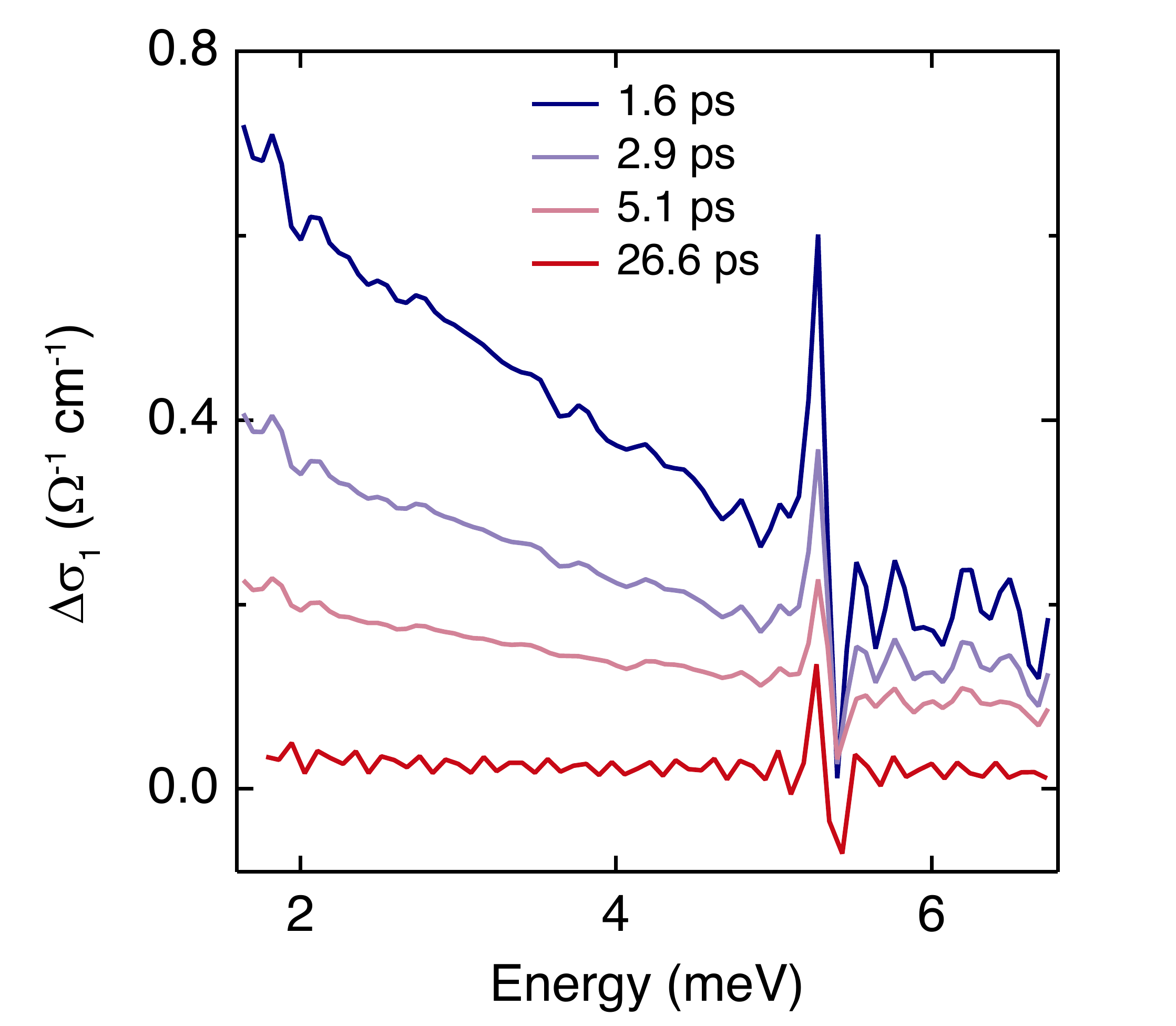}
\caption{\textbf{Comparison of the conductivity over time.} Pump-induced change in the real part of the optical conductivity ($\Delta\sigma_1$) as a function of frequency at several representative pump-probe delay times: 1.6~ps (at the peak of the Drude response), 2.9 and 5.1~ps (during the decay of the Drude), and 26.6~ps (after the Drude has disappeared). These traces are cuts taken from the 2D maps shown in Fig.~\ref{fig:Fig2}b in the main text (1.6, 2.9, and 5.1~ps) and Fig.~\ref{fig:FigS13} (26.6~ps). The first-derivative-like shape around the magnon energy remains sharp at all times, indicating that the mobile carriers do not affect the width of the magnon lineshape.}
\label{fig:FigS14}
\end{figure}

First, we consider the lineshape of the magnon mode. The sharpness of the magnon feature displayed in Fig.~\ref{fig:Fig2}b in the main text and its lack of any noticeable broadening indicates that the magnon remains a well-defined mode after the arrival of the pump pulse and it does not evolve into a paramagnon \cite{gretarsson2017raman}. If we suppose that the pump creates paramagnetic metallic patches, these regions would act like impurities in the insulating antiferromagnetic system, thereby disrupting the long-range antiferromagnetic order and decreasing its correlation length. Consequently, the presence of the metallic puddles would lead to a broadening of the long-wavelength ($q=0$) magnon that would increase with fluence as the volume of the metallic regions increases \cite{battisti2017universality,gretarsson2017raman}. However, the damping of our magnon oscillations does not show any noticeable change as a function of fluence. Moreover, the magnon lineshape retains the same sharp width at all pump-probe time delays both while the mobile carriers are present and after they disappear once the Drude decays (after $\sim\,$15~ps). The former is shown in Fig.~\ref{fig:Fig2}b in the main text and the latter is presented in Fig.~\ref{fig:FigS13}. From Fig.~\ref{fig:FigS13}, we see that the 2D map of $\Delta\sigma_1$ contains only the sharp, first-derivative-like shape of the magnon without the Drude response, and the magnon is oscillating coherently as a function of pump-probe delay. The same behavior was observed up to 80~ps after the pump pulse arrival. To better visualize the spectra before and after the Drude decays, in Fig.~\ref{fig:FigS14} we plot $\Delta\sigma_1$ as a function of frequency at several representative pump-probe delay times. Comparing the traces, we see that the magnon retains its narrow lineshape and therefore is not affected by the presence or absence of the mobile carriers. Thus, our observations are inconsistent with the scenario of phase-separated, paramagnetic metallic patches.

Next, we examine the fluence dependence of the magnon oscillation amplitude. Since the coherent magnon is generated by the pump excitation due to the coupling of the spin--orbit-entangled excitons with the magnetic order, its amplitude at times $<\,$1~ps should be linearly proportional to the number of excitons created. Afterwards, exciton dissociation into free carriers through mutual collisions produces a Drude conductivity in the system. Due to the time resolution of our ultrafast THz experiment (see Supplementary Note 3D), the magnon amplitude can be evaluated only by fitting the oscillation that emerges on top of the Drude response, i.e. after the exciton dissociation process is complete ($>\,$1~ps). If the dissociated excitons formed separate, paramagnetic metallic patches, then the size of the insulating antiferromagnetic region would be reduced and hence the magnon precession amplitude would become smaller than that initially generated by the pump. With increasing fluence, the initial magnon amplitude would rise linearly but the volume of metallic patches would increase quadratically (as determined by the quadratic dependence of the exponential amplitude in Fig.~\ref{fig:Fig2}f in the main text), leading to a reduction in the magnon amplitude that scales quadratically as a function of fluence. Hence, the observed magnon amplitude ($>\,$1~ps) would show a sublinear trend with fluence. In contrast, as presented in Fig.~\ref{fig:Fig2}e in the main text, the amplitude of the magnon evaluated after 1~ps still scales linearly with fluence, as if no changes occur when the excitons dissociate compared to when the excitons are initially generated. This feature is expected when the itinerant carriers coexist with the underlying long-range antiferromagnetism and do not interfere with the long-wavelength precession of the localized spins.

Thus, combining these aspects of our data, an interpretation involving phase separation of paramagnetic metallic patches embedded in the antiferromagnetic insulating state can be readily ruled out, and we conclude that our data indeed demonstrates a coexistence of a finite itinerant conductivity and antiferromagnetism.

\subsection{\normalsize Supplementary Note 7: Theoretical calculations}

\subsubsection{\normalsize A. Model Hamiltonian for NiPS$_3$}

As mentioned in the main text, the crystal structure of NiPS$_3$ consists of two dimensional (2D) layers of Ni atoms arranged in a honeycomb lattice in the $ab$-plane that are coupled in the $c$ direction by the van der Waals interaction. Each individual layer is invariant under a $C_3$ rotation but the stacking pattern breaks this symmetry and induces an anisotropy along the $a$-axis.

In the resulting effective spin model, the Ni sites have spin-1 and no orbital degeneracy. Assuming the interlayer coupling is small, for each layer the effective spin Hamiltonian consists of $XXZ$ terms up to third-nearest neighbors and single-ion anisotropy along the $a$-axis and along the $c$-axis \cite{kim2019suppression}:
\begin{equation}
    \begin{split}
        H_{spin} =& J_1 \sum_{\langle i,j \rangle} (S_{i,x} S_{j,x}+S_{i,y} S_{j,y}+\alpha S_{i,z} S_{j,z})\\
       &+J_2 \sum_{\langle\langle i,j \rangle\rangle} (S_{i,x} S_{j,x}+S_{i,y} S_{j,y}+\alpha S_{i,z} S_{j,z})\\
       &+J_3 \sum_{\langle\langle\langle i,j \rangle\rangle\rangle} (S_{i,x} S_{j,x}+S_{i,y} S_{j,y}+\alpha S_{i,z} S_{j,z})\\
        &+ D_x \sum_i S_{i,x}^2
        + D_z \sum_i S_{i,z}^2,
    \end{split}
\end{equation}
where $J_1$, $J_2$, $J_3$ are the nearest neighbor, second-nearest neighbor, and third-nearest neighbor coupling, $\alpha$ is the anisotropic spin-spin interaction parameter, and $D_{x,z}$ are the single-ion anisotropic coefficients. A small $D_x$ breaks $U(1)$ down to $Z_2$, resulting in an Ising transition at finite temperature. In the single layer limit, the $U(1)$ symmetry is restored so there will be a Kosterlitz-Thouless (KT) transition at finite temperature. We ignore interlayer couplings here since they are much smaller than the intralayer interactions. 
\vspace{10pt}

\subsubsection{\normalsize B. Classical ground state}

When the spin operators are treated as classical variables, we can diagonalize the spin Hamiltonian in momentum space. Note that $S_x$, $S_y$, and $S_z$ are decoupled and there are two atoms ($A$ and $B$) in a honeycomb unit cell. We only need to diagonalize three $2\times 2$ matrices. We denote each block by $h_x(\Vec{k})$, $h_y(\Vec{k})$, and $h_z(\Vec{k})$:
\begin{widetext}
\begin{equation}
    h_x(\Vec{k})=(S_{x,A}(\Vec{k})~S_{x,B}(\Vec{k}))\left(
    \begin{array}{cc}
        Q_2(k) + D_x & Q_1(k)+Q_3(k) \\
        Q_1^*(k)+Q_3^*(k) & Q_2(k) + D_x
    \end{array}
    \right)\left(
    \begin{array}{c}
        S_{x,A}(-\Vec{k})  \\
        S_{x,B}(-\Vec{k})
    \end{array}
    \right)
\end{equation}
\begin{equation}
    h_y(\Vec{k})=(S_{y,A}(\Vec{k})~S_{y,B}(\Vec{k}))\left(
    \begin{array}{cc}
        Q_2(k) & Q_1(k)+Q_3(k) \\
        Q_1^*(k)+Q_3^*(k) & Q_2(k)
    \end{array}
    \right)\left(
    \begin{array}{c}
        S_{y,A}(-\Vec{k})  \\
        S_{y,B}(-\Vec{k})
    \end{array}
    \right)
\end{equation}
\begin{equation}
    h_z(\Vec{k})=(S_{z,A}(\Vec{k})~S_{z,B}(\Vec{k}))\left(
    \begin{array}{cc}
        \alpha Q_2(k) + D_z & \alpha (Q_1(k)+Q_3(k)) \\
       \alpha(Q_1^*(k)+Q_3^*(k)) & \alpha Q_2(k) + D_z
    \end{array}
    \right)\left(
    \begin{array}{c}
        S_{z,A}(-\Vec{k})  \\
        S_{z,B}(-\Vec{k})
    \end{array}
    \right),
\end{equation}
\end{widetext}
where $Q_1(k) = J_1 (e^{i k_y}+ e^{-i (\sqrt{3} k_x/2 + k_y/2)}+e^{-i (-\sqrt{3} k_x/2 + k_y/2)})$, $Q_2(k) = 2 J_2 [\cos(\sqrt{3}k_x)+\cos(\sqrt{3}k_x/2 +3 k_y/2)+\cos(-\sqrt{3}k_x/2+3k_y/2)]$, and $Q_3 = J_3 (e^{-2 i k_y}+ e^{i (\sqrt{3} k_x + k_y)}+e^{i (-\sqrt{3} k_x + k_y)})$. Here, we set the distance between the nearest $A$ and $B$ atoms to be 1 and the honeycomb $a$-axis to be aligned with the $x$-axis. 

Since $D_x<0$ and all three matrices have a similar $k$ dependence, it follows that the lowest-energy state will be fully polarized along the $a$-axis. For the given parameters, one can show that the band bottoms are at $(0, \pm 2\pi/3)$ and all the points are related by a $C_3$ rotation. The eigenvector at $k_+ = (0,  2\pi/3)$ for the lower band is $(e^{i \pi/3}, 1)^T$ with some normalization factor. Performing an inverse Fourier transformation, we obtain $S_{A,x}(\Vec{r}) = 2S \cos(\frac{2\pi}{3} (y+\frac{1}{2}))$ and $S_{B,x}(\Vec{r}) = 2S \cos(\frac{2\pi}{3} y)$. This is the zigzag order.

\subsubsection{\normalsize C. Symmetries of the zigzag order}

\begin{figure}[htb!]
\includegraphics[width=0.98\columnwidth]{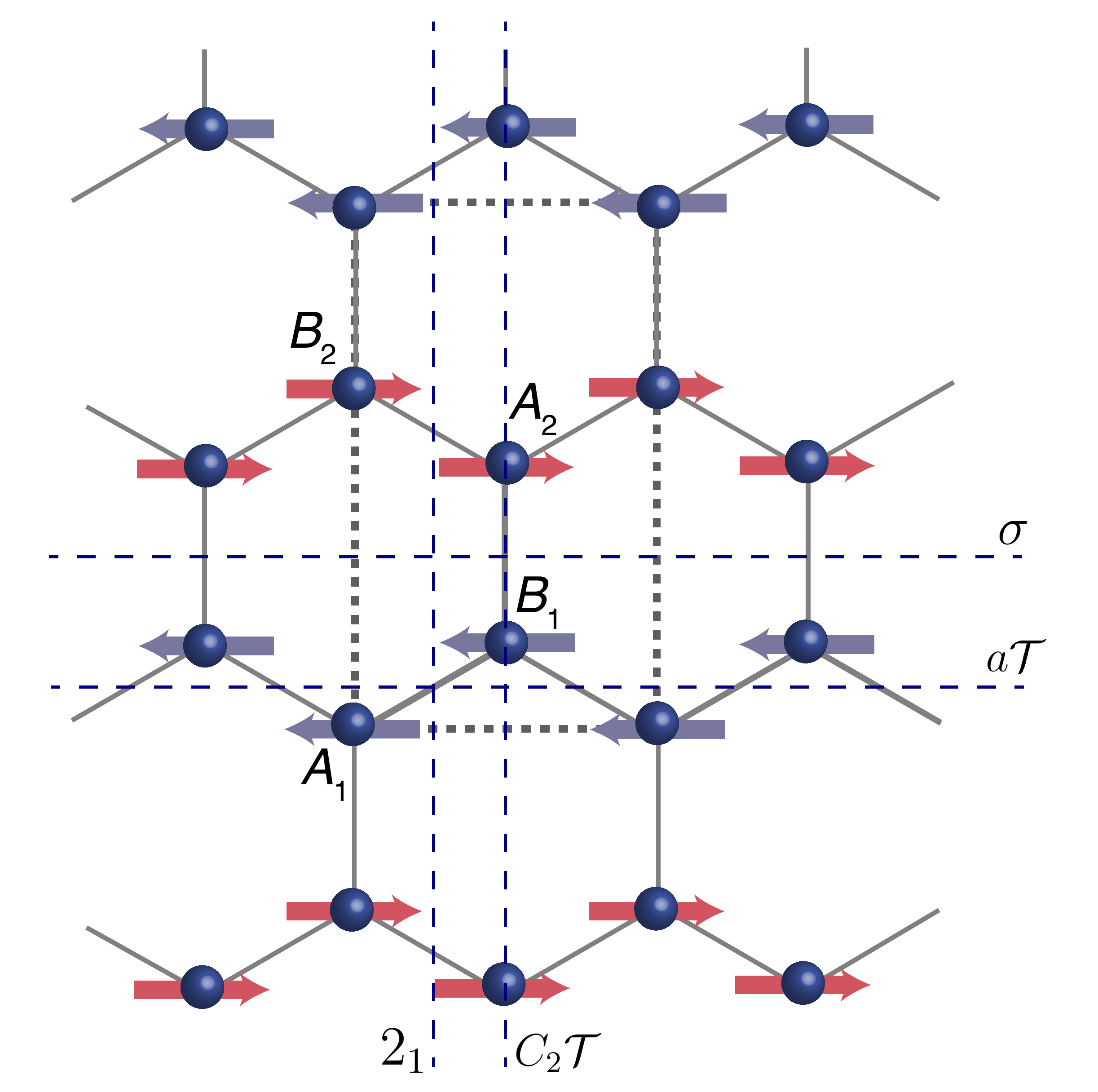}
\caption{\textbf{Magnetic unit cell of NiPS$_3$.} The thick gray dashed rectangle denotes the magnetic unit cell, which consists of four Ni atoms labeled $A_1$, $A_2$, $B_1$, and $B_2$. The blue spheres denote Ni atoms and the red and blue arrows represent the spins, showing the zigzag antiferromagnetic order. The thin blue dashed lines represent the symmetries $\sigma$, $a\mathcal{T}$, $2_1$, and $C_2\mathcal{T}$ of the zigzag order as discussed in the text.}
\label{fig:FigS15}
\end{figure}

The magnetic unit cell is twice as large as the original honeycomb unit cell and it contains four atoms (see Fig.~\ref{fig:FigS15}). Apart from the lattice translation symmetry, we have:
\begin{itemize}
    \item Magnetic translation $T_{1/2}\mathcal{T}$: translation along the $A_1-A_2$ direction followed by time reversal $\mathcal{T}$;
    \item Inversion $\mathcal{I}$: inversion centers at the midpoint of the $A_1-B_1$ bond and the $A_2-B_2$ bond;
    \item $C_2\mathcal{T}$: $\pi$ rotation along the $A_1-B_2$ axis followed by time reversal.
\end{itemize}
All other symmetries can be generated by the above transformations. In order to facilitate the discussion of degeneracy in the magnon spectrum, it is useful to consider the symmetry actions in terms of gliding planes and screw axes. There is a screw axis $2_1$ along the $y$-axis and a gliding plane $a$ lying in the $xz$-plane followed by time reversal, denoted by $a\mathcal{T}$. Furthermore, $C_2\mathcal{T}$ and $\mathcal{I}$ can be combined, which yields a mirror symmetry, labeled $\sigma$ (see Fig.~\ref{fig:FigS15}).

\subsubsection{\normalsize D. Magnon dispersion}

We perform the Holstein-Primakoff (HP) transformation to find the magnon dispersion. The general form of the HP transformation is:
\begin{equation}
    \begin{split}
        S_+ &= \hbar \sqrt{2S-a^\dag a}a\\
        S_- &= \hbar a^\dag\sqrt{2S-a^\dag a}\\
        S_z &=\hbar (S-a^\dag a).
    \end{split}
\end{equation}
In the zigzag order, there are four atoms per magnetic unit cell labeled $A_1$, $A_2$, $B_1$, and $B_2$ (see Fig.~\ref{fig:FigS15}). Let the spins on the $A_1$ and $B_1$ sites point in the positive $x$-direction and the spins on the $A_2$ and $B_2$ sites point in the negative $x$-direction, and we change the basis for the HP transformation. We ignore the higher order terms and set $\hbar =1$. Then we have
\begin{equation}
    \begin{split}
        &S_{A_1,i,+}\equiv S_{A_1,i,y}+ i S_{A_1,i,z} \approx\sqrt{2S} b_{A_1,i}\\
        &S_{A_1,i,-}\equiv S_{A_1,i,y}- i S_{A_1,i,z}\approx \sqrt{2S} b_{A_1,i}^\dag\\
        &S_{A_1,i,x} = S-b_{A_1,i}^\dag b_{A_1,i}\\
        &S_{B_1,i,+}\equiv S_{B_1,i,y}+ i S_{B_1,i,z} \approx\sqrt{2S} b_{B_1,i}\\
        &S_{B_1,i,-}\equiv S_{B_1,i,y}- i S_{B_1,i,z}\approx \sqrt{2S} b_{B_1,i}^\dag\\
        &S_{B_1,i,x} = S-b_{B_1,i}^\dag b_{B_1,i}\\
        &S_{A_2,i,+}\equiv S_{A_2,i,y}+ i S_{A_2,i,z} \approx\sqrt{2S} b_{A_2,i}^\dag\\
        &S_{A_2,i,-}\equiv S_{A_2,i,y}- i S_{A_2,i,z}\approx \sqrt{2S} b_{A_2,i}\\
        &S_{A_2,i,x} = -S+b_{A_2,i}^\dag b_{A_2,i}\\
        &S_{B_2,i,+}\equiv S_{B_2,i,y}+ i S_{B_2,i,z} \approx\sqrt{2S} b_{B_2,i}^\dag\\
        &S_{B_2,i,-}\equiv S_{B_2,i,y}- i S_{B_2,i,z}\approx \sqrt{2S} b_{B_2,i}\\
        &S_{B_2,i,x} = -S+b_{B_2,i}^\dag b_{B_2,i}.
    \end{split}
\end{equation}
Plugging in the HP transformation and taking the limit $\langle b^\dag b\rangle\ll S$, we can ignore the higher order terms and only keep the boson bilinear terms. Then we obtain
\begin{widetext}
\begin{equation}
\begin{split}
    H_{Spin} &= (-2J_1+4J_2+6J_3-2D_x+D_z+2J_2P_2(k) (1+\alpha))S\sum_a b_{a,k}^\dag b_{a,k}\\
    &+ J_1 S(1-\alpha)P_1^*(k) (b_{A_1,k} b_{B_1,-k}+b_{A_2,k} b_{B_2,-k})+ J_1 S(1+\alpha)P_1(k) (b_{A_1,k}^\dag b_{B_1,k}+b_{A_2,k}^\dag b_{B_2,k})+h.c.\\
    &+(J_2 S(1-\alpha)P_2(k)-D_z S/2) (b_{A_1,k} b_{A_1,-k}+b_{A_2,k} b_{A_2,-k}+b_{B_1,k} b_{B_1,-k}+b_{B_2,k} b_{B_2,-k})+h.c.\\
    &+J_2 S (1+\alpha) P_3(k)(b_{A_1,k} b_{A_2,-k}+b_{B_1,k} b_{B_2,-k}) + J_2 S (1-\alpha) P_3(k)(b_{A_1,k}^\dag b_{A_2,k} + b_{B_1,k}^\dag b_{B_2,k})+h.c.\\
    &+[J_1 S(1+\alpha)P_4^*(k)+J_3 S(1+\alpha)P_5^*(k)](b_{A_1,k}b_{B_2,-k}+b_{A_2,k}b_{B_1,-k})
    \\& +[J_1 S(1-\alpha)P_4(k)+J_3 S(1-\alpha)P_5(k)](b_{A_1,k}^\dag b_{B_2,k}+b_{A_2,k}^\dag b_{B_1,k})+h.c.,
\end{split}
\end{equation}
where $P_1(k) = e^{i(\sqrt{3} k_x/2 +k_y/2)}+e^{i(-\sqrt{3} k_x/2 +k_y/2)}$, $P_2(k) = 2\cos(\sqrt{3}k_x)$, $P_3(k) = 2\cos(\sqrt{3}k_x/2 + 3k_y/2)+ 2\cos(-\sqrt{3}k_x/2 + 3k_y/2)$, $P_4(k) = e^{-ik_y}$, and $P_5(k) = e^{2 i k_y}+ e^{-i (\sqrt{3} k_x + k_y)}+e^{i (\sqrt{3} k_x - k_y)}$, and we ignore the constant term.

\noindent For simplicity, let us consider the $\alpha=1$, $J_1=J_2=0$ case:
\begin{equation}
\begin{split}
    H_{Spin} &= (6J_3-2D_x+D_z)S\sum_a b_{a,k}^\dag b_{a,k}+(-D_z S/2) (b_{A_1,k} b_{A_1,-k}+b_{A_2,k} b_{A_2,-k}\\ &+b_{B_1,k} b_{B_1,-k}+b_{B_2,k} b_{B_2,-k})+h.c.+2J_3SP_5^*(k)(b_{A_1,k}b_{B_2,-k}+b_{A_2,k}b_{B_1,-k}).
\end{split}
\end{equation}
The Hamiltonian is block-diagonalized into two $4\times 4$ matrices: $A_1$ and $B_2$ are coupled and $A_2$ and $B_1$ are coupled. The lowest-energy magnon dispersion is
\begin{equation}
    \epsilon_k = \sqrt{36J_3^2-4J_3^2|P_5(k)|^2+12J_3D_z-4J_3D_z|P_5(k)|-4D_x(D_z+6J_3)+4D_x^2}
\end{equation}
and the magnon gap is given by $\epsilon_0 = \sqrt{-4D_x(D_z+6J_3)+4D_x^2}$.
\end{widetext}

\begin{figure}[htb!]
\includegraphics[width=0.96\columnwidth]{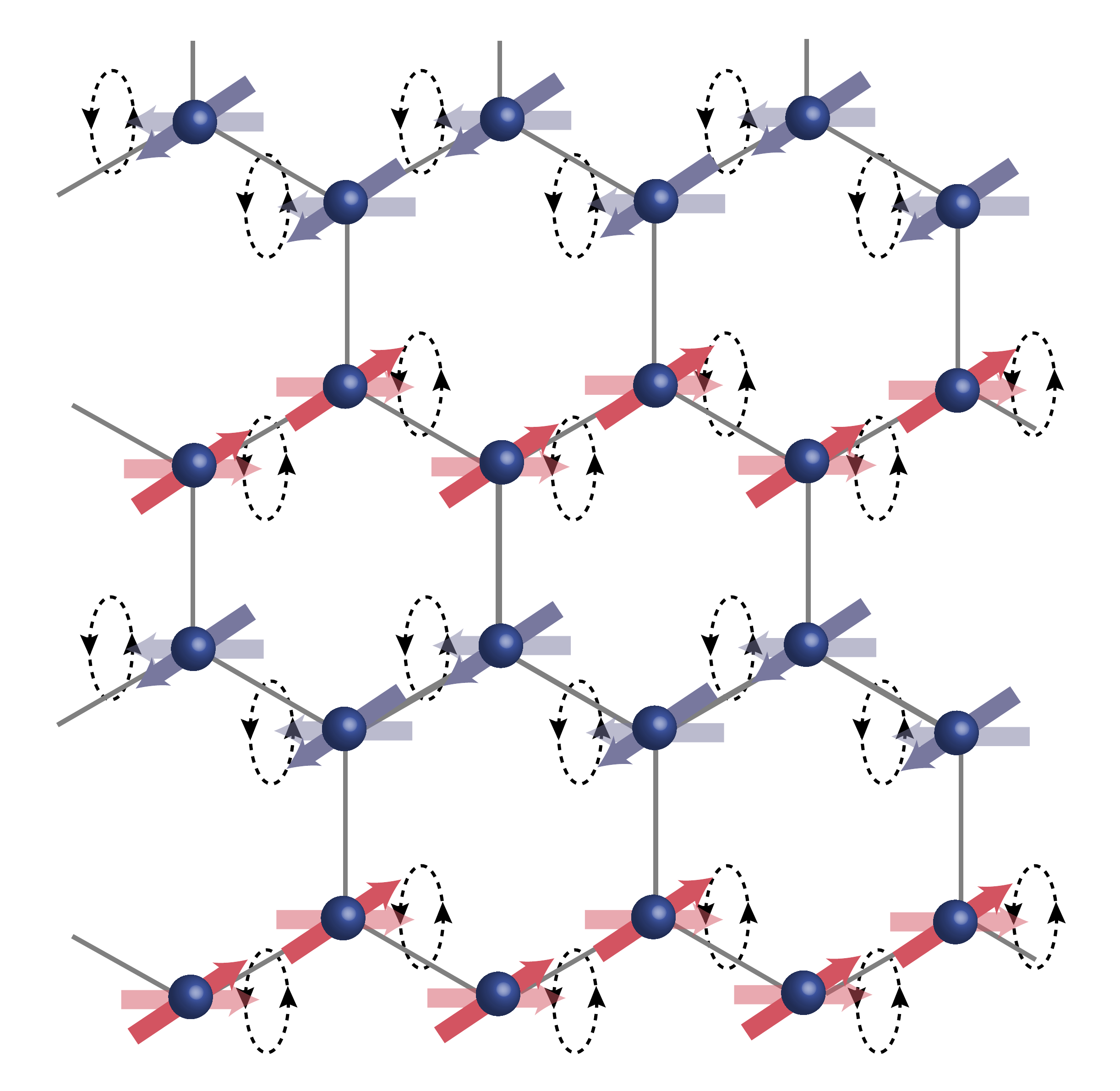}
\caption{\textbf{Spin precession of the lowest-energy magnon.} Pictorial representation of the spin precession corresponding to the lowest-energy magnon mode. This magnon has zero wavevector, which implies that all the spins precess in phase with each other. The net magnetization remains zero since opposite spins precess with opposite chirality.}
\label{fig:FigS16}
\end{figure}

Let us consider the lowest-energy magnon at $k=0$, labeled by the bosonic field $c_0$. Projecting the $b$ fields to the lowest mode, we obtain
\begin{equation}
    \begin{split}
        &b_{A_1} = -\sqrt{\frac{J_0 - \epsilon_0}{4\epsilon_0}}c_0^\dag -\sqrt{\frac{J_0 + \epsilon_0}{4\epsilon_0}}c_0\\
        &b_{B_2} = -b_{A_1}
    \end{split}
\end{equation}
where $J_0 = 6J_3-2D_x+D_z$. In the Heisenberg picture, we have $c_0(t) = c_0 e^{-i\epsilon_0 t}$. The precession of the spins in real space is therefore given by
\begin{equation}
    \begin{split}
        &S_{A_1,y} = A\cos{(\epsilon_0 t+\phi)}\\
        &S_{A_1,z} = A'\sin{(\epsilon_0 t+\phi)}\\
         &S_{B_2,y} =- S_{A_1,y}\\
        &S_{B_2,z} = -S_{A_1,z}.
    \end{split}
\end{equation}
This spin precession is illustrated in Fig.~\ref{fig:FigS16}. There is no spatial dependence since we consider the $k=0$ mode. We note that $A$, $A'$, and $\phi$ depend on $J_0$, $\epsilon_0$, and $c_0$. The matrices $A_2$ and $B_1$ have the same relations as $A_1$ and $B_2$. The minus signs in the above equations indicate that the spins on sites $A_1$ and $B_2$ precess with opposite chirality, and thus the net magnetization of the sample remains zero.

\begin{figure}[htb!]
\includegraphics[width=\columnwidth]{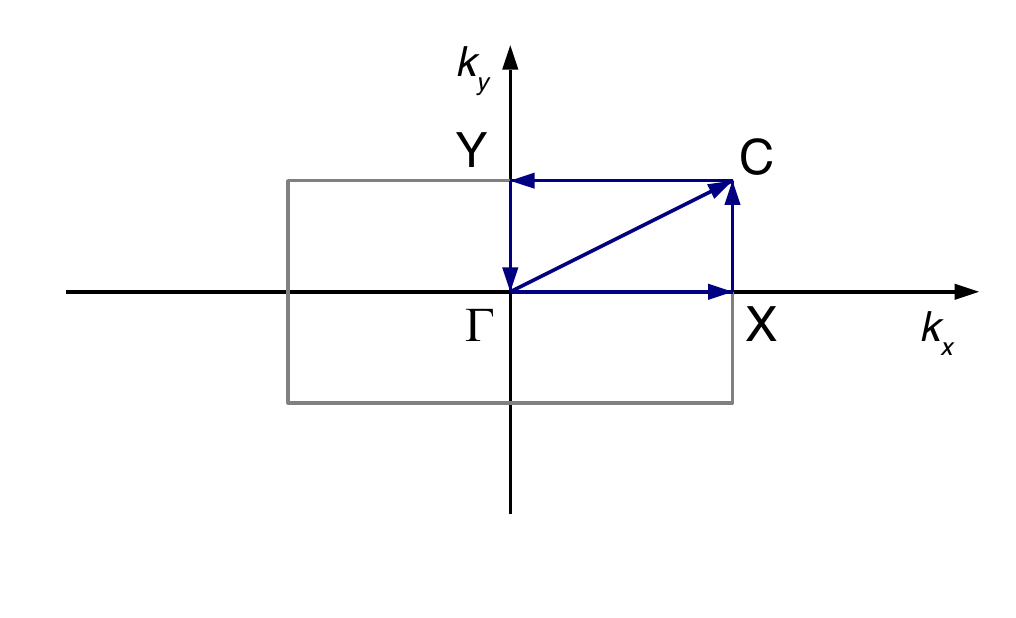}
\caption{\textbf{Magnetic Brillouin zone.} The gray rectangle denotes the magnetic Brillouin zone. The blue arrows represent cuts along which the magnon dispersion is determined (see Figs.~\ref{fig:FigS18} and~\ref{fig:FigS19}).}
\label{fig:FigS17}
\end{figure}

\begin{figure}[htb!]
\includegraphics[width=\columnwidth]{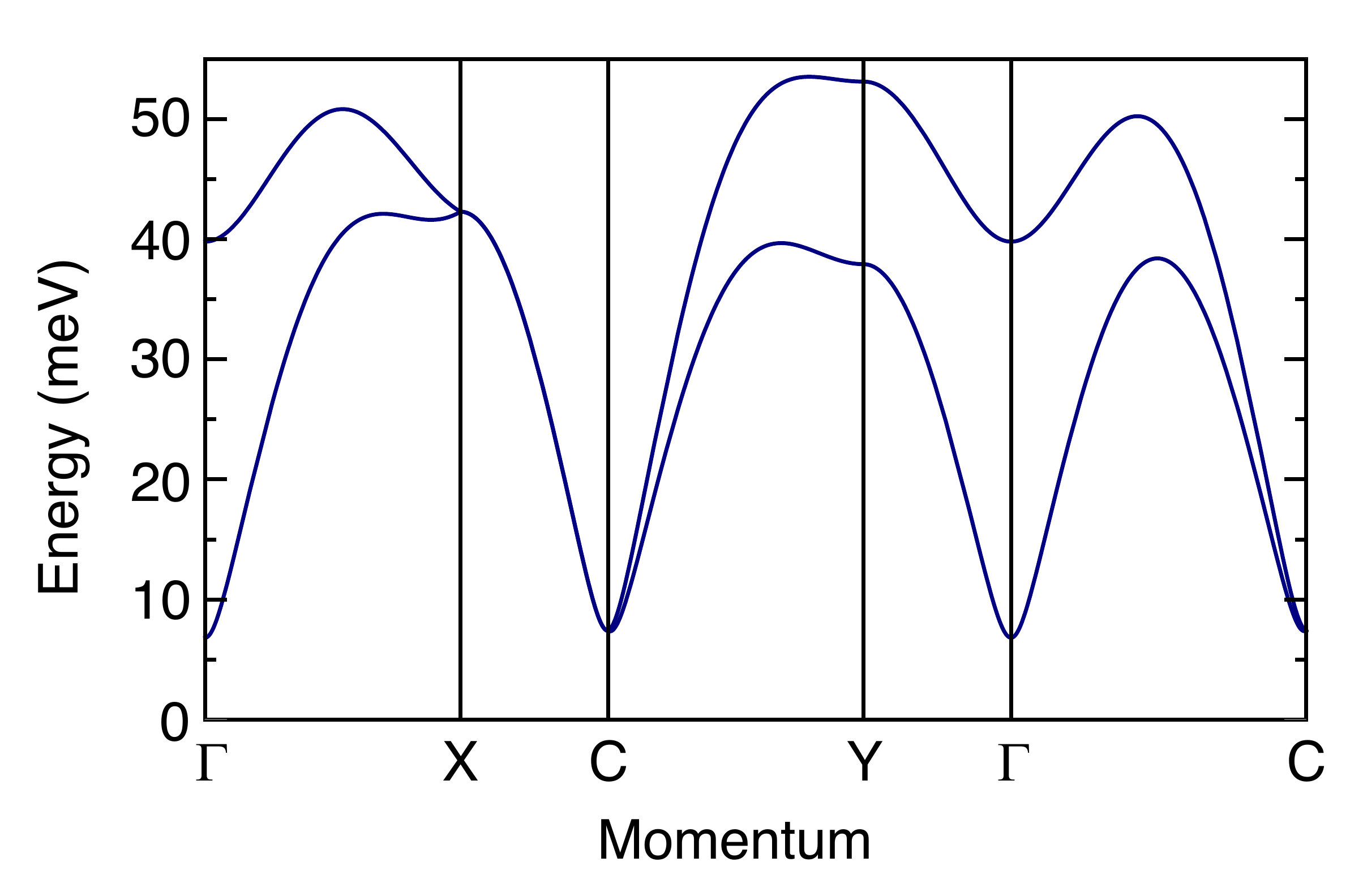}
\caption{\textbf{Magnon dispersion with $SO(2)$ symmetry.} Magnon dispersion along certain cuts in the magnetic Brillouin zone (see Fig.~\ref{fig:FigS17}) when $SO(2)$ symmetry is preserved. There is a four-fold degeneracy along X-C. The band minimum is at the zone center with $E_\Gamma \approx 6.81$~meV. There is another magnon at the zone corner with energy $E_C\approx 7.39$~meV, which is close to $E_\Gamma$.}
\label{fig:FigS18}
\end{figure}

By diagonalizing the full spin Hamiltonian, we obtain the magnon dispersion. Following the model in Ref. \cite{lanccon2018magnetic}, we have $J_1=-1.9$~meV, $J_2=0.1$~meV, $J_3=6.9$~meV, $\alpha=1$, $D_x=-0.3$~meV, and $D_z=0$~meV. Note that the model has $SO(2)$ symmetry corresponding to a spin rotation along the $x$-axis. As a result, the magnon dispersion has at least a two-fold degeneracy. However, the $SO(2)$ symmetry is only approximate and we expect the symmetry breaking terms in the spin Hamiltonian to be nonzero in general so that the two-fold degeneracy will be lifted. The dispersion along a certain cut in the magnetic Brillouin zone (see Fig.~\ref{fig:FigS17}) is shown in Fig.~\ref{fig:FigS18}. There is a four-fold degeneracy along X-C in the magnetic Brillouin zone. The band minimum is at the zone center with $E_\Gamma \approx 6.81$~meV. There is another magnon at the zone corner with energy $E_C\approx 7.39$~meV, which is close to $E_\Gamma$.

The four-fold degeneracy can be explained by the space group symmetry of NiPS$_3$. We first examine the two-fold degeneracy of each magnon band. The $SO(2)$ spin rotation acts on the HP boson as a $U(1)$ rotation, i.e. $b_{A_1,B_1} \rightarrow e^{i\theta} b_{A_1,B_1}$ and $b_{A_2,B_2}\rightarrow e^{-i\theta} b_{A_2,B_2}$. The Hamiltonian is block-diagonalized into two $4\times4$ matrices, and the magnetic translation $T_{1/2}\mathcal{T}$ acts as $(b_{A_1},b_{A_2},b_{B_1},b_{B_2})\rightarrow (b_{A_2},b_{A_1},b_{B_2},b_{B_1})$, which ensures that the two blocks in the spin Hamiltonian are identical. The odd and even parity under exchanging $A_1,A_2$ and $B_1,B_2$ are essentially degenerate.

Now, we consider the screw axis $2_1$ and magnetic translation $T_{1/2}\mathcal{T}$. Note that acting on the lattice, we have $2_1 =\{ \begin{psmallmatrix}-1&0\\0&1\end{psmallmatrix}| (\frac{1}{2},\frac{1}{2})\}$ and $T_{1/2}\mathcal{T} =\{ \begin{psmallmatrix}1&0\\0&1\end{psmallmatrix}| (\frac{1}{2},\frac{1}{2})\}$. Here, the notation $\{M|\Vec{b}\}$ means that a symmetry acts on the coordinates in the 2D plane as $\Vec{x}\rightarrow M\Vec{x}+\Vec{b}$, where $M$ is a $2\times 2$ matrix and $\Vec{b}$ is a 2D vector. We find that $2_1$ and $T_{1/2}\mathcal{T}$ do not commute with each other. Also, we have $T_x(2_1)( T_{1/2}\mathcal{T}) = (T_{1/2}\mathcal{T})(2_1)$, where $T_x$ denotes a translation along the $x$-axis by one unit cell. At $k_x=\pi$ (the Brillouin zone boundary along the $x$-axis) $T_x = -1$ so $-(2_1)( T_{1/2}\mathcal{T}) = (T_{1/2}\mathcal{T})( 2_1)$. Note that $2_1$ does not alter the parity of exchanging $A_1A_2$ and $B_1B_2$ so its action is within one $4\times 4$ block of the spin Hamiltonian. There will be an extra degeneracy, i.e. a four-fold degeneracy, along the X-C path.

\begin{figure}[htb!]
\includegraphics[width=1.03\columnwidth]{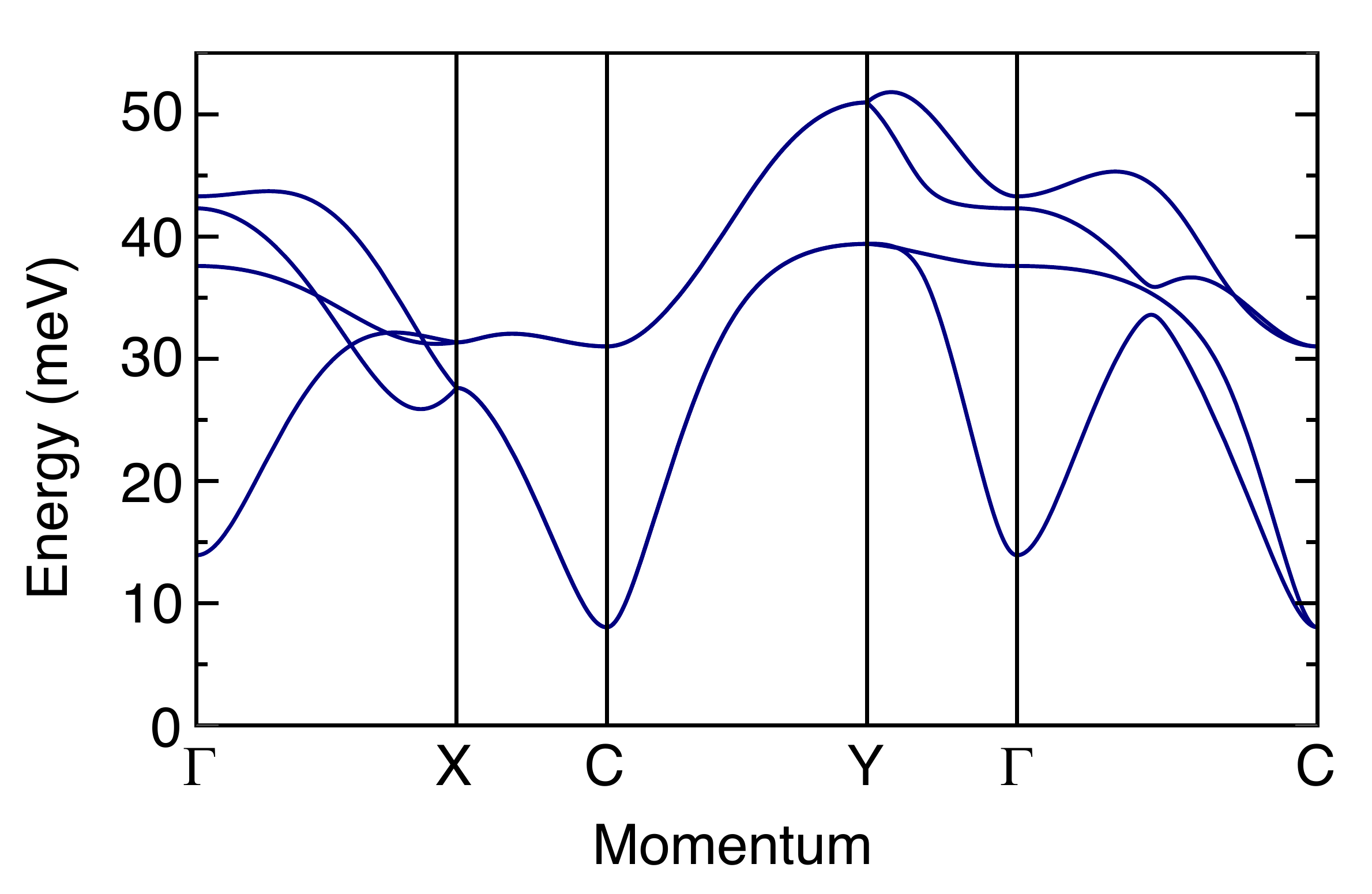}
\caption{\textbf{Magnon dispersion without $SO(2)$ symmetry.} Magnon dispersion along certain cuts in the magnetic Brillouin zone (see Fig.~\ref{fig:FigS17}) when $SO(2)$ symmetry is broken. In general, the magnon bands split into four bands but there are two-fold degeneracies along X-C and C-Y. These degeneracies can be explained by the space group symmetry of NiPS$_3$.}
\label{fig:FigS19}
\end{figure}

Let us further consider what happens to the magnon spectrum if the $SO(2)$ spin rotation is broken. We use the parameters in Ref. \cite{kim2019suppression}: $J_1=1.59$~meV, $J_2 =2.41$~meV, $J_3=4.54$~meV, $\alpha=0.66$, $D_x = -0.89$~meV, and $D_z=2.85$~meV. We can see that in general the magnon bands split into four bands but there are two-fold degeneracies along X-C and C-Y (Fig.~\ref{fig:FigS19}). These degeneracies can be explained by the space group symmetry. First, let us consider $2_1$ and $C_2\mathcal{T}$. These symmetry actions yield $2_1 =\{ \begin{psmallmatrix}-1&0\\0&1\end{psmallmatrix}| (\frac{1}{2},\frac{1}{2})\}$ and $C_2\mathcal{T} =\{ \begin{psmallmatrix}-1&0\\0&1\end{psmallmatrix}| (\frac{1}{2},\frac{1}{2})\}$, and they do not commute with each other. We have $T_x(2_1)(C_2\mathcal{T}) = (C_2\mathcal{T})( 2_1)$ so at $k_x = \pi$, we have $-(2_1)(C_2\mathcal{T}) = (C_2\mathcal{T})( 2_1)$. This yields at least two-fold degeneracy along X-C in the Brillouin zone. Similar arguments can be carried out for $k_y=\pi$ (the Brillouin zone boundary along the $y$-axis) if we consider the symmetries $a\mathcal{T}$ and $\sigma$. Note that $a\mathcal{T} = \{ \begin{psmallmatrix}1&0\\0&-1\end{psmallmatrix}| (\frac{1}{2},\frac{1}{6})\}$ and $\sigma = \{ \begin{psmallmatrix}1&0\\0&-1\end{psmallmatrix}| (0,\frac{2}{3})\}$ and we have $T_y(a\mathcal{T})(\sigma) = (\sigma)(a\mathcal{T})$, where $T_y$ denotes a translation along the $y$-axis by one unit vector. The anticommutation of the symmetries $a\mathcal{T}$ and $\sigma$ along the Brillouin zone boundary in the $y$-direction gives the two-fold degeneracy along C-Y.

Note that a previous study on NiPS$_3$ has also observed the magnetization axis slightly tilted away from the honeycomb plane \cite{kim2019suppression}. This will not change the symmetry properties and our arguments about the extra degeneracies still hold.

\subsubsection{\normalsize E. The effect of photoexcitation}

In the main text, we discuss the dynamics of the photoexcitation mechanism in which the pump photon energy is in the vicinity of the spin--orbit-entangled excitons. Here, we provide a rough estimate of the scale of the magnon energy redshift. The magnon gap is given by $\epsilon_0 = \sqrt{-4D_x(D_z+6J_3)+4D_x^2}S$ in the simplified model of the previous section. The reduction in the magnetization is $\Delta S = \Delta n/N$, where $\Delta n$ is the photogenerated exciton density, which is proportional to the absorbed pump laser fluence, and $N$ is the total number of Ni sites. Therefore, qualitatively the redshift of the magnon energy is given by $\Delta \epsilon_0 / \epsilon_0 = \Delta n/N$. Note that we have neglected the potential change in the spin-spin exchange coupling and the quantum fluctuation of spin-1/2, so the result we obtain above only provides a qualitative understanding of the linear dependence of the magnon energy on the absorbed fluence.

\clearpage
\bibliography{NiPS3}

\providecommand{\noopsort}[1]{}\providecommand{\singleletter}[1]{#1}%
\begin{thebibliography}{71}%
\makeatletter
\providecommand \@ifxundefined [1]{%
 \@ifx{#1\undefined}
}%
\providecommand \@ifnum [1]{%
 \ifnum #1\expandafter \@firstoftwo
 \else \expandafter \@secondoftwo
 \fi
}%
\providecommand \@ifx [1]{%
 \ifx #1\expandafter \@firstoftwo
 \else \expandafter \@secondoftwo
 \fi
}%
\providecommand \natexlab [1]{#1}%
\providecommand \enquote  [1]{``#1''}%
\providecommand \bibnamefont  [1]{#1}%
\providecommand \bibfnamefont [1]{#1}%
\providecommand \citenamefont [1]{#1}%
\providecommand \href@noop [0]{\@secondoftwo}%
\providecommand \href [0]{\begingroup \@sanitize@url \@href}%
\providecommand \@href[1]{\@@startlink{#1}\@@href}%
\providecommand \@@href[1]{\endgroup#1\@@endlink}%
\providecommand \@sanitize@url [0]{\catcode `\\12\catcode `\$12\catcode
  `\&12\catcode `\#12\catcode `\^12\catcode `\_12\catcode `\%12\relax}%
\providecommand \@@startlink[1]{}%
\providecommand \@@endlink[0]{}%
\providecommand \url  [0]{\begingroup\@sanitize@url \@url }%
\providecommand \@url [1]{\endgroup\@href {#1}{\urlprefix }}%
\providecommand \urlprefix  [0]{URL }%
\providecommand \Eprint [0]{\href }%
\providecommand \doibase [0]{http://dx.doi.org/}%
\providecommand \selectlanguage [0]{\@gobble}%
\providecommand \bibinfo  [0]{\@secondoftwo}%
\providecommand \bibfield  [0]{\@secondoftwo}%
\providecommand \translation [1]{[#1]}%
\providecommand \BibitemOpen [0]{}%
\providecommand \bibitemStop [0]{}%
\providecommand \bibitemNoStop [0]{.\EOS\space}%
\providecommand \EOS [0]{\spacefactor3000\relax}%
\providecommand \BibitemShut  [1]{\csname bibitem#1\endcsname}%
\let\auto@bib@innerbib\@empty
\bibitem [{\citenamefont {Khomskii}(2014)}]{khomskii2014transition}%
  \BibitemOpen
  \bibfield  {author} {\bibinfo {author} {\bibfnamefont {D.~I.}\ \bibnamefont
  {Khomskii}},\ }\href@noop {} {\emph {\bibinfo {title} {{Transition Metal
  Compounds}}}}\ (\bibinfo  {publisher} {Cambridge Univ. Press, Cambridge},\
  \bibinfo {year} {2014})\BibitemShut {NoStop}%
\bibitem [{\citenamefont {Kim}\ \emph {et~al.}(2014)\citenamefont {Kim},
  \citenamefont {Daghofer}, \citenamefont {Said}, \citenamefont {Gog},
  \citenamefont {Van~den Brink}, \citenamefont {Khaliullin},\ and\
  \citenamefont {Kim}}]{kim2014excitonic}%
  \BibitemOpen
  \bibfield  {author} {\bibinfo {author} {\bibfnamefont {J.}~\bibnamefont
  {Kim}}, \bibinfo {author} {\bibfnamefont {M.}~\bibnamefont {Daghofer}},
  \bibinfo {author} {\bibfnamefont {A.}~\bibnamefont {Said}}, \bibinfo {author}
  {\bibfnamefont {T.}~\bibnamefont {Gog}}, \bibinfo {author} {\bibfnamefont
  {J.}~\bibnamefont {Van~den Brink}}, \bibinfo {author} {\bibfnamefont
  {G.}~\bibnamefont {Khaliullin}}, \ and\ \bibinfo {author} {\bibfnamefont
  {B.}~\bibnamefont {Kim}},\ }\href@noop {} {\bibfield  {journal} {\bibinfo
  {journal} {Nat. Commun.}\ }\textbf {\bibinfo {volume} {5}},\ \bibinfo {pages}
  {4453} (\bibinfo {year} {2014})}\BibitemShut {NoStop}%
\bibitem [{\citenamefont {Warzanowski}\ \emph {et~al.}(2020)\citenamefont
  {Warzanowski}, \citenamefont {Borgwardt}, \citenamefont {Hopfer},
  \citenamefont {Attig}, \citenamefont {Koethe}, \citenamefont {Becker},
  \citenamefont {Tsurkan}, \citenamefont {Loidl}, \citenamefont {Hermanns},
  \citenamefont {van Loosdrecht} \emph {et~al.}}]{warzanowski2020multiple}%
  \BibitemOpen
  \bibfield  {author} {\bibinfo {author} {\bibfnamefont {P.}~\bibnamefont
  {Warzanowski}}, \bibinfo {author} {\bibfnamefont {N.}~\bibnamefont
  {Borgwardt}}, \bibinfo {author} {\bibfnamefont {K.}~\bibnamefont {Hopfer}},
  \bibinfo {author} {\bibfnamefont {J.}~\bibnamefont {Attig}}, \bibinfo
  {author} {\bibfnamefont {T.~C.}\ \bibnamefont {Koethe}}, \bibinfo {author}
  {\bibfnamefont {P.}~\bibnamefont {Becker}}, \bibinfo {author} {\bibfnamefont
  {V.}~\bibnamefont {Tsurkan}}, \bibinfo {author} {\bibfnamefont
  {A.}~\bibnamefont {Loidl}}, \bibinfo {author} {\bibfnamefont
  {M.}~\bibnamefont {Hermanns}}, \bibinfo {author} {\bibfnamefont {P.~H.~M.}\
  \bibnamefont {van Loosdrecht}},  \emph {et~al.},\ }\href@noop {} {\bibfield
  {journal} {\bibinfo  {journal} {Phys. Rev. Research}\ }\textbf {\bibinfo
  {volume} {2}},\ \bibinfo {pages} {042007} (\bibinfo {year}
  {2020})}\BibitemShut {NoStop}%
\bibitem [{\citenamefont {Jeckelmann}(2003)}]{jeckelmann2003optical}%
  \BibitemOpen
  \bibfield  {author} {\bibinfo {author} {\bibfnamefont {E.}~\bibnamefont
  {Jeckelmann}},\ }\href@noop {} {\bibfield  {journal} {\bibinfo  {journal}
  {Phys. Rev. B}\ }\textbf {\bibinfo {volume} {67}},\ \bibinfo {pages} {075106}
  (\bibinfo {year} {2003})}\BibitemShut {NoStop}%
\bibitem [{\citenamefont {Kang}\ \emph {et~al.}(2020)\citenamefont {Kang},
  \citenamefont {Kim}, \citenamefont {Kim}, \citenamefont {Kim}, \citenamefont
  {Sim}, \citenamefont {Lee}, \citenamefont {Lee}, \citenamefont {Park},
  \citenamefont {Yun}, \citenamefont {Kim} \emph {et~al.}}]{kang2020coherent}%
  \BibitemOpen
  \bibfield  {author} {\bibinfo {author} {\bibfnamefont {S.}~\bibnamefont
  {Kang}}, \bibinfo {author} {\bibfnamefont {K.}~\bibnamefont {Kim}}, \bibinfo
  {author} {\bibfnamefont {B.~H.}\ \bibnamefont {Kim}}, \bibinfo {author}
  {\bibfnamefont {J.}~\bibnamefont {Kim}}, \bibinfo {author} {\bibfnamefont
  {K.~I.}\ \bibnamefont {Sim}}, \bibinfo {author} {\bibfnamefont {J.-U.}\
  \bibnamefont {Lee}}, \bibinfo {author} {\bibfnamefont {S.}~\bibnamefont
  {Lee}}, \bibinfo {author} {\bibfnamefont {K.}~\bibnamefont {Park}}, \bibinfo
  {author} {\bibfnamefont {S.}~\bibnamefont {Yun}}, \bibinfo {author}
  {\bibfnamefont {T.}~\bibnamefont {Kim}},  \emph {et~al.},\ }\href@noop {}
  {\bibfield  {journal} {\bibinfo  {journal} {Nature}\ }\textbf {\bibinfo
  {volume} {583}},\ \bibinfo {pages} {785} (\bibinfo {year}
  {2020})}\BibitemShut {NoStop}%
\bibitem [{\citenamefont {Wildes}\ \emph {et~al.}(2015)\citenamefont {Wildes},
  \citenamefont {Simonet}, \citenamefont {Ressouche}, \citenamefont {Mcintyre},
  \citenamefont {Avdeev}, \citenamefont {Suard}, \citenamefont {Kimber},
  \citenamefont {Lan{\c{c}}on}, \citenamefont {Pepe}, \citenamefont {Moubaraki}
  \emph {et~al.}}]{wildes2015magnetic}%
  \BibitemOpen
  \bibfield  {author} {\bibinfo {author} {\bibfnamefont {A.~R.}\ \bibnamefont
  {Wildes}}, \bibinfo {author} {\bibfnamefont {V.}~\bibnamefont {Simonet}},
  \bibinfo {author} {\bibfnamefont {E.}~\bibnamefont {Ressouche}}, \bibinfo
  {author} {\bibfnamefont {G.~J.}\ \bibnamefont {Mcintyre}}, \bibinfo {author}
  {\bibfnamefont {M.}~\bibnamefont {Avdeev}}, \bibinfo {author} {\bibfnamefont
  {E.}~\bibnamefont {Suard}}, \bibinfo {author} {\bibfnamefont {S.~A.~J.}\
  \bibnamefont {Kimber}}, \bibinfo {author} {\bibfnamefont {D.}~\bibnamefont
  {Lan{\c{c}}on}}, \bibinfo {author} {\bibfnamefont {G.}~\bibnamefont {Pepe}},
  \bibinfo {author} {\bibfnamefont {B.}~\bibnamefont {Moubaraki}},  \emph
  {et~al.},\ }\href@noop {} {\bibfield  {journal} {\bibinfo  {journal} {Phys.
  Rev. B}\ }\textbf {\bibinfo {volume} {92}},\ \bibinfo {pages} {224408}
  (\bibinfo {year} {2015})}\BibitemShut {NoStop}%
\bibitem [{\citenamefont {Kim}\ \emph {et~al.}(2018)\citenamefont {Kim},
  \citenamefont {Kim}, \citenamefont {Sandilands}, \citenamefont {Sinn},
  \citenamefont {Lee}, \citenamefont {Son}, \citenamefont {Lee}, \citenamefont
  {Choi}, \citenamefont {Kim}, \citenamefont {Park} \emph
  {et~al.}}]{kim2018charge}%
  \BibitemOpen
  \bibfield  {author} {\bibinfo {author} {\bibfnamefont {S.~Y.}\ \bibnamefont
  {Kim}}, \bibinfo {author} {\bibfnamefont {T.~Y.}\ \bibnamefont {Kim}},
  \bibinfo {author} {\bibfnamefont {L.~J.}\ \bibnamefont {Sandilands}},
  \bibinfo {author} {\bibfnamefont {S.}~\bibnamefont {Sinn}}, \bibinfo {author}
  {\bibfnamefont {M.-C.}\ \bibnamefont {Lee}}, \bibinfo {author} {\bibfnamefont
  {J.}~\bibnamefont {Son}}, \bibinfo {author} {\bibfnamefont {S.}~\bibnamefont
  {Lee}}, \bibinfo {author} {\bibfnamefont {K.-Y.}\ \bibnamefont {Choi}},
  \bibinfo {author} {\bibfnamefont {W.}~\bibnamefont {Kim}}, \bibinfo {author}
  {\bibfnamefont {B.-G.}\ \bibnamefont {Park}},  \emph {et~al.},\ }\href@noop
  {} {\bibfield  {journal} {\bibinfo  {journal} {Phys. Rev. Lett.}\ }\textbf
  {\bibinfo {volume} {120}},\ \bibinfo {pages} {136402} (\bibinfo {year}
  {2018})}\BibitemShut {NoStop}%
\bibitem [{\citenamefont {Zhang}\ and\ \citenamefont
  {Rice}(1988)}]{zhang1988effective}%
  \BibitemOpen
  \bibfield  {author} {\bibinfo {author} {\bibfnamefont {F.~C.}\ \bibnamefont
  {Zhang}}\ and\ \bibinfo {author} {\bibfnamefont {T.~M.}\ \bibnamefont
  {Rice}},\ }\href@noop {} {\bibfield  {journal} {\bibinfo  {journal} {Phys.
  Rev. B}\ }\textbf {\bibinfo {volume} {37}},\ \bibinfo {pages} {3759}
  (\bibinfo {year} {1988})}\BibitemShut {NoStop}%
\bibitem [{\citenamefont {Lan{\c{c}}on}\ \emph {et~al.}(2018)\citenamefont
  {Lan{\c{c}}on}, \citenamefont {Ewings}, \citenamefont {Guidi}, \citenamefont
  {Formisano},\ and\ \citenamefont {Wildes}}]{lanccon2018magnetic}%
  \BibitemOpen
  \bibfield  {author} {\bibinfo {author} {\bibfnamefont {D.}~\bibnamefont
  {Lan{\c{c}}on}}, \bibinfo {author} {\bibfnamefont {R.~A.}\ \bibnamefont
  {Ewings}}, \bibinfo {author} {\bibfnamefont {T.}~\bibnamefont {Guidi}},
  \bibinfo {author} {\bibfnamefont {F.}~\bibnamefont {Formisano}}, \ and\
  \bibinfo {author} {\bibfnamefont {A.~R.}\ \bibnamefont {Wildes}},\
  }\href@noop {} {\bibfield  {journal} {\bibinfo  {journal} {Phys. Rev. B}\
  }\textbf {\bibinfo {volume} {98}},\ \bibinfo {pages} {134414} (\bibinfo
  {year} {2018})}\BibitemShut {NoStop}%
\bibitem [{\citenamefont {Gretarsson}\ \emph {et~al.}(2017)\citenamefont
  {Gretarsson}, \citenamefont {Sauceda}, \citenamefont {Sung}, \citenamefont
  {H{\"o}ppner}, \citenamefont {Minola}, \citenamefont {Kim}, \citenamefont
  {Keimer},\ and\ \citenamefont {Le~Tacon}}]{gretarsson2017raman}%
  \BibitemOpen
  \bibfield  {author} {\bibinfo {author} {\bibfnamefont {H.}~\bibnamefont
  {Gretarsson}}, \bibinfo {author} {\bibfnamefont {J.}~\bibnamefont {Sauceda}},
  \bibinfo {author} {\bibfnamefont {N.~H.}\ \bibnamefont {Sung}}, \bibinfo
  {author} {\bibfnamefont {M.}~\bibnamefont {H{\"o}ppner}}, \bibinfo {author}
  {\bibfnamefont {M.}~\bibnamefont {Minola}}, \bibinfo {author} {\bibfnamefont
  {B.~J.}\ \bibnamefont {Kim}}, \bibinfo {author} {\bibfnamefont
  {B.}~\bibnamefont {Keimer}}, \ and\ \bibinfo {author} {\bibfnamefont
  {M.}~\bibnamefont {Le~Tacon}},\ }\href@noop {} {\bibfield  {journal}
  {\bibinfo  {journal} {Phys. Rev. B}\ }\textbf {\bibinfo {volume} {96}},\
  \bibinfo {pages} {115138} (\bibinfo {year} {2017})}\BibitemShut {NoStop}%
\bibitem [{\citenamefont {Zielbauer}\ and\ \citenamefont
  {Wegener}(1996)}]{zielbauer1996ultrafast}%
  \BibitemOpen
  \bibfield  {author} {\bibinfo {author} {\bibfnamefont {J.}~\bibnamefont
  {Zielbauer}}\ and\ \bibinfo {author} {\bibfnamefont {M.}~\bibnamefont
  {Wegener}},\ }\href@noop {} {\bibfield  {journal} {\bibinfo  {journal} {Appl.
  Phys. Lett.}\ }\textbf {\bibinfo {volume} {68}},\ \bibinfo {pages} {1223}
  (\bibinfo {year} {1996})}\BibitemShut {NoStop}%
\bibitem [{\citenamefont {Castro}(1971)}]{castro1971photoconduction}%
  \BibitemOpen
  \bibfield  {author} {\bibinfo {author} {\bibfnamefont {G.}~\bibnamefont
  {Castro}},\ }\href@noop {} {\bibfield  {journal} {\bibinfo  {journal} {IBM J.
  Res. Develop.}\ }\textbf {\bibinfo {volume} {15}},\ \bibinfo {pages} {27}
  (\bibinfo {year} {1971})}\BibitemShut {NoStop}%
\bibitem [{\citenamefont {Boyd}(2003)}]{boyd2003nonlinear}%
  \BibitemOpen
  \bibfield  {author} {\bibinfo {author} {\bibfnamefont {R.~W.}\ \bibnamefont
  {Boyd}},\ }\href@noop {} {\emph {\bibinfo {title} {{Nonlinear Optics}}}}\
  (\bibinfo  {publisher} {Elsevier, London},\ \bibinfo {year}
  {2003})\BibitemShut {NoStop}%
\bibitem [{\citenamefont {Braun}(1968)}]{braun1968singlet}%
  \BibitemOpen
  \bibfield  {author} {\bibinfo {author} {\bibfnamefont {C.~L.}\ \bibnamefont
  {Braun}},\ }\href@noop {} {\bibfield  {journal} {\bibinfo  {journal} {Phys.
  Rev. Lett.}\ }\textbf {\bibinfo {volume} {21}},\ \bibinfo {pages} {215}
  (\bibinfo {year} {1968})}\BibitemShut {NoStop}%
\bibitem [{\citenamefont {Bergman}\ and\ \citenamefont
  {Jortner}(1974)}]{bergman1974photoconductivity}%
  \BibitemOpen
  \bibfield  {author} {\bibinfo {author} {\bibfnamefont {A.}~\bibnamefont
  {Bergman}}\ and\ \bibinfo {author} {\bibfnamefont {J.}~\bibnamefont
  {Jortner}},\ }\href@noop {} {\bibfield  {journal} {\bibinfo  {journal} {Phys.
  Rev. B}\ }\textbf {\bibinfo {volume} {9}},\ \bibinfo {pages} {4560} (\bibinfo
  {year} {1974})}\BibitemShut {NoStop}%
\bibitem [{\citenamefont {Lee}\ \emph {et~al.}(1993)\citenamefont {Lee},
  \citenamefont {Yu}, \citenamefont {Moses}, \citenamefont {Srdanov},
  \citenamefont {Wei},\ and\ \citenamefont {Vardeny}}]{lee1993transient}%
  \BibitemOpen
  \bibfield  {author} {\bibinfo {author} {\bibfnamefont {C.~H.}\ \bibnamefont
  {Lee}}, \bibinfo {author} {\bibfnamefont {G.}~\bibnamefont {Yu}}, \bibinfo
  {author} {\bibfnamefont {D.}~\bibnamefont {Moses}}, \bibinfo {author}
  {\bibfnamefont {V.~I.}\ \bibnamefont {Srdanov}}, \bibinfo {author}
  {\bibfnamefont {X.}~\bibnamefont {Wei}}, \ and\ \bibinfo {author}
  {\bibfnamefont {Z.~V.}\ \bibnamefont {Vardeny}},\ }\href@noop {} {\bibfield
  {journal} {\bibinfo  {journal} {Phys. Rev. B}\ }\textbf {\bibinfo {volume}
  {48}},\ \bibinfo {pages} {8506} (\bibinfo {year} {1993})}\BibitemShut
  {NoStop}%
\bibitem [{\citenamefont {Sun}\ \emph {et~al.}(2014)\citenamefont {Sun},
  \citenamefont {Rao}, \citenamefont {Reider}, \citenamefont {Chen},
  \citenamefont {You}, \citenamefont {Brézin}, \citenamefont {Harutyunyan},\
  and\ \citenamefont {Heinz}}]{sun2014observation}%
  \BibitemOpen
  \bibfield  {author} {\bibinfo {author} {\bibfnamefont {D.}~\bibnamefont
  {Sun}}, \bibinfo {author} {\bibfnamefont {Y.}~\bibnamefont {Rao}}, \bibinfo
  {author} {\bibfnamefont {G.~A.}\ \bibnamefont {Reider}}, \bibinfo {author}
  {\bibfnamefont {G.}~\bibnamefont {Chen}}, \bibinfo {author} {\bibfnamefont
  {Y.}~\bibnamefont {You}}, \bibinfo {author} {\bibfnamefont {L.}~\bibnamefont
  {Brézin}}, \bibinfo {author} {\bibfnamefont {A.~R.}\ \bibnamefont
  {Harutyunyan}}, \ and\ \bibinfo {author} {\bibfnamefont {T.~F.}\ \bibnamefont
  {Heinz}},\ }\href@noop {} {\bibfield  {journal} {\bibinfo  {journal} {Nano
  Lett.}\ }\textbf {\bibinfo {volume} {14}},\ \bibinfo {pages} {5625} (\bibinfo
  {year} {2014})}\BibitemShut {NoStop}%
\bibitem [{\citenamefont {Mikhaylovskiy}\ \emph {et~al.}(2015)\citenamefont
  {Mikhaylovskiy}, \citenamefont {Hendry}, \citenamefont {Secchi},
  \citenamefont {Mentink}, \citenamefont {Eckstein}, \citenamefont {Wu},
  \citenamefont {Pisarev}, \citenamefont {Kruglyak}, \citenamefont
  {Katsnelson}, \citenamefont {Rasing} \emph
  {et~al.}}]{mikhaylovskiy2015ultrafast}%
  \BibitemOpen
  \bibfield  {author} {\bibinfo {author} {\bibfnamefont {R.~V.}\ \bibnamefont
  {Mikhaylovskiy}}, \bibinfo {author} {\bibfnamefont {E.}~\bibnamefont
  {Hendry}}, \bibinfo {author} {\bibfnamefont {A.}~\bibnamefont {Secchi}},
  \bibinfo {author} {\bibfnamefont {J.~H.}\ \bibnamefont {Mentink}}, \bibinfo
  {author} {\bibfnamefont {M.}~\bibnamefont {Eckstein}}, \bibinfo {author}
  {\bibfnamefont {A.}~\bibnamefont {Wu}}, \bibinfo {author} {\bibfnamefont
  {R.~V.}\ \bibnamefont {Pisarev}}, \bibinfo {author} {\bibfnamefont {V.~V.}\
  \bibnamefont {Kruglyak}}, \bibinfo {author} {\bibfnamefont {M.}~\bibnamefont
  {Katsnelson}}, \bibinfo {author} {\bibfnamefont {T.}~\bibnamefont {Rasing}},
  \emph {et~al.},\ }\href@noop {} {\bibfield  {journal} {\bibinfo  {journal}
  {Nat. Commun.}\ }\textbf {\bibinfo {volume} {6}},\ \bibinfo {pages} {8190}
  (\bibinfo {year} {2015})}\BibitemShut {NoStop}%
\bibitem [{\citenamefont {Kirilyuk}\ \emph {et~al.}(2010)\citenamefont
  {Kirilyuk}, \citenamefont {Kimel},\ and\ \citenamefont
  {Rasing}}]{kirilyuk2010ultrafast}%
  \BibitemOpen
  \bibfield  {author} {\bibinfo {author} {\bibfnamefont {A.}~\bibnamefont
  {Kirilyuk}}, \bibinfo {author} {\bibfnamefont {A.~V.}\ \bibnamefont {Kimel}},
  \ and\ \bibinfo {author} {\bibfnamefont {T.}~\bibnamefont {Rasing}},\
  }\href@noop {} {\bibfield  {journal} {\bibinfo  {journal} {Rev. Mod. Phys.}\
  }\textbf {\bibinfo {volume} {82}},\ \bibinfo {pages} {2731} (\bibinfo {year}
  {2010})}\BibitemShut {NoStop}%
\bibitem [{\citenamefont {Battisti}\ \emph {et~al.}(2017)\citenamefont
  {Battisti}, \citenamefont {Bastiaans}, \citenamefont {Fedoseev},
  \citenamefont {De~La~Torre}, \citenamefont {Iliopoulos}, \citenamefont
  {Tamai}, \citenamefont {Hunter}, \citenamefont {Perry}, \citenamefont
  {Zaanen}, \citenamefont {Baumberger} \emph
  {et~al.}}]{battisti2017universality}%
  \BibitemOpen
  \bibfield  {author} {\bibinfo {author} {\bibfnamefont {I.}~\bibnamefont
  {Battisti}}, \bibinfo {author} {\bibfnamefont {K.~M.}\ \bibnamefont
  {Bastiaans}}, \bibinfo {author} {\bibfnamefont {V.}~\bibnamefont {Fedoseev}},
  \bibinfo {author} {\bibfnamefont {A.}~\bibnamefont {De~La~Torre}}, \bibinfo
  {author} {\bibfnamefont {N.}~\bibnamefont {Iliopoulos}}, \bibinfo {author}
  {\bibfnamefont {A.}~\bibnamefont {Tamai}}, \bibinfo {author} {\bibfnamefont
  {E.~C.}\ \bibnamefont {Hunter}}, \bibinfo {author} {\bibfnamefont {R.~S.}\
  \bibnamefont {Perry}}, \bibinfo {author} {\bibfnamefont {J.}~\bibnamefont
  {Zaanen}}, \bibinfo {author} {\bibfnamefont {F.}~\bibnamefont {Baumberger}},
  \emph {et~al.},\ }\href@noop {} {\bibfield  {journal} {\bibinfo  {journal}
  {Nat. Phys.}\ }\textbf {\bibinfo {volume} {13}},\ \bibinfo {pages} {21}
  (\bibinfo {year} {2017})}\BibitemShut {NoStop}%
\bibitem [{\citenamefont {Dean}\ \emph {et~al.}(2016)\citenamefont {Dean},
  \citenamefont {Cao}, \citenamefont {Liu}, \citenamefont {Wall}, \citenamefont
  {Zhu}, \citenamefont {Mankowsky}, \citenamefont {Thampy}, \citenamefont
  {Chen}, \citenamefont {Vale}, \citenamefont {Casa} \emph
  {et~al.}}]{dean2016ultrafast}%
  \BibitemOpen
  \bibfield  {author} {\bibinfo {author} {\bibfnamefont {M.~P.~M.}\
  \bibnamefont {Dean}}, \bibinfo {author} {\bibfnamefont {Y.}~\bibnamefont
  {Cao}}, \bibinfo {author} {\bibfnamefont {X.}~\bibnamefont {Liu}}, \bibinfo
  {author} {\bibfnamefont {S.}~\bibnamefont {Wall}}, \bibinfo {author}
  {\bibfnamefont {D.}~\bibnamefont {Zhu}}, \bibinfo {author} {\bibfnamefont
  {R.}~\bibnamefont {Mankowsky}}, \bibinfo {author} {\bibfnamefont
  {V.}~\bibnamefont {Thampy}}, \bibinfo {author} {\bibfnamefont {X.~M.}\
  \bibnamefont {Chen}}, \bibinfo {author} {\bibfnamefont {J.~G.}\ \bibnamefont
  {Vale}}, \bibinfo {author} {\bibfnamefont {D.}~\bibnamefont {Casa}},  \emph
  {et~al.},\ }\href@noop {} {\bibfield  {journal} {\bibinfo  {journal} {Nat.
  Mater.}\ }\textbf {\bibinfo {volume} {15}},\ \bibinfo {pages} {601} (\bibinfo
  {year} {2016})}\BibitemShut {NoStop}%
\bibitem [{\citenamefont {Afanasiev}\ \emph {et~al.}(2019)\citenamefont
  {Afanasiev}, \citenamefont {Gatilova}, \citenamefont {Groenendijk},
  \citenamefont {Ivanov}, \citenamefont {Gibert}, \citenamefont {Gariglio},
  \citenamefont {Mentink}, \citenamefont {Li}, \citenamefont {Dasari},
  \citenamefont {Eckstein} \emph {et~al.}}]{afanasiev2019ultrafast}%
  \BibitemOpen
  \bibfield  {author} {\bibinfo {author} {\bibfnamefont {D.}~\bibnamefont
  {Afanasiev}}, \bibinfo {author} {\bibfnamefont {A.}~\bibnamefont {Gatilova}},
  \bibinfo {author} {\bibfnamefont {D.~J.}\ \bibnamefont {Groenendijk}},
  \bibinfo {author} {\bibfnamefont {B.~A.}\ \bibnamefont {Ivanov}}, \bibinfo
  {author} {\bibfnamefont {M.}~\bibnamefont {Gibert}}, \bibinfo {author}
  {\bibfnamefont {S.}~\bibnamefont {Gariglio}}, \bibinfo {author}
  {\bibfnamefont {J.}~\bibnamefont {Mentink}}, \bibinfo {author} {\bibfnamefont
  {J.}~\bibnamefont {Li}}, \bibinfo {author} {\bibfnamefont {N.}~\bibnamefont
  {Dasari}}, \bibinfo {author} {\bibfnamefont {M.}~\bibnamefont {Eckstein}},
  \emph {et~al.},\ }\href@noop {} {\bibfield  {journal} {\bibinfo  {journal}
  {Phys. Rev. X}\ }\textbf {\bibinfo {volume} {9}},\ \bibinfo {pages} {021020}
  (\bibinfo {year} {2019})}\BibitemShut {NoStop}%
\bibitem [{\citenamefont {Yang}\ \emph {et~al.}(2020)\citenamefont {Yang},
  \citenamefont {Pellatz}, \citenamefont {Wolf}, \citenamefont {Nandkishore},\
  and\ \citenamefont {Reznik}}]{yang2020ultrafast}%
  \BibitemOpen
  \bibfield  {author} {\bibinfo {author} {\bibfnamefont {J.-A.}\ \bibnamefont
  {Yang}}, \bibinfo {author} {\bibfnamefont {N.}~\bibnamefont {Pellatz}},
  \bibinfo {author} {\bibfnamefont {T.}~\bibnamefont {Wolf}}, \bibinfo {author}
  {\bibfnamefont {R.}~\bibnamefont {Nandkishore}}, \ and\ \bibinfo {author}
  {\bibfnamefont {D.}~\bibnamefont {Reznik}},\ }\href@noop {} {\bibfield
  {journal} {\bibinfo  {journal} {Nat. Commun.}\ }\textbf {\bibinfo {volume}
  {11}},\ \bibinfo {pages} {2548} (\bibinfo {year} {2020})}\BibitemShut
  {NoStop}%
\bibitem [{\citenamefont {Balzer}\ \emph {et~al.}(2015)\citenamefont {Balzer},
  \citenamefont {Wolf}, \citenamefont {McCulloch}, \citenamefont {Werner},\
  and\ \citenamefont {Eckstein}}]{balzer2015nonthermal}%
  \BibitemOpen
  \bibfield  {author} {\bibinfo {author} {\bibfnamefont {K.}~\bibnamefont
  {Balzer}}, \bibinfo {author} {\bibfnamefont {F.~A.}\ \bibnamefont {Wolf}},
  \bibinfo {author} {\bibfnamefont {I.~P.}\ \bibnamefont {McCulloch}}, \bibinfo
  {author} {\bibfnamefont {P.}~\bibnamefont {Werner}}, \ and\ \bibinfo {author}
  {\bibfnamefont {M.}~\bibnamefont {Eckstein}},\ }\href@noop {} {\bibfield
  {journal} {\bibinfo  {journal} {Phys. Rev. X}\ }\textbf {\bibinfo {volume}
  {5}},\ \bibinfo {pages} {031039} (\bibinfo {year} {2015})}\BibitemShut
  {NoStop}%
\bibitem [{\citenamefont {G{\"o}ssling}\ \emph {et~al.}(2008)\citenamefont
  {G{\"o}ssling}, \citenamefont {Schmitz}, \citenamefont {Roth}, \citenamefont
  {Haverkort}, \citenamefont {Lorenz}, \citenamefont {Mydosh}, \citenamefont
  {M{\"u}ller-Hartmann},\ and\ \citenamefont
  {Gr{\"u}ninger}}]{gossling2008mott}%
  \BibitemOpen
  \bibfield  {author} {\bibinfo {author} {\bibfnamefont {A.}~\bibnamefont
  {G{\"o}ssling}}, \bibinfo {author} {\bibfnamefont {R.}~\bibnamefont
  {Schmitz}}, \bibinfo {author} {\bibfnamefont {H.}~\bibnamefont {Roth}},
  \bibinfo {author} {\bibfnamefont {M.}~\bibnamefont {Haverkort}}, \bibinfo
  {author} {\bibfnamefont {T.}~\bibnamefont {Lorenz}}, \bibinfo {author}
  {\bibfnamefont {J.}~\bibnamefont {Mydosh}}, \bibinfo {author} {\bibfnamefont
  {E.}~\bibnamefont {M{\"u}ller-Hartmann}}, \ and\ \bibinfo {author}
  {\bibfnamefont {M.}~\bibnamefont {Gr{\"u}ninger}},\ }\href@noop {} {\bibfield
   {journal} {\bibinfo  {journal} {Phys. Rev. B}\ }\textbf {\bibinfo {volume}
  {78}},\ \bibinfo {pages} {075122} (\bibinfo {year} {2008})}\BibitemShut
  {NoStop}%
\bibitem [{\citenamefont {Novelli}\ \emph {et~al.}(2012)\citenamefont
  {Novelli}, \citenamefont {Fausti}, \citenamefont {Reul}, \citenamefont
  {Cilento}, \citenamefont {Van~Loosdrecht}, \citenamefont {Nugroho},
  \citenamefont {Palstra}, \citenamefont {Gr{\"u}ninger},\ and\ \citenamefont
  {Parmigiani}}]{novelli2012ultrafast}%
  \BibitemOpen
  \bibfield  {author} {\bibinfo {author} {\bibfnamefont {F.}~\bibnamefont
  {Novelli}}, \bibinfo {author} {\bibfnamefont {D.}~\bibnamefont {Fausti}},
  \bibinfo {author} {\bibfnamefont {J.}~\bibnamefont {Reul}}, \bibinfo {author}
  {\bibfnamefont {F.}~\bibnamefont {Cilento}}, \bibinfo {author} {\bibfnamefont
  {P.~H.}\ \bibnamefont {Van~Loosdrecht}}, \bibinfo {author} {\bibfnamefont
  {A.~A.}\ \bibnamefont {Nugroho}}, \bibinfo {author} {\bibfnamefont {T.~T.}\
  \bibnamefont {Palstra}}, \bibinfo {author} {\bibfnamefont {M.}~\bibnamefont
  {Gr{\"u}ninger}}, \ and\ \bibinfo {author} {\bibfnamefont {F.}~\bibnamefont
  {Parmigiani}},\ }\href@noop {} {\bibfield  {journal} {\bibinfo  {journal}
  {Phys. Rev. B}\ }\textbf {\bibinfo {volume} {86}},\ \bibinfo {pages} {165135}
  (\bibinfo {year} {2012})}\BibitemShut {NoStop}%
\bibitem [{\citenamefont {Kuo}\ \emph {et~al.}(2016)\citenamefont {Kuo},
  \citenamefont {Neumann}, \citenamefont {Balamurugan}, \citenamefont {Park},
  \citenamefont {Kang}, \citenamefont {Shiu}, \citenamefont {Kang},
  \citenamefont {Hong}, \citenamefont {Han}, \citenamefont {Noh} \emph
  {et~al.}}]{kuo2016exfoliation}%
  \BibitemOpen
  \bibfield  {author} {\bibinfo {author} {\bibfnamefont {C.-T.}\ \bibnamefont
  {Kuo}}, \bibinfo {author} {\bibfnamefont {M.}~\bibnamefont {Neumann}},
  \bibinfo {author} {\bibfnamefont {K.}~\bibnamefont {Balamurugan}}, \bibinfo
  {author} {\bibfnamefont {H.~J.}\ \bibnamefont {Park}}, \bibinfo {author}
  {\bibfnamefont {S.}~\bibnamefont {Kang}}, \bibinfo {author} {\bibfnamefont
  {H.~W.}\ \bibnamefont {Shiu}}, \bibinfo {author} {\bibfnamefont {J.~H.}\
  \bibnamefont {Kang}}, \bibinfo {author} {\bibfnamefont {B.~H.}\ \bibnamefont
  {Hong}}, \bibinfo {author} {\bibfnamefont {M.}~\bibnamefont {Han}}, \bibinfo
  {author} {\bibfnamefont {T.~W.}\ \bibnamefont {Noh}},  \emph {et~al.},\
  }\href@noop {} {\bibfield  {journal} {\bibinfo  {journal} {Sci. Rep.}\
  }\textbf {\bibinfo {volume} {6}},\ \bibinfo {pages} {20904} (\bibinfo {year}
  {2016})}\BibitemShut {NoStop}%
\bibitem [{\citenamefont {Kim}\ \emph {et~al.}(2019)\citenamefont {Kim},
  \citenamefont {Lim}, \citenamefont {Lee}, \citenamefont {Lee}, \citenamefont
  {Kim}, \citenamefont {Park}, \citenamefont {Jeon}, \citenamefont {Park},
  \citenamefont {Park},\ and\ \citenamefont {Cheong}}]{kim2019suppression}%
  \BibitemOpen
  \bibfield  {author} {\bibinfo {author} {\bibfnamefont {K.}~\bibnamefont
  {Kim}}, \bibinfo {author} {\bibfnamefont {S.~Y.}\ \bibnamefont {Lim}},
  \bibinfo {author} {\bibfnamefont {J.-U.}\ \bibnamefont {Lee}}, \bibinfo
  {author} {\bibfnamefont {S.}~\bibnamefont {Lee}}, \bibinfo {author}
  {\bibfnamefont {T.~Y.}\ \bibnamefont {Kim}}, \bibinfo {author} {\bibfnamefont
  {K.}~\bibnamefont {Park}}, \bibinfo {author} {\bibfnamefont {G.~S.}\
  \bibnamefont {Jeon}}, \bibinfo {author} {\bibfnamefont {C.-H.}\ \bibnamefont
  {Park}}, \bibinfo {author} {\bibfnamefont {J.-G.}\ \bibnamefont {Park}}, \
  and\ \bibinfo {author} {\bibfnamefont {H.}~\bibnamefont {Cheong}},\
  }\href@noop {} {\bibfield  {journal} {\bibinfo  {journal} {Nat. Commun.}\
  }\textbf {\bibinfo {volume} {10}},\ \bibinfo {pages} {345} (\bibinfo {year}
  {2019})}\BibitemShut {NoStop}%
\bibitem [{\citenamefont {Duvillaret}\ \emph {et~al.}(1996)\citenamefont
  {Duvillaret}, \citenamefont {Garet},\ and\ \citenamefont
  {Coutaz}}]{duvillaret1996reliable}%
  \BibitemOpen
  \bibfield  {author} {\bibinfo {author} {\bibfnamefont {L.}~\bibnamefont
  {Duvillaret}}, \bibinfo {author} {\bibfnamefont {F.}~\bibnamefont {Garet}}, \
  and\ \bibinfo {author} {\bibfnamefont {J.-L.}\ \bibnamefont {Coutaz}},\
  }\href@noop {} {\bibfield  {journal} {\bibinfo  {journal} {IEEE J. Sel. Top.
  Quantum Electron.}\ }\textbf {\bibinfo {volume} {2}},\ \bibinfo {pages} {739}
  (\bibinfo {year} {1996})}\BibitemShut {NoStop}%
\bibitem [{\citenamefont {Beard}\ \emph {et~al.}(2000)\citenamefont {Beard},
  \citenamefont {Turner},\ and\ \citenamefont
  {Schmuttenmaer}}]{beard2000transient}%
  \BibitemOpen
  \bibfield  {author} {\bibinfo {author} {\bibfnamefont {M.~C.}\ \bibnamefont
  {Beard}}, \bibinfo {author} {\bibfnamefont {G.~M.}\ \bibnamefont {Turner}}, \
  and\ \bibinfo {author} {\bibfnamefont {C.~A.}\ \bibnamefont
  {Schmuttenmaer}},\ }\href@noop {} {\bibfield  {journal} {\bibinfo  {journal}
  {Phys. Rev. B}\ }\textbf {\bibinfo {volume} {62}},\ \bibinfo {pages} {15764}
  (\bibinfo {year} {2000})}\BibitemShut {NoStop}%
\bibitem [{\citenamefont {Wildes}\ \emph {et~al.}(1998)\citenamefont {Wildes},
  \citenamefont {Roessli}, \citenamefont {Lebech},\ and\ \citenamefont
  {Godfrey}}]{wildes1998spin}%
  \BibitemOpen
  \bibfield  {author} {\bibinfo {author} {\bibfnamefont {A.~R.}\ \bibnamefont
  {Wildes}}, \bibinfo {author} {\bibfnamefont {B.}~\bibnamefont {Roessli}},
  \bibinfo {author} {\bibfnamefont {B.}~\bibnamefont {Lebech}}, \ and\ \bibinfo
  {author} {\bibfnamefont {K.~W.}\ \bibnamefont {Godfrey}},\ }\href@noop {}
  {\bibfield  {journal} {\bibinfo  {journal} {J. Phys.: Condens. Matter}\
  }\textbf {\bibinfo {volume} {10}},\ \bibinfo {pages} {6417} (\bibinfo {year}
  {1998})}\BibitemShut {NoStop}%
\bibitem [{\citenamefont {Lan{\c{c}}on}\ \emph {et~al.}(2016)\citenamefont
  {Lan{\c{c}}on}, \citenamefont {Walker}, \citenamefont {Ressouche},
  \citenamefont {Ouladdiaf}, \citenamefont {Rule}, \citenamefont {McIntyre},
  \citenamefont {Hicks}, \citenamefont {R{\o}nnow},\ and\ \citenamefont
  {Wildes}}]{lanccon2016magnetic}%
  \BibitemOpen
  \bibfield  {author} {\bibinfo {author} {\bibfnamefont {D.}~\bibnamefont
  {Lan{\c{c}}on}}, \bibinfo {author} {\bibfnamefont {H.~C.}\ \bibnamefont
  {Walker}}, \bibinfo {author} {\bibfnamefont {E.}~\bibnamefont {Ressouche}},
  \bibinfo {author} {\bibfnamefont {B.}~\bibnamefont {Ouladdiaf}}, \bibinfo
  {author} {\bibfnamefont {K.~C.}\ \bibnamefont {Rule}}, \bibinfo {author}
  {\bibfnamefont {G.~J.}\ \bibnamefont {McIntyre}}, \bibinfo {author}
  {\bibfnamefont {T.~J.}\ \bibnamefont {Hicks}}, \bibinfo {author}
  {\bibfnamefont {H.~M.}\ \bibnamefont {R{\o}nnow}}, \ and\ \bibinfo {author}
  {\bibfnamefont {A.~R.}\ \bibnamefont {Wildes}},\ }\href@noop {} {\bibfield
  {journal} {\bibinfo  {journal} {Phys. Rev. B}\ }\textbf {\bibinfo {volume}
  {94}},\ \bibinfo {pages} {214407} (\bibinfo {year} {2016})}\BibitemShut
  {NoStop}%
\bibitem [{\citenamefont {Brec}\ \emph {et~al.}(1979)\citenamefont {Brec},
  \citenamefont {Schleich}, \citenamefont {Ouvrard}, \citenamefont {Louisy},\
  and\ \citenamefont {Rouxel}}]{brec1979physical}%
  \BibitemOpen
  \bibfield  {author} {\bibinfo {author} {\bibfnamefont {R.}~\bibnamefont
  {Brec}}, \bibinfo {author} {\bibfnamefont {D.~M.}\ \bibnamefont {Schleich}},
  \bibinfo {author} {\bibfnamefont {G.}~\bibnamefont {Ouvrard}}, \bibinfo
  {author} {\bibfnamefont {A.}~\bibnamefont {Louisy}}, \ and\ \bibinfo {author}
  {\bibfnamefont {J.}~\bibnamefont {Rouxel}},\ }\href@noop {} {\bibfield
  {journal} {\bibinfo  {journal} {Inorg. Chem.}\ }\textbf {\bibinfo {volume}
  {18}},\ \bibinfo {pages} {1814} (\bibinfo {year} {1979})}\BibitemShut
  {NoStop}%
\bibitem [{\citenamefont {Grasso}\ \emph {et~al.}(1986)\citenamefont {Grasso},
  \citenamefont {Santangelo},\ and\ \citenamefont
  {Piacentini}}]{grasso1986optical}%
  \BibitemOpen
  \bibfield  {author} {\bibinfo {author} {\bibfnamefont {V.}~\bibnamefont
  {Grasso}}, \bibinfo {author} {\bibfnamefont {S.}~\bibnamefont {Santangelo}},
  \ and\ \bibinfo {author} {\bibfnamefont {M.}~\bibnamefont {Piacentini}},\
  }\href@noop {} {\bibfield  {journal} {\bibinfo  {journal} {Solid State Ion.}\
  }\textbf {\bibinfo {volume} {20}},\ \bibinfo {pages} {9} (\bibinfo {year}
  {1986})}\BibitemShut {NoStop}%
\bibitem [{\citenamefont {Banda}(1986)}]{banda1986opticalFePS3}%
  \BibitemOpen
  \bibfield  {author} {\bibinfo {author} {\bibfnamefont {E.~J. K.~B.}\
  \bibnamefont {Banda}},\ }\href@noop {} {\bibfield  {journal} {\bibinfo
  {journal} {Phys. Status Solidi B}\ }\textbf {\bibinfo {volume} {138}},\
  \bibinfo {pages} {K125} (\bibinfo {year} {1986})}\BibitemShut {NoStop}%
\bibitem [{\citenamefont {Grasso}\ \emph {et~al.}(1991)\citenamefont {Grasso},
  \citenamefont {Neri}, \citenamefont {Perillo}, \citenamefont {Silipigni},\
  and\ \citenamefont {Piacentini}}]{grasso1991optical}%
  \BibitemOpen
  \bibfield  {author} {\bibinfo {author} {\bibfnamefont {V.}~\bibnamefont
  {Grasso}}, \bibinfo {author} {\bibfnamefont {F.}~\bibnamefont {Neri}},
  \bibinfo {author} {\bibfnamefont {P.}~\bibnamefont {Perillo}}, \bibinfo
  {author} {\bibfnamefont {L.}~\bibnamefont {Silipigni}}, \ and\ \bibinfo
  {author} {\bibfnamefont {M.}~\bibnamefont {Piacentini}},\ }\href@noop {}
  {\bibfield  {journal} {\bibinfo  {journal} {Phys. Rev. B}\ }\textbf {\bibinfo
  {volume} {44}},\ \bibinfo {pages} {11060} (\bibinfo {year}
  {1991})}\BibitemShut {NoStop}%
\bibitem [{\citenamefont {Joy}\ and\ \citenamefont
  {Vasudevan}(1992)}]{joy1992optical}%
  \BibitemOpen
  \bibfield  {author} {\bibinfo {author} {\bibfnamefont {P.~A.}\ \bibnamefont
  {Joy}}\ and\ \bibinfo {author} {\bibfnamefont {S.}~\bibnamefont
  {Vasudevan}},\ }\href@noop {} {\bibfield  {journal} {\bibinfo  {journal}
  {Phys. Rev. B}\ }\textbf {\bibinfo {volume} {46}},\ \bibinfo {pages} {5134}
  (\bibinfo {year} {1992})}\BibitemShut {NoStop}%
\bibitem [{\citenamefont {Gnatchenko}\ \emph {et~al.}(2011)\citenamefont
  {Gnatchenko}, \citenamefont {Kachur}, \citenamefont {Piryatinskaya},
  \citenamefont {Vysochanskii},\ and\ \citenamefont
  {Gurzan}}]{gnatchenko2011exciton}%
  \BibitemOpen
  \bibfield  {author} {\bibinfo {author} {\bibfnamefont {S.~L.}\ \bibnamefont
  {Gnatchenko}}, \bibinfo {author} {\bibfnamefont {I.~S.}\ \bibnamefont
  {Kachur}}, \bibinfo {author} {\bibfnamefont {V.~G.}\ \bibnamefont
  {Piryatinskaya}}, \bibinfo {author} {\bibfnamefont {Y.~M.}\ \bibnamefont
  {Vysochanskii}}, \ and\ \bibinfo {author} {\bibfnamefont {M.~I.}\
  \bibnamefont {Gurzan}},\ }\href@noop {} {\bibfield  {journal} {\bibinfo
  {journal} {Low Temp. Phys.}\ }\textbf {\bibinfo {volume} {37}},\ \bibinfo
  {pages} {144} (\bibinfo {year} {2011})}\BibitemShut {NoStop}%
\bibitem [{\citenamefont {Furman}\ and\ \citenamefont
  {Tikhonravov}(1992)}]{furman1992basics}%
  \BibitemOpen
  \bibfield  {author} {\bibinfo {author} {\bibfnamefont {S.~A.}\ \bibnamefont
  {Furman}}\ and\ \bibinfo {author} {\bibfnamefont {A.~V.}\ \bibnamefont
  {Tikhonravov}},\ }\href@noop {} {\emph {\bibinfo {title} {{Basics of Optics
  of Multilayer Systems}}}}\ (\bibinfo  {publisher} {Editions Fronti{\`e}res,
  Gif-sur-Yvette},\ \bibinfo {year} {1992})\BibitemShut {NoStop}%
\bibitem [{\citenamefont {Karlsen}\ and\ \citenamefont
  {Hendry}(2019)}]{karlsen2019approximations}%
  \BibitemOpen
  \bibfield  {author} {\bibinfo {author} {\bibfnamefont {P.}~\bibnamefont
  {Karlsen}}\ and\ \bibinfo {author} {\bibfnamefont {E.}~\bibnamefont
  {Hendry}},\ }\href@noop {} {\ ,\ \bibinfo {pages} {Preprint at
  https://arXiv.org/abs/1902.08619} (\bibinfo {year} {2019})}\BibitemShut
  {NoStop}%
\bibitem [{\citenamefont {Beard}\ \emph {et~al.}(2001)\citenamefont {Beard},
  \citenamefont {Turner},\ and\ \citenamefont
  {Schmuttenmaer}}]{beard2001subpicosecond}%
  \BibitemOpen
  \bibfield  {author} {\bibinfo {author} {\bibfnamefont {M.~C.}\ \bibnamefont
  {Beard}}, \bibinfo {author} {\bibfnamefont {G.~M.}\ \bibnamefont {Turner}}, \
  and\ \bibinfo {author} {\bibfnamefont {C.~A.}\ \bibnamefont
  {Schmuttenmaer}},\ }\href@noop {} {\bibfield  {journal} {\bibinfo  {journal}
  {J. Appl. Phys.}\ }\textbf {\bibinfo {volume} {90}},\ \bibinfo {pages} {5915}
  (\bibinfo {year} {2001})}\BibitemShut {NoStop}%
\bibitem [{\citenamefont {Ulbricht}\ \emph {et~al.}(2011)\citenamefont
  {Ulbricht}, \citenamefont {Hendry}, \citenamefont {Shan}, \citenamefont
  {Heinz},\ and\ \citenamefont {Bonn}}]{ulbricht2011carrier}%
  \BibitemOpen
  \bibfield  {author} {\bibinfo {author} {\bibfnamefont {R.}~\bibnamefont
  {Ulbricht}}, \bibinfo {author} {\bibfnamefont {E.}~\bibnamefont {Hendry}},
  \bibinfo {author} {\bibfnamefont {J.}~\bibnamefont {Shan}}, \bibinfo {author}
  {\bibfnamefont {T.~F.}\ \bibnamefont {Heinz}}, \ and\ \bibinfo {author}
  {\bibfnamefont {M.}~\bibnamefont {Bonn}},\ }\href@noop {} {\bibfield
  {journal} {\bibinfo  {journal} {Rev. Mod. Phys.}\ }\textbf {\bibinfo {volume}
  {83}},\ \bibinfo {pages} {543} (\bibinfo {year} {2011})}\BibitemShut
  {NoStop}%
\bibitem [{\citenamefont {Larsen}\ \emph {et~al.}(2011)\citenamefont {Larsen},
  \citenamefont {Cooke},\ and\ \citenamefont {Jepsen}}]{larsen2011finite}%
  \BibitemOpen
  \bibfield  {author} {\bibinfo {author} {\bibfnamefont {C.}~\bibnamefont
  {Larsen}}, \bibinfo {author} {\bibfnamefont {D.~G.}\ \bibnamefont {Cooke}}, \
  and\ \bibinfo {author} {\bibfnamefont {P.~U.}\ \bibnamefont {Jepsen}},\
  }\href@noop {} {\bibfield  {journal} {\bibinfo  {journal} {J. Opt. Soc. Am.
  B}\ }\textbf {\bibinfo {volume} {28}},\ \bibinfo {pages} {1308} (\bibinfo
  {year} {2011})}\BibitemShut {NoStop}%
\bibitem [{\citenamefont {Kalashnikova}\ \emph {et~al.}(2007)\citenamefont
  {Kalashnikova}, \citenamefont {Kimel}, \citenamefont {Pisarev}, \citenamefont
  {Gridnev}, \citenamefont {Kirilyuk},\ and\ \citenamefont
  {Rasing}}]{kalashnikova2007impulsive}%
  \BibitemOpen
  \bibfield  {author} {\bibinfo {author} {\bibfnamefont {A.~M.}\ \bibnamefont
  {Kalashnikova}}, \bibinfo {author} {\bibfnamefont {A.~V.}\ \bibnamefont
  {Kimel}}, \bibinfo {author} {\bibfnamefont {R.~V.}\ \bibnamefont {Pisarev}},
  \bibinfo {author} {\bibfnamefont {V.~N.}\ \bibnamefont {Gridnev}}, \bibinfo
  {author} {\bibfnamefont {A.}~\bibnamefont {Kirilyuk}}, \ and\ \bibinfo
  {author} {\bibfnamefont {T.}~\bibnamefont {Rasing}},\ }\href@noop {}
  {\bibfield  {journal} {\bibinfo  {journal} {Phys. Rev. Lett.}\ }\textbf
  {\bibinfo {volume} {99}},\ \bibinfo {pages} {167205} (\bibinfo {year}
  {2007})}\BibitemShut {NoStop}%
\bibitem [{\citenamefont {Kimel}\ \emph {et~al.}(2005)\citenamefont {Kimel},
  \citenamefont {Kirilyuk}, \citenamefont {Usachev}, \citenamefont {Pisarev},
  \citenamefont {Balbashov},\ and\ \citenamefont
  {Rasing}}]{kimel2005ultrafast}%
  \BibitemOpen
  \bibfield  {author} {\bibinfo {author} {\bibfnamefont {A.~V.}\ \bibnamefont
  {Kimel}}, \bibinfo {author} {\bibfnamefont {A.}~\bibnamefont {Kirilyuk}},
  \bibinfo {author} {\bibfnamefont {P.~A.}\ \bibnamefont {Usachev}}, \bibinfo
  {author} {\bibfnamefont {R.~V.}\ \bibnamefont {Pisarev}}, \bibinfo {author}
  {\bibfnamefont {A.~M.}\ \bibnamefont {Balbashov}}, \ and\ \bibinfo {author}
  {\bibfnamefont {T.}~\bibnamefont {Rasing}},\ }\href@noop {} {\bibfield
  {journal} {\bibinfo  {journal} {Nature}\ }\textbf {\bibinfo {volume} {435}},\
  \bibinfo {pages} {655} (\bibinfo {year} {2005})}\BibitemShut {NoStop}%
\bibitem [{\citenamefont {Kaindl}\ \emph {et~al.}(2003)\citenamefont {Kaindl},
  \citenamefont {Carnahan}, \citenamefont {H{\"a}gele}, \citenamefont
  {L{\"o}venich},\ and\ \citenamefont {Chemla}}]{kaindl2003ultrafast}%
  \BibitemOpen
  \bibfield  {author} {\bibinfo {author} {\bibfnamefont {R.~A.}\ \bibnamefont
  {Kaindl}}, \bibinfo {author} {\bibfnamefont {M.~A.}\ \bibnamefont
  {Carnahan}}, \bibinfo {author} {\bibfnamefont {D.}~\bibnamefont
  {H{\"a}gele}}, \bibinfo {author} {\bibfnamefont {R.}~\bibnamefont
  {L{\"o}venich}}, \ and\ \bibinfo {author} {\bibfnamefont {D.~S.}\
  \bibnamefont {Chemla}},\ }\href@noop {} {\bibfield  {journal} {\bibinfo
  {journal} {Nature}\ }\textbf {\bibinfo {volume} {423}},\ \bibinfo {pages}
  {734} (\bibinfo {year} {2003})}\BibitemShut {NoStop}%
\bibitem [{\citenamefont {Huber}\ \emph {et~al.}(2006)\citenamefont {Huber},
  \citenamefont {Schmid}, \citenamefont {Shen}, \citenamefont {Chemla},\ and\
  \citenamefont {Kaindl}}]{huber2006stimulated}%
  \BibitemOpen
  \bibfield  {author} {\bibinfo {author} {\bibfnamefont {R.}~\bibnamefont
  {Huber}}, \bibinfo {author} {\bibfnamefont {B.~A.}\ \bibnamefont {Schmid}},
  \bibinfo {author} {\bibfnamefont {Y.~R.}\ \bibnamefont {Shen}}, \bibinfo
  {author} {\bibfnamefont {D.~S.}\ \bibnamefont {Chemla}}, \ and\ \bibinfo
  {author} {\bibfnamefont {R.~A.}\ \bibnamefont {Kaindl}},\ }\href@noop {}
  {\bibfield  {journal} {\bibinfo  {journal} {Phys. Rev. Lett.}\ }\textbf
  {\bibinfo {volume} {96}},\ \bibinfo {pages} {017402} (\bibinfo {year}
  {2006})}\BibitemShut {NoStop}%
\bibitem [{\citenamefont {Kuzmenko}(2005)}]{kuzmenko2005kramers}%
  \BibitemOpen
  \bibfield  {author} {\bibinfo {author} {\bibfnamefont {A.~B.}\ \bibnamefont
  {Kuzmenko}},\ }\href@noop {} {\bibfield  {journal} {\bibinfo  {journal} {Rev.
  Sci. Instrum.}\ }\textbf {\bibinfo {volume} {76}},\ \bibinfo {pages} {083108}
  (\bibinfo {year} {2005})}\BibitemShut {NoStop}%
\bibitem [{\citenamefont {Lane}\ and\ \citenamefont
  {Zhu}(2020)}]{lane2020thickness}%
  \BibitemOpen
  \bibfield  {author} {\bibinfo {author} {\bibfnamefont {C.}~\bibnamefont
  {Lane}}\ and\ \bibinfo {author} {\bibfnamefont {J.-X.}\ \bibnamefont {Zhu}},\
  }\href@noop {} {\bibfield  {journal} {\bibinfo  {journal} {Phys. Rev. B}\
  }\textbf {\bibinfo {volume} {102}},\ \bibinfo {pages} {075124} (\bibinfo
  {year} {2020})}\BibitemShut {NoStop}%
\bibitem [{\citenamefont {Kampfrath}\ \emph {et~al.}(2005)\citenamefont
  {Kampfrath}, \citenamefont {Perfetti}, \citenamefont {Schapper},
  \citenamefont {Frischkorn},\ and\ \citenamefont
  {Wolf}}]{kampfrath2005strongly}%
  \BibitemOpen
  \bibfield  {author} {\bibinfo {author} {\bibfnamefont {T.}~\bibnamefont
  {Kampfrath}}, \bibinfo {author} {\bibfnamefont {L.}~\bibnamefont {Perfetti}},
  \bibinfo {author} {\bibfnamefont {F.}~\bibnamefont {Schapper}}, \bibinfo
  {author} {\bibfnamefont {C.}~\bibnamefont {Frischkorn}}, \ and\ \bibinfo
  {author} {\bibfnamefont {M.}~\bibnamefont {Wolf}},\ }\href@noop {} {\bibfield
   {journal} {\bibinfo  {journal} {Phys. Rev. Lett.}\ }\textbf {\bibinfo
  {volume} {95}},\ \bibinfo {pages} {187403} (\bibinfo {year}
  {2005})}\BibitemShut {NoStop}%
\bibitem [{\citenamefont {Hall}(1952)}]{hall1952electron}%
  \BibitemOpen
  \bibfield  {author} {\bibinfo {author} {\bibfnamefont {R.~N.}\ \bibnamefont
  {Hall}},\ }\href@noop {} {\bibfield  {journal} {\bibinfo  {journal} {Phys.
  Rev.}\ }\textbf {\bibinfo {volume} {87}},\ \bibinfo {pages} {387} (\bibinfo
  {year} {1952})}\BibitemShut {NoStop}%
\bibitem [{\citenamefont {Shockley}\ and\ \citenamefont
  {Read~Jr}(1952)}]{shockley1952statistics}%
  \BibitemOpen
  \bibfield  {author} {\bibinfo {author} {\bibfnamefont {W.}~\bibnamefont
  {Shockley}}\ and\ \bibinfo {author} {\bibfnamefont {W.~T.}\ \bibnamefont
  {Read~Jr}},\ }\href@noop {} {\bibfield  {journal} {\bibinfo  {journal} {Phys.
  Rev.}\ }\textbf {\bibinfo {volume} {87}},\ \bibinfo {pages} {835} (\bibinfo
  {year} {1952})}\BibitemShut {NoStop}%
\bibitem [{\citenamefont {Linnros}(1998)}]{linnros1998carrier}%
  \BibitemOpen
  \bibfield  {author} {\bibinfo {author} {\bibfnamefont {J.}~\bibnamefont
  {Linnros}},\ }\href@noop {} {\bibfield  {journal} {\bibinfo  {journal} {J.
  Appl. Phys.}\ }\textbf {\bibinfo {volume} {84}},\ \bibinfo {pages} {275}
  (\bibinfo {year} {1998})}\BibitemShut {NoStop}%
\bibitem [{\citenamefont {Lui}\ and\ \citenamefont
  {Hegmann}(2003)}]{lui2003fluence}%
  \BibitemOpen
  \bibfield  {author} {\bibinfo {author} {\bibfnamefont {K.~P.~H.}\
  \bibnamefont {Lui}}\ and\ \bibinfo {author} {\bibfnamefont {F.~A.}\
  \bibnamefont {Hegmann}},\ }\href@noop {} {\bibfield  {journal} {\bibinfo
  {journal} {J. Appl. Phys.}\ }\textbf {\bibinfo {volume} {93}},\ \bibinfo
  {pages} {9012} (\bibinfo {year} {2003})}\BibitemShut {NoStop}%
\bibitem [{\citenamefont {Okamoto}\ \emph {et~al.}(2011)\citenamefont
  {Okamoto}, \citenamefont {Miyagoe}, \citenamefont {Kobayashi}, \citenamefont
  {Uemura}, \citenamefont {Nishioka}, \citenamefont {Matsuzaki}, \citenamefont
  {Sawa},\ and\ \citenamefont {Tokura}}]{okamoto2011photoinduced}%
  \BibitemOpen
  \bibfield  {author} {\bibinfo {author} {\bibfnamefont {H.}~\bibnamefont
  {Okamoto}}, \bibinfo {author} {\bibfnamefont {T.}~\bibnamefont {Miyagoe}},
  \bibinfo {author} {\bibfnamefont {K.}~\bibnamefont {Kobayashi}}, \bibinfo
  {author} {\bibfnamefont {H.}~\bibnamefont {Uemura}}, \bibinfo {author}
  {\bibfnamefont {H.}~\bibnamefont {Nishioka}}, \bibinfo {author}
  {\bibfnamefont {H.}~\bibnamefont {Matsuzaki}}, \bibinfo {author}
  {\bibfnamefont {A.}~\bibnamefont {Sawa}}, \ and\ \bibinfo {author}
  {\bibfnamefont {Y.}~\bibnamefont {Tokura}},\ }\href@noop {} {\bibfield
  {journal} {\bibinfo  {journal} {Phys. Rev. B}\ }\textbf {\bibinfo {volume}
  {83}},\ \bibinfo {pages} {125102} (\bibinfo {year} {2011})}\BibitemShut
  {NoStop}%
\bibitem [{\citenamefont {Hendry}\ \emph {et~al.}(2004)\citenamefont {Hendry},
  \citenamefont {Wang}, \citenamefont {Shan}, \citenamefont {Heinz},\ and\
  \citenamefont {Bonn}}]{hendry2004electron}%
  \BibitemOpen
  \bibfield  {author} {\bibinfo {author} {\bibfnamefont {E.}~\bibnamefont
  {Hendry}}, \bibinfo {author} {\bibfnamefont {F.}~\bibnamefont {Wang}},
  \bibinfo {author} {\bibfnamefont {J.}~\bibnamefont {Shan}}, \bibinfo {author}
  {\bibfnamefont {T.~F.}\ \bibnamefont {Heinz}}, \ and\ \bibinfo {author}
  {\bibfnamefont {M.}~\bibnamefont {Bonn}},\ }\href@noop {} {\bibfield
  {journal} {\bibinfo  {journal} {Phys. Rev. B}\ }\textbf {\bibinfo {volume}
  {69}},\ \bibinfo {pages} {081101} (\bibinfo {year} {2004})}\BibitemShut
  {NoStop}%
\bibitem [{\citenamefont {Dean}\ \emph {et~al.}(2011)\citenamefont {Dean},
  \citenamefont {Petersen}, \citenamefont {Fausti}, \citenamefont {Tobey},
  \citenamefont {Kaiser}, \citenamefont {Gasparov}, \citenamefont {Berger},\
  and\ \citenamefont {Cavalleri}}]{dean2011polaronic}%
  \BibitemOpen
  \bibfield  {author} {\bibinfo {author} {\bibfnamefont {N.}~\bibnamefont
  {Dean}}, \bibinfo {author} {\bibfnamefont {J.~C.}\ \bibnamefont {Petersen}},
  \bibinfo {author} {\bibfnamefont {D.}~\bibnamefont {Fausti}}, \bibinfo
  {author} {\bibfnamefont {R.~I.}\ \bibnamefont {Tobey}}, \bibinfo {author}
  {\bibfnamefont {S.}~\bibnamefont {Kaiser}}, \bibinfo {author} {\bibfnamefont
  {L.~V.}\ \bibnamefont {Gasparov}}, \bibinfo {author} {\bibfnamefont
  {H.}~\bibnamefont {Berger}}, \ and\ \bibinfo {author} {\bibfnamefont
  {A.}~\bibnamefont {Cavalleri}},\ }\href@noop {} {\bibfield  {journal}
  {\bibinfo  {journal} {Phys. Rev. Lett.}\ }\textbf {\bibinfo {volume} {106}},\
  \bibinfo {pages} {016401} (\bibinfo {year} {2011})}\BibitemShut {NoStop}%
\bibitem [{\citenamefont {Emin}(2013)}]{emin2013polarons}%
  \BibitemOpen
  \bibfield  {author} {\bibinfo {author} {\bibfnamefont {D.}~\bibnamefont
  {Emin}},\ }\href@noop {} {\emph {\bibinfo {title} {Polarons}}}\ (\bibinfo
  {publisher} {Cambridge Univ. Press, Cambridge},\ \bibinfo {year}
  {2013})\BibitemShut {NoStop}%
\bibitem [{\citenamefont {Piacentini}\ \emph {et~al.}(1982)\citenamefont
  {Piacentini}, \citenamefont {Khumalo}, \citenamefont {Olson}, \citenamefont
  {Anderegg},\ and\ \citenamefont {Lynch}}]{piacentini1982optical}%
  \BibitemOpen
  \bibfield  {author} {\bibinfo {author} {\bibfnamefont {M.}~\bibnamefont
  {Piacentini}}, \bibinfo {author} {\bibfnamefont {F.~S.}\ \bibnamefont
  {Khumalo}}, \bibinfo {author} {\bibfnamefont {C.~G.}\ \bibnamefont {Olson}},
  \bibinfo {author} {\bibfnamefont {J.~W.}\ \bibnamefont {Anderegg}}, \ and\
  \bibinfo {author} {\bibfnamefont {D.~W.}\ \bibnamefont {Lynch}},\ }\href@noop
  {} {\bibfield  {journal} {\bibinfo  {journal} {Chem. Phys.}\ }\textbf
  {\bibinfo {volume} {65}},\ \bibinfo {pages} {289} (\bibinfo {year}
  {1982})}\BibitemShut {NoStop}%
\bibitem [{\citenamefont {Catalano}\ \emph {et~al.}(1975)\citenamefont
  {Catalano}, \citenamefont {Cingolani}, \citenamefont {Ferrara},\ and\
  \citenamefont {Minafra}}]{catalano1975luminescence}%
  \BibitemOpen
  \bibfield  {author} {\bibinfo {author} {\bibfnamefont {I.~M.}\ \bibnamefont
  {Catalano}}, \bibinfo {author} {\bibfnamefont {A.}~\bibnamefont {Cingolani}},
  \bibinfo {author} {\bibfnamefont {M.}~\bibnamefont {Ferrara}}, \ and\
  \bibinfo {author} {\bibfnamefont {A.}~\bibnamefont {Minafra}},\ }\href@noop
  {} {\bibfield  {journal} {\bibinfo  {journal} {Phys. Status Solidi B}\
  }\textbf {\bibinfo {volume} {68}},\ \bibinfo {pages} {341} (\bibinfo {year}
  {1975})}\BibitemShut {NoStop}%
\bibitem [{\citenamefont {Hase}\ \emph {et~al.}(2003)\citenamefont {Hase},
  \citenamefont {Kitajima}, \citenamefont {Constantinescu},\ and\ \citenamefont
  {Petek}}]{hase2003birth}%
  \BibitemOpen
  \bibfield  {author} {\bibinfo {author} {\bibfnamefont {M.}~\bibnamefont
  {Hase}}, \bibinfo {author} {\bibfnamefont {M.}~\bibnamefont {Kitajima}},
  \bibinfo {author} {\bibfnamefont {A.~M.}\ \bibnamefont {Constantinescu}}, \
  and\ \bibinfo {author} {\bibfnamefont {H.}~\bibnamefont {Petek}},\
  }\href@noop {} {\bibfield  {journal} {\bibinfo  {journal} {Nature}\ }\textbf
  {\bibinfo {volume} {426}},\ \bibinfo {pages} {51} (\bibinfo {year}
  {2003})}\BibitemShut {NoStop}%
\bibitem [{\citenamefont {Almeida}\ \emph {et~al.}(2004)\citenamefont
  {Almeida}, \citenamefont {Barrios}, \citenamefont {Panepucci},\ and\
  \citenamefont {Lipson}}]{almeida2004all}%
  \BibitemOpen
  \bibfield  {author} {\bibinfo {author} {\bibfnamefont {V.~R.}\ \bibnamefont
  {Almeida}}, \bibinfo {author} {\bibfnamefont {C.~A.}\ \bibnamefont
  {Barrios}}, \bibinfo {author} {\bibfnamefont {R.~R.}\ \bibnamefont
  {Panepucci}}, \ and\ \bibinfo {author} {\bibfnamefont {M.}~\bibnamefont
  {Lipson}},\ }\href@noop {} {\bibfield  {journal} {\bibinfo  {journal}
  {Nature}\ }\textbf {\bibinfo {volume} {431}},\ \bibinfo {pages} {1081}
  (\bibinfo {year} {2004})}\BibitemShut {NoStop}%
\bibitem [{\citenamefont {Rivas}\ \emph {et~al.}(2006)\citenamefont {Rivas},
  \citenamefont {S{\'a}nchez-Gil}, \citenamefont {Kuttge}, \citenamefont
  {Bolivar},\ and\ \citenamefont {Kurz}}]{rivas2006optically}%
  \BibitemOpen
  \bibfield  {author} {\bibinfo {author} {\bibfnamefont {J.~G.}\ \bibnamefont
  {Rivas}}, \bibinfo {author} {\bibfnamefont {J.~A.}\ \bibnamefont
  {S{\'a}nchez-Gil}}, \bibinfo {author} {\bibfnamefont {M.}~\bibnamefont
  {Kuttge}}, \bibinfo {author} {\bibfnamefont {P.~H.}\ \bibnamefont {Bolivar}},
  \ and\ \bibinfo {author} {\bibfnamefont {H.}~\bibnamefont {Kurz}},\
  }\href@noop {} {\bibfield  {journal} {\bibinfo  {journal} {Phys. Rev. B}\
  }\textbf {\bibinfo {volume} {74}},\ \bibinfo {pages} {245324} (\bibinfo
  {year} {2006})}\BibitemShut {NoStop}%
\bibitem [{\citenamefont {Haug}\ and\ \citenamefont
  {Schmitt-Rink}(1985)}]{haug1985basic}%
  \BibitemOpen
  \bibfield  {author} {\bibinfo {author} {\bibfnamefont {H.}~\bibnamefont
  {Haug}}\ and\ \bibinfo {author} {\bibfnamefont {S.}~\bibnamefont
  {Schmitt-Rink}},\ }\href@noop {} {\bibfield  {journal} {\bibinfo  {journal}
  {J. Opt. Soc. Am. B}\ }\textbf {\bibinfo {volume} {2}},\ \bibinfo {pages}
  {1135} (\bibinfo {year} {1985})}\BibitemShut {NoStop}%
\bibitem [{\citenamefont {Ishioka}\ and\ \citenamefont
  {Misochko}(2010)}]{ishioka2010coherent}%
  \BibitemOpen
  \bibfield  {author} {\bibinfo {author} {\bibfnamefont {K.}~\bibnamefont
  {Ishioka}}\ and\ \bibinfo {author} {\bibfnamefont {O.~V.}\ \bibnamefont
  {Misochko}},\ }in\ \href@noop {} {\emph {\bibinfo {booktitle} {Progress in
  Ultrafast Intense Laser Science}}}\ (\bibinfo  {publisher} {Springer},\
  \bibinfo {year} {2010})\ pp.\ \bibinfo {pages} {23--46}\BibitemShut {NoStop}%
\bibitem [{\citenamefont {Garrett}\ \emph {et~al.}(1997)\citenamefont
  {Garrett}, \citenamefont {Whitaker}, \citenamefont {Sood},\ and\
  \citenamefont {Merlin}}]{garrett1997ultrafast}%
  \BibitemOpen
  \bibfield  {author} {\bibinfo {author} {\bibfnamefont {G.~A.}\ \bibnamefont
  {Garrett}}, \bibinfo {author} {\bibfnamefont {J.~F.}\ \bibnamefont
  {Whitaker}}, \bibinfo {author} {\bibfnamefont {A.~K.}\ \bibnamefont {Sood}},
  \ and\ \bibinfo {author} {\bibfnamefont {R.}~\bibnamefont {Merlin}},\
  }\href@noop {} {\bibfield  {journal} {\bibinfo  {journal} {Opt. Express}\
  }\textbf {\bibinfo {volume} {1}},\ \bibinfo {pages} {385} (\bibinfo {year}
  {1997})}\BibitemShut {NoStop}%
\bibitem [{\citenamefont {Liao}\ \emph {et~al.}(2016)\citenamefont {Liao},
  \citenamefont {Maznev}, \citenamefont {Nelson},\ and\ \citenamefont
  {Chen}}]{liao2016photo}%
  \BibitemOpen
  \bibfield  {author} {\bibinfo {author} {\bibfnamefont {B.}~\bibnamefont
  {Liao}}, \bibinfo {author} {\bibfnamefont {A.~A.}\ \bibnamefont {Maznev}},
  \bibinfo {author} {\bibfnamefont {K.~A.}\ \bibnamefont {Nelson}}, \ and\
  \bibinfo {author} {\bibfnamefont {G.}~\bibnamefont {Chen}},\ }\href@noop {}
  {\bibfield  {journal} {\bibinfo  {journal} {Nat. Commun.}\ }\textbf {\bibinfo
  {volume} {7}},\ \bibinfo {pages} {13174} (\bibinfo {year}
  {2016})}\BibitemShut {NoStop}%
\bibitem [{\citenamefont {Zeiger}\ \emph {et~al.}(1992)\citenamefont {Zeiger},
  \citenamefont {Vidal}, \citenamefont {Cheng}, \citenamefont {Ippen},
  \citenamefont {Dresselhaus},\ and\ \citenamefont
  {Dresselhaus}}]{zeiger1992theory}%
  \BibitemOpen
  \bibfield  {author} {\bibinfo {author} {\bibfnamefont {H.~J.}\ \bibnamefont
  {Zeiger}}, \bibinfo {author} {\bibfnamefont {J.}~\bibnamefont {Vidal}},
  \bibinfo {author} {\bibfnamefont {T.~K.}\ \bibnamefont {Cheng}}, \bibinfo
  {author} {\bibfnamefont {E.~P.}\ \bibnamefont {Ippen}}, \bibinfo {author}
  {\bibfnamefont {G.}~\bibnamefont {Dresselhaus}}, \ and\ \bibinfo {author}
  {\bibfnamefont {M.~S.}\ \bibnamefont {Dresselhaus}},\ }\href@noop {}
  {\bibfield  {journal} {\bibinfo  {journal} {Phys. Rev. B}\ }\textbf {\bibinfo
  {volume} {45}},\ \bibinfo {pages} {768} (\bibinfo {year} {1992})}\BibitemShut
  {NoStop}%
\bibitem [{\citenamefont {Stevens}\ \emph {et~al.}(2002)\citenamefont
  {Stevens}, \citenamefont {Kuhl},\ and\ \citenamefont
  {Merlin}}]{stevens2002coherent}%
  \BibitemOpen
  \bibfield  {author} {\bibinfo {author} {\bibfnamefont {T.~E.}\ \bibnamefont
  {Stevens}}, \bibinfo {author} {\bibfnamefont {J.}~\bibnamefont {Kuhl}}, \
  and\ \bibinfo {author} {\bibfnamefont {R.}~\bibnamefont {Merlin}},\
  }\href@noop {} {\bibfield  {journal} {\bibinfo  {journal} {Phys. Rev. B}\
  }\textbf {\bibinfo {volume} {65}},\ \bibinfo {pages} {144304} (\bibinfo
  {year} {2002})}\BibitemShut {NoStop}%
\bibitem [{\citenamefont {Ruello}\ and\ \citenamefont
  {Gusev}(2015)}]{ruello2015physical}%
  \BibitemOpen
  \bibfield  {author} {\bibinfo {author} {\bibfnamefont {P.}~\bibnamefont
  {Ruello}}\ and\ \bibinfo {author} {\bibfnamefont {V.~E.}\ \bibnamefont
  {Gusev}},\ }\href@noop {} {\bibfield  {journal} {\bibinfo  {journal}
  {Ultrasonics}\ }\textbf {\bibinfo {volume} {56}},\ \bibinfo {pages} {21}
  (\bibinfo {year} {2015})}\BibitemShut {NoStop}%
\bibitem [{\citenamefont {Ishioka}\ \emph {et~al.}(2008)\citenamefont
  {Ishioka}, \citenamefont {Hase}, \citenamefont {Kitajima}, \citenamefont
  {Wirtz}, \citenamefont {Rubio},\ and\ \citenamefont
  {Petek}}]{ishioka2008ultrafast}%
  \BibitemOpen
  \bibfield  {author} {\bibinfo {author} {\bibfnamefont {K.}~\bibnamefont
  {Ishioka}}, \bibinfo {author} {\bibfnamefont {M.}~\bibnamefont {Hase}},
  \bibinfo {author} {\bibfnamefont {M.}~\bibnamefont {Kitajima}}, \bibinfo
  {author} {\bibfnamefont {L.}~\bibnamefont {Wirtz}}, \bibinfo {author}
  {\bibfnamefont {A.}~\bibnamefont {Rubio}}, \ and\ \bibinfo {author}
  {\bibfnamefont {H.}~\bibnamefont {Petek}},\ }\href@noop {} {\bibfield
  {journal} {\bibinfo  {journal} {Phys. Rev. B}\ }\textbf {\bibinfo {volume}
  {77}},\ \bibinfo {pages} {121402(R)} (\bibinfo {year} {2008})}\BibitemShut
  {NoStop}%
\end{thebibliography}%

\end{document}